\documentclass[11pt]{article}
\usepackage{amssymb}
\usepackage{graphicx}
\usepackage{amsmath}
\newtheorem{theorem}{Theorem}
\newtheorem{acknowledgement}[theorem]{Acknowledgement}

\input{tcilatex}
\makeatletter \@addtoreset{equation}{section}

\newcommand{\be}{\begin{equation}}
\newcommand{\ee}{\end{equation}}
\newcommand{\bea}{\begin{eqnarray}}
\newcommand{\eea}{\end{eqnarray}}

\begin{document}

\title{%
\rightline{\mbox {\normalsize
{Lab/UFR-HEP0504/GNPHE/0504/VACBT/0504}}} 
\textbf{Superstring Theory on pp
Waves\\ with ADE Geometries}}
\author{R Abounasr$^{1,2}$,\ A. Belhaj$^{3,5,}$\thanks{%
abelhaj@uottawa.ca}\ ,\ J. Rasmussen$^{4,}$\thanks{%
rasmusse@crm.umontreal.ca}\ ,\ E.H. Saidi$^{2,5,}$\thanks{%
esaidi@ictp.trieste.it} \\[.3cm]
{$^{1}$\textit{\small National Grouping of High Energy Physics,
GNPHE, Faculty of Sciences, Rabat, Morocco }}\\[.1cm]
{$^2$\textit{\small Lab/UFR HEP, Department of Physics, Faculty of Sciences,
Rabat, Morocco}} \\[.1cm]
{$^3$\textit{\small Department of Mathematics and Statistics, Ottawa
University}}\\
\textit{\small 585 King Edward Ave., Ottawa, ON, Canada, K1N 6N5} \\[.1cm]
{$^4$\textit{\small Department of Mathematics and Statistics, Concordia
University}} \\
\textit{\small 7141 Sherbrooke St. W, Montr\'eal, PQ, Canada H4B 1R6}\\[.1cm]
{$^{5}$\textit{\small Virtual Centre for basic Sciences \& Technology, 
VACBT,}}\\
{\textit{{\small \ focal point: Lab/UFR-Physique des Haute Energies, 
Facult\'{e} des Sciences, Rabat, Morocco}}}}
\maketitle

\begin{abstract}
We study the BMN correspondence between certain Penrose limits of type IIB
superstrings on pp-wave orbifolds with $ADE$ geometries, and the set of
four-dimensional $\mathcal{N}=2$ superconformal field theories constructed
as quiver gauge models classified by finite $ADE$ Lie algebras and affine 
$\widehat{ADE}$ Kac-Moody algebras. These models have 16 preserved
supercharges and are based on systems of D3-branes and wrapped D5- and
D7-branes. We derive explicitly the metrics of these pp-wave orbifolds and
show that the BMN extension requires, in addition to D5-D5 open strings in
bi-fundamental representations, D5-D7 open strings involving orientifolds
with $Sp(N)$ gauge symmetry. We also give the correspondence rule between
leading string states and gauge-invariant operators in the $\mathcal{N}=2$
quiver gauge models. 
\\[.2cm]
\textit{Keywords}: Superstring theory, AdS/CFT correspondence, pp waves, $ADE$
orbifolds, toric geometry, mirror geometry, four-dimensional $\mathcal{N}=2$
superconformal quiver models.
\end{abstract}
\newpage
\tableofcontents

\newpage

\section{Introduction}

\quad\ The AdS/CFT correspondence relates the spectrum of type IIB
superstring theory on $AdS_{5}\times S^{5}$
to the spectrum of single-trace operators in the large-$N$ limit of 
$\mathcal{N}=4$ $U(N)$ SYM$_{4}$ gauge theory on the boundary of $AdS_{5}$
\cite{M1,GKP,W1,Pet,KW,AGMOO,Ve,K,M2}. The proposal by Berenstein, Maldacena
and Nastase (BMN) predicts even more \cite{BMN}, as it offers one to derive
the spectrum of the superstring theory from the gauge-theory point of view,
not only on flat space as in the AdS/CFT correspondence, but also on a
plane-wave background obtained as a Penrose limit of $AdS_{5}\times S^{5}$.
The type IIB superstring states on plane-polarized (pp) waves preserving all
32 supercharges are thus related to gauge-invariant conformal operators of 
$\mathcal{N}=4$ CFT$_4$ in the world-volume of $N$ D3-branes. This
correspondence has already been studied extensively in the literature, and
we refer to \cite{BFHP1,BFHP2,IKM,GO,RT,CLP,GNS,DGR,DP,S,HS} for more
details. The BMN proposal has recently been extended successfully to models
preserving 16 supercharges \cite{KPRT,AFM,BGMNN,AS,SS}. These ideas are
pursued further in the present work, where we present an explicit analysis
of the correspondence between type IIB superstring theory on pp-wave
orbifolds, and the dual $\mathcal{N}=2$ quiver CFT$_4$s classified by affine
$\widehat{ADE}$ Kac-Moody algebras or by finite $ADE$ Lie algebras with a
specific open-string sector.

As already indicated, the BMN proposal offers a prescription for relating
chiral primary operators from the conformal model to states in the
superstring theory. Let $\Delta$ denote the conformal weight of the complex
field $Z$, i.e., its canonical dimension in CFT$_4$, and let $J$ be the
charge under an $SO(2)$ R-symmetry rotating two of the six scalars of the 
$\mathcal{N}=4$ multiplet. For large $N$, a gauge-invariant complex field
operator like $\text{Tr}\left[Z^{J}\right]$ with charge $J\sim \sqrt{N}$ and
$\Delta-J=0$, for example, is thus proposed to be associated with the vacuum
string state $|{0,p^{+}}\rangle_{lc}$ in the light-cone gauge with large
momentum $p^{+}$. In the string model, the parameters $\Delta$ and $J$ are
the eigenvalues of the energy operator $i\partial/\partial t$ and the
angular-momentum operator $-i\partial /\partial \psi $ (with respect to the
great circle of $S^{5}$), respectively, in which case $\Delta -J$
corresponds to the energy of a light-cone state. The proposed correspondence
can also be worked out explicitly for small (integer) values of $\Delta -J=n$, 
and in principle for the full tower of string states $|{n,p^{+}}\rangle$
and conformal operators 
$\mathcal{A}_{\Delta ,J}\sim\text{Tr}\left[\mathrm{A}_{\Delta ,J}Z^{J}\right]$.

The BMN proposal in its original form concerns a scenario with 32 preserved
supercharges, and has been extended recently to cases preserving 16
conformal supercharges. An objective of the present work is to indicate how
this may be extended further to models preserving an integral fraction of
the original 32 conformal supercharges. Our method is developed along the
lines of \cite{LNV} and \cite{KS}, and the results are obtained by soft
truncations of $\mathcal{N}=4$ $U(N)$ SYM$_{4}$ gauge theory. In view of our
construction, the studies of \cite{KPRT} and \cite{BGMNN} merely become
particular cases preserving 16 conformal supercharges. The analysis in 
\cite{KPRT} corresponds to an orbifold group $\Gamma \subset SU(2)$, while the
work \cite{BGMNN} deals with closely related models with a symplectic gauge
group. As in the case of \cite{KPRT}, the orbifold groups $\Gamma$ of
interest to our analysis are subgroups of an $SU(2)$ subgroup of the 
$SU(4)\sim SO(6)$ R-symmetry associated to the space $\mathbb{R}^{6}$
transverse to the aforementioned D3-branes containing the CFT$_4$ in their
world-volume. A substantial part of the present work is therefore devoted to
the case with 16 preserved supercharges. The following naive picture
summarizes the paths we will follow to achieve our goals:

\begin{equation}
\begin{tabular}{|llcll|}
\hline
&  &  &  &  \\
Type IIB on $AdS_5\times S^5$ pp waves &  & $\Longleftrightarrow$ &  & 
$\mathcal{N}=4\ $ $U(N)\ $ SYM$_4$/CFT$_4$ \\
&  &  &  &  \\
\emph{Orbifolding by} $\Gamma \subset SU(2)$, $\left|\Gamma \right| =k$ &  &
$\Downarrow \Downarrow$ &  & \emph{Orbifolding by} $\Gamma \subset SU(2)$, 
$\left| \Gamma \right| =k$ \\
&  &  &  &  \\
Type IIB on pp-wave orbifolds &  &  &  & $U(kN)$ SCFT$_4$ on D-branes \\
away from the fixed point &  &  &  & away from the fixed point \\
&  &  &  &  \\
\emph{Deformation of orbifold singularity} &  & $\Downarrow \Downarrow$ &  &
\emph{Deformation of orbifold singularity} \\
&  &  &  &  \\
Type IIB on deformed pp-wave orbifolds &  & $\Longleftrightarrow $ &  & 
$\mathcal{N}=2\ $ $\left[\otimes_{i=1}^{k}U(N_{i})\right]$ CFT$_4$ \\
&  &  &  &  \\ \hline
\end{tabular}
\label{pic}
\end{equation}
\newline
Details on the various steps will be given in the main body of the present
paper. At this point, we merely wish to point out that in the presence of
Chan-Paton factors, the brane engineering of supersymmetric CFT$_4$ will
involve not only D3-branes and wrapped D5-branes, but also D7-branes wrapped
around four-cycles with large volume. This property makes the brane
engineering of the models more subtle than described in \cite{KPRT}. It also
offers clues to the case $\Gamma\subset SU(3)$ (and even $\Gamma\subset
SU(4)$), and indicates how one may descend to models with 8 conformal
supercharges. Our explicit analysis is partly motivated by this last issue,
especially in light of the recent developments on $\mathcal{N}=1$ quiver
gauge theories inheriting basic properties of $\mathcal{N}=2$ quiver 
SYM$_{4}$ theories \cite{LNV}. We intend to explore these ideas further in
order to extend the picture above to dual models preserving 8 conformal
supercharges \cite{work}.

The remaining part of this paper is organized as follows. In Section 2, we
review results on the BMN proposal and on pp-wave geometries preserving 16
supercharges, and give a brief outline of our construction. In Section 3, we
study type IIB superstrings on pp waves with $ADE$ orbifold structures, with
particular emphasis on the various metrics. In Section 4, we discuss 
$\mathcal{N}=2$ CFT$_4$s classified by finite $ADE$ and affine $\widehat{ADE}$
Dynkin diagrams. Details on the non-abelian cases are deferred to 
Appendix A. Since most known results are on the affine case, we take this
opportunity to supplement the literature with explicit discussions on the
finite cases. In Section 5, we give the brane interpretation of these
models. In Section 6, we discuss the correspondence between type IIB
superstring on pp-wave orbifolds and $\mathcal{N}=2$ quiver CFT$_4$s.
Section 7 is devoted to some concluding remarks. An interpretation of conformal
invariance in terms of toric geometry is given in Appendix B.

\section{Strings on pp waves}

\subsection{The BMN proposal}

\quad\ We first recall the usual metric of the Anti-de Sitter space 
$AdS_{5}\times S^{5}$ \cite{BMN},
\begin{equation}
ds^{2}=R^{2}\left[ -\cosh ^{2}(\rho)dt^{2}+d\rho ^{2}+\sinh ^{2}(\rho)
d\Omega _{3}^{2}+\cos ^{2}(\theta)d\psi ^{2} +d\theta ^{2}+\sin ^{2}(\theta)
d\Omega_{3}^{\prime 2}\right] ,  \label{adsmetric}
\end{equation}
and consider the coordinate transformation
\begin{equation}
x^+=\frac{t+\psi}{2},\ \ \ \ \ x^-=R^2\frac{t-\psi}{2},\ \ \ \ \ x=R\rho,\ \
\ \ \ y=R\theta.  \label{trans}
\end{equation}
In the Penrose limit where
\begin{equation}
R\rightarrow\infty  \label{limit}
\end{equation}
the geometry is altered, and the new metric may be written as
\begin{equation}
ds^2=-4dx^-dx^+-(\mathbf{x}^2+\mathbf{y}^2)(dx^+)^2+d\mathbf{x}^2 
+d\mathbf{y}^2.  
\label{ppmetric1}
\end{equation}
Here $\mathbf{x}=(x^1,x^2,x^3,x^4)$ and $\mathbf{y}=(y^1,y^2,y^3,y^4)$ each
parameterizes $\mathbb{R}^4$, and $|\mathbf{x}|=x$, $|\mathbf{y}|=y$. After
the simple re-scaling $x^-\rightarrow x^-/\mu$, $x^+\rightarrow\mu x^+$, and
by combining the two four-dimensional coordinates into a single one, 
$\mathbf{w}=(w^1,\dots,w^8)$, parameterizing $\mathbb{R}^8$, this metric is
recognized as a plane-wave metric:
\begin{equation}
ds^{2}=-4dx^{-}dx^{+}-\mu^{2}\mathbf{w}^2(dx^+)^{2}+d\mathbf{w}^2 .
\label{ppmetric}
\end{equation}
In the following, though, we shall mainly use the coordinates 
$(\mathbf{x},\mathbf{y})$. The mass parameter, $\mu$, is related to the 
five-form field strength
\begin{equation}
F=\mu dx^{+}\wedge\left(dx^1\wedge dx^2\wedge dx^3\wedge dx^4 +dy^1\wedge
dy^2\wedge dy^3\wedge dy^4\right).  
\label{F}
\end{equation}

Following \cite{BMN}, one may consider the energy and angular-momentum
generators as given by $E=i\partial_t$ and $J=-i\partial_\psi$,
respectively, in terms of the original $AdS_5\times S^5$ coordinates 
(\ref{adsmetric}). One then finds the following expressions for the light-cone
momenta:
\begin{eqnarray}
&&2p^{-} =i\partial _{x^{+}}=i\left( \partial _{t}+\partial _{\psi }\right)
=\Delta -J,  \notag \\
&&2p^{+} =i\partial _{x^{-}} =\frac{i}{R^{2}}(\partial_{t}-\partial _\psi) 
=\frac{\Delta +J}{R^{2}},  
\label{deltaj}
\end{eqnarray}
where the BPS condition reads $\Delta\geq|J|$. The Penrose limit (\ref{limit}) 
of the string model now corresponds to the large-$N$ limit of the dual 
$\mathcal{N}=4$ $U(N)$ gauge theory with focus on operators for which $J\sim
N^{1/2}$, while $\Delta-J$ and the gauge coupling remain fixed. With these
properties in mind, BMN \cite{BMN} have worked out the correspondence
between the spectrum of string states and the spectrum of gauge-model
operators. For the minimum value $\Delta -J=0$, for instance, the
correspondence is between the single-trace operator $\text{Tr}[Z^J]$, with $Z$ a
complex scalar, and the vacuum state in light-cone gauge $|0,p^{+}\rangle
_{lc}$. At the next level, where $\Delta -J=1$, the gauge-invariant field
operators of the form $\sum_{l=0}^{J}\text{Tr}\left[Z^l\phi ^{r}Z^{J-l}\right]$ 
and $\sum_{l=0}^{J}\text{Tr}\left[Z^l\psi_{\frac{1}{2}}^{b}Z^{J-l}\right]$ 
are associated with $a_{0}^{\dagger k}|0,p^{+}\rangle_{lc}$ and 
$S_{0}^{\dagger b}|0,p^{+}\rangle _{lc}$, where 
$\phi ^{r}$, $r=1,2,3,4$, are neutral bosonic scalars with respect to $J$,
while $\psi_{\frac{1}{2}}^{b}$, $b=1,\ldots,8$ are fermionic fields, and 
$a_{0}^{\dagger i}$ and $S_{0}^{\dagger b}$ with $i,b=1,\ldots,8$ are bosonic
and fermionic zero-momentum oscillators, respectively. For $\Delta -J=2$,
operators of the form $\sum_{l=0}^{J}\text{Tr} \left[\phi^{r}Z^l
\psi_{\frac{1}{2}}^{b}Z^{J-l}\right]$ are associated with states of the form 
$a_{0}^{\dagger r}S_{0}^{\dagger b}|0,p^{+}\rangle_{lc}$. The general
correspondence rule is outlined in \cite{BMN}.

Shortly after the BMN proposal appeared, extensions to plane-wave
backgrounds preserving 16 space-time supercharges were considered. These
models are based on orbifolds or orientifolds. Since similar constructions
are at the core of the present work, a brief summary now follows.

\subsubsection{Orbifolds}

\quad\ Recently, Kim, Pankiewicz, Rey and Theisen (KPRT) extended the BMN
proposal to type IIB superstrings propagating on pp-waves with one of the 
$\mathbb{R}^{4}$ factors in $\mathbb{R}^8\sim\mathbb{R}^{4}\times 
\mathbb{R}^{4}$ replaced by an $\mathbb{R}^{4}/\mathbb{Z}_{k}$ orbifold 
\cite{KPRT}.
It is convenient to describe the orbifold structure in terms of complex
coordinates, replacing $(\mathbf{x},\mathbf{y})\in\mathbb{R}^4\times\mathbb{R}^4$ 
by $(\mathbf{x},z_1,z_2)\in\mathbb{R}^4\times\mathbb{C}^2$. The
orbifold action is then defined as
\begin{equation}
\mathbb{Z}_k:\ \ \ (z_1,z_2) \mapsto(\omega z_1,\overline{\omega}z_2),\ \ \
\ \ \ \ \ \ \ \omega =\exp\left(\frac{2\pi i}{k}\right),  \label{gac}
\end{equation}
where $z_1=(y^1+iy^2)$ and $z_2=(y^3+iy^4)$. Due to the orbifold structure,
the string theory will have twisted sectors indexed by the charge, $q$, of
the orbifold group. Physical string states are obtained by applying the
bosonic and fermionic creation operators to the light-cone vacuum 
$|0,p^{+}\rangle_{q}$ of each twisted sector labeled by $q=1,\ldots,k$.

According to \cite{KPRT}, the associated superstring theory is described by
the large-$N$ limit of $\mathcal{N}=2$ $\left[U(N)\right]^k$ quiver gauge
theory with fixed gauge coupling and $k$ hypermultiplets in bi-fundamental
representations but without fundamental matter. They have also proposed a
precise map between the gauge-theory operators and the string states for
both untwisted and twisted sectors. The orbifold structure presented  here
reduces the 32 supercharges of the BMN pp-wave background to 16 supercharges.

As in the BMN correspondence, we here focus on states with conformal weight 
$\Delta$ and $U(1)$ charge $J\sim N^{1/2}$ whose difference $\Delta-J$
remains finite in the large-$N$ limit. Due to the $\mathbb{Z}_k$
orbifolding, some of the fields get promoted to $kN\times kN$ matrices. This
applies to the gauge field $A_{\mu }$, the complex scalar $Z$ associated to
the coordinates $x^3$ and $x^4$, and the complex scalars 
$\phi^m=(\phi^1,\phi^2)$ associated to the coordinates $\mathbf{y}$ as well
as their superpartners $\chi$ and $\xi$, which are promoted to 
$\mathcal{A}_{\mu }$, $\mathcal{Z}$, $\Phi^{m}$, $\Lambda$ and $\Xi $, 
respectively. They satisfy the following conditions
\begin{eqnarray}
&&S\mathcal{A}_{\mu }S^{-1} =\mathcal{A}_{\mu },\qquad\ \ \ 
S\mathcal{Z}S^{-1}= \mathcal{Z},\qquad S\Lambda S^{-1}=\Lambda  \notag \\
&&S\Phi^{m}S^{-1} =\omega\Phi^{m},\qquad S\Xi S^{-1}=\omega \Xi ,
\label{st1}
\end{eqnarray}
where $S=\mathrm{diag}\left(1,\omega ^{-1},\omega ^{-2},\ldots,
\omega^{-k+1}\right)$. It was then proposed that gauge-invariant field operators
satisfying $\Delta-J=0$ are associated with the vacuum in the $q$-th twisted
sector
\begin{equation}
\text{Tr}\left[ S^{q}\mathcal{Z}^{J}\right] \quad \leftrightarrow \quad
|0,p^{+}\rangle _{q},  
\label{st2}
\end{equation}
while for $\Delta-J=1$ we have
\begin{eqnarray}
\text{Tr}\left[ S^{q}\mathcal{Z}^{J}\mathcal{D}_{\mu }\mathcal{Z}\right]
\quad &\leftrightarrow&\quad a_{0}^{\dagger \mu }|0,p^{+}\rangle _{q},
\notag \\
\text{Tr}\left[ S^{q}\mathcal{Z}^{J}\mathbf{\chi }_{\frac{1}{2}}\right] \quad
&\leftrightarrow &\quad \chi _{0}^{\dagger }|0,p^{+}\rangle _{q},  \notag \\
\text{Tr}\left[ S^{q}\mathcal{Z}^{J}\overline{\mathbf{\chi }}_{\frac{1}{2}}\right]
\quad &\leftrightarrow &\quad \overline{\chi }_{0}^{\dagger}|0,p^{+}\rangle_{q},  
\label{st3}
\end{eqnarray}
where $\mathcal{D}_{\mu}\mathcal{Z}=
\partial_{\mu}\mathcal{Z} +\left[\mathcal{A}_{\mu},\mathcal{Z}\right]$. 
Operators corresponding to higher
string states may also be constructed. For $\Delta-J=2$, for instance, we
have for the untwisted sector
\begin{equation}
\sum_{l=0}^J\text{Tr}\left[ S^{q}\mathcal{Z}^{l}(\mathcal{D}_{\mu } 
\mathcal{Z})\mathcal{Z}^{J-l}(\mathcal{D}_{\nu }\mathcal{Z})\right] 
e^{\frac{2\pi il}{J}n}\quad
\leftrightarrow \quad a_{n}^{\dagger \mu } a_{-n}^{\dagger
\nu}|0,p^{+}\rangle_{q=0},  \label{st4}
\end{equation}
where $n$ is the level of the oscillator $a^\dagger_n$. For the twisted
sectors the correspondence is given by
\begin{equation}
\sum_{l=0}^J\text{Tr}\left[ S^{q}\mathcal{Z}^{l}\Phi^{r} 
\mathcal{Z}^{J-l}\overline{\Phi}^{s}\right] e^{\frac{2\pi il}{J}n(q)}\quad 
\leftrightarrow
\quad \alpha _{n(q)}^{\dagger r}\overline{\alpha}_{-n(q)}^{\dagger s}
|0,p^{+}\rangle_{q},  
\label{st5}
\end{equation}
with $n(q)=n+\frac{q}{k}$.

We shall continue the KPRT analysis \cite{KPRT} in the following, with
particular emphasis on the brane engineering of their quiver CFT$_4$.
Moreover, the KPRT orbifold model is not unique. We shall thus extend their
construction to the entire class of $\mathcal{N}=2$ quiver CFT$_4$s, thereby
covering also the cases with fundamental matter described by D7-branes.

\subsubsection{Orientifold}

\quad\ The other important extension of the BMN construction concerns
strings moving on a pp-wave orientifold background \cite{BGMNN}. There, one
considers an $\mathcal{N}=2$ $Sp(N)$ gauge theory with a hypermultiplet in
the anti-symmetric representation and four fundamental hypermultiplets with
open strings in the 't Hooft limit \cite{M1}. This is dual to type IIB
superstring theory on $AdS_{5}\times S^{5}/\mathbb{Z}_{2}$, where 
$\mathbb{Z}_{2}$ is an orientifold action \cite{AFM,BGMNN}. 
As before, one looks for
operators that carry large charges with respect to $J$ while their conformal
dimensions are such that $\Delta -J$ is small. They are then identified with
string states propagating on an orientifold of a maximally supersymmetric
plane-wave background \cite{BFHP1,BFHP2}. One thereby relates closed strings
to single-trace gauge-invariant operators, and open strings to
gauge-invariant operators with two quarks at the ends. Underlying the
construction in \cite{BGMNN} is the $O(7)$ orientifold of the pp-wave
solution of ten-dimensional type IIB supergravity discussed in \cite%
{BFHP1,BFHP2}. An $O(7)$ plane carries $-4$ units of D7-brane charge, so the
introduction of four D7-branes can cancel this charge, locally producing the
gauge group $SO(8)$. The metric of this orientifold background corresponds
to (\ref{ppmetric}) after the substitution $w_{7,8}\rightarrow -w_{7,8}$.

In a field theory with gauge group $Sp(N)$, one has a vector-multiplet 
$\left(0^{2},1,\frac{1}{2}^{2}\right)$ whose scalars, denoted by the complex
fields $W_{ab}$ in the symmetric representation of the gauge group, describe
the movement of D3-branes in the (7-8) plane. Matter fields belonging to
anti-symmetric hypermultiplets $\left(0^{4},\frac{1}{2}^{2}\right)$,
splitting into $\mathcal{N}=1$ multiplets as 
$\left( 0^{2},\frac{1}{2}\right)\oplus\left( 0^{2},\frac{1}{2}\right)$, 
are denoted by $Z$ and $Z^{\prime}$. 
They describe the movement of D3-branes in the $3,4,5,6$
directions. The fundamental representations of the gauge group associated to
these directions are denoted $q_{i}$ and $\widetilde{q}_{i}$ and encode open
strings stretching between D3- and D7-branes. The chiral operator that
should be identified with the closed-string ground states for $\Delta-J=0$
is
\begin{equation}
\text{Tr}\left[ \left( Z\Omega \right) ^{J}\right] =\left(
Z_{ab}\Omega^{ba}\right)^{J},
\end{equation}
where $\Omega$ is an $Sp(N)$ invariant tensor. In this way, the
gauge-invariant field operator corresponding to the open-string ground state
for $\Delta-J=1$ reads
\begin{equation}
\text{Tr}\left [Q_{i}\Omega(Z\Omega)^{J}Q_{j}\right],
\end{equation}
where the chiral multiplets $Q_{i}$, $i=1,\ldots,8$, are any of the four
fundamentals $q$ and $\widetilde{q}$. Thus, we have the following heuristic
rule:

\begin{equation}
\begin{tabular}{|l|c|}
\hline
Coordinates & Field insertions \\ \hline
$w^i=x^i,$ \ \ \ \ \ \ $i=1,\ldots,4$ & $\partial _{i}Z$ \\
$w^{5,6}=y^{1,2}$ & $Z$, $\overline{Z}^{\prime }$ \\
$w^{7,8}=y^{3,4}$ & $W$, $\overline{W}$ \\ \hline
\end{tabular}%
\end{equation}
\newline
For non-zero modes we have
\begin{equation}
\sum_{l=0}^Je^{\frac{i2\pi l}{J}n}\text{Tr}\left[ W\Omega
(Z\Omega)^lZ^{\prime }\Omega (Z\Omega)^{J-l}\right] \leftrightarrow \left(
a_{-n}^{(7+i8)\dagger }a_{n}^{(5+i6)\dagger }
-a_{n}^{(7+i8)\dagger}a_{-n}^{(5+i6)\dagger }\right) |0,p^{+}\rangle_{lc},
\end{equation}
where $a^{(k+il)}=a^k+ia^l$.

\subsection{Outline of construction}

\quad\ As already mentioned, we wish to study extensions of the BMN
correspondence in which type IIB superstrings on a Penrose limit of $%
AdS_{5}\times S^{5}$ orbifolded by $\Gamma$ have four-dimensional $\mathcal{N%
}=2$ superconformal field theories as their dual gauge models. We shall
proceed in steps by first studying the Penrose limit of type IIB
superstrings on $AdS_{5}\times S^{5}/\Gamma$, where $\Gamma\subset SU(2)$.
We then use the AdS/CFT correspondence to explore their various 
$\mathcal{N}=2$ CFT$_4$ duals. Following this, we give brane interpretations 
of these dual models. We finally consider extensions of the BMN proposal for this
class of supersymmetric models.

Our analysis obviously depends on the discrete orbifold group $\Gamma$. To
indicate how and in order to outline our approach, we shall comment briefly
on some of the steps below. More detailed discussions are found in the
subsequent sections and in the appendices.

\subsubsection{Comments on string-theory side}

\quad\ To derive pp-wave backgrounds in the Penrose limit of $AdS_{5}\times
S^{5}/\Gamma$, we must specify the nature of the orbifold group $\Gamma$.
This is in general contained in the $SO(6) \sim SU(4)$ R-symmetry of 
$\mathcal{N}=4$ SYM$_4$. Depending on the number of preserved supercharges,
in particular, we have the following situations \cite{LNV}:
\begin{eqnarray}
&&\Gamma \subset SU(4),\ \ \ \ \ \ \ \ \ \ \ \ \ \ \ \ \ \ \ \ \
\Gamma\not\subset SU(3)\subset SU(4),  \notag \\
&&\Gamma \subset SU(3)\subset SU(4),\ \ \ \ \ \ \ \ \ \Gamma\not\subset
SU(2)\subset SU(4),  \notag \\
&&\Gamma \subset SU(2) \subset SU(4) .
\end{eqnarray}
For orbifold models with 16 supercharges, one should consider $\Gamma\subset
SU(2)$. Using further group-theoretical arguments based on the
classification of discrete subgroups of $SU(2)$, it follows that $\Gamma$
may be any of the discrete $\widetilde{ADE}$ 
subgroups\footnote{This $\widetilde{ADE}$ appellation for discrete 
groups should not be
confused with the usual notation for ordinary $ADE$ and affine $\widehat{ADE}
$ Lie algebras.} of $SU(2)$ \cite{FHHP1,FHHP2}. These finite groups are
either abelian or non abelian, and are classified as follows:

\begin{equation}
\begin{tabular}{|l|l|l|l|}
\hline
$\Gamma $ & Name & Order $|\Gamma|$ & Group generators \\ \hline
$\widetilde{A}_{k-1}\simeq\mathbb{Z}_{k}$ & Cyclic & $k$ & $\left\{
a|a^{k}=Id\right\} $ \\ \hline
$\widetilde{D}_{2k}$ & Dihedral & $4k$ & $\left\{
a,b|b^{2}=a^{k};ab=ba^{-1};a^{2k}=Id\right\} $ \\ \hline
$\widetilde{E}_{6}$ & Tetrahedral & $24$ & $\left\{ a,b|a^{3}=b^{3}
=(ab)^{3}\right\} $ \\ \hline
$\widetilde{E}_{7}$ & Octahedral & $48$ & $\left\{ a,b|a^{4}=b^{2}
=(ab)^{2}\right\} $ \\ \hline
$\widetilde{E}_{8}$ & Icosahedral & $120$ & $\left\{ a,b|a^{5}=b^{3}
=(ab)^{2}\right\} $ \\ \hline
\end{tabular}
\label{classGamma}
\end{equation}
\newline
{}From this classification, one sees that 
$\widetilde{A}_{k-1}\simeq \mathbb{Z}_{k}$ is an abelian group with one 
generator $a$, while $\widetilde{D}_{2k} $, $\widetilde{E}_{6}$, 
$\widetilde{E}_{7}$ and $\widetilde{E}_{8}$ are non-abelian discrete groups 
with two non-commuting generators $a$ and $b$.

There are as many pp-wave geometries with the properties specified 
by this classification as there are orbifold groups in the table 
(\ref{classGamma}). These geometries are naturally referred to as 
$\widetilde{ADE}$ pp-wave geometries.

In the presence of D5-D5 and D3-D5 open-string states, the closed-string
geometries have a more general form following from the resolution of the
orbifold singularities. Details on the D5-D5 and D3-D5 systems, as well as
an implementation of D5-D7 open strings, will be discussed in Section 5.

\subsubsection{Comments on field-theory side}

\quad\ In order to describe the duality between $\mathcal{N}=2$ CFT$_4$s and
type IIB superstring theory on $AdS_{5}\times S^{5}/\Gamma$, we now identify
the various $\mathcal{N}=2$ CFT$_4$s.

\paragraph{$\mathcal{N}=2$ CFTs in four dimensions:}

To get the appropriate $\mathcal{N}=2$ CFT$_4$s we are interested in, we use
results on conformal field theories in four dimensions obtained by taking
orbifolds of $\mathcal{N}=4$ $U(N)$ SYM$_4$ \cite{LNV}. As indicated above,
we restrict ourselves to the case where $\Gamma$ is a discrete subgroup of 
$SU(2)$, itself a subgroup of the $SU(4)$ R-symmetry. In this way, 
$\mathcal{N}=4$ supersymmetry is broken down to $\mathcal{N}=2$ 
supersymmetry, while the original $U(N)$ gauge group gets promoted to 
$U(|\Gamma|N)$. The $\mathcal{N}=4$ SYM$_{4}$ multiplet
\begin{equation}
\left( 0^{6},\frac{1}{2}^{4},1\right)\otimes \left(|\Gamma|N,
\overline{|\Gamma|N}\right)
\end{equation}
has $6(|\Gamma|N)^2+2(|\Gamma|N)^2=8(|\Gamma|N)^2$ bosonic degrees of
freedom and the same number of fermionic ones. It splits into $\mathcal{N}=2$
representations as follows
\begin{equation}
\left(\sum_{i=1}^{|\Gamma|}\left( 0^{2},\frac{1}{2}^{2},1\right) \otimes
\left(N_i,\overline{N}_i\right) \right)\oplus
\left(\sum_{i,j=1}^{|\Gamma|}\left(0^{4},\frac{1}{2}^{2}
\right)\otimes\left(N_i,\overline{N}_j\right) \right).
\end{equation}
In this decomposition, the $U(|\Gamma|N)$ gauge group is broken down to $\left[\otimes_{i=1}^{|\Gamma|}U(N_i)\right]$ where 
\begin{equation}
|\Gamma|N=\sum_{i=1}^{|\Gamma|}N_i,
\end{equation} 
and the one-loop beta function for each gauge coupling $g_{i}$ is proportional to
\begin{equation}  \label{beta}
\beta_{i}=\frac{1}{6}\left( 22N_{i}-\sum_{j=1}^{|\Gamma|} \left[ 2\left( a_{ij}^{4}
+\overline{a}_{ij}^{4}\right) +a_{ij}^{6}\right] N_{j}\right).
\end{equation}
Here $a_{ij}^{4}$ and $a_{ij}^{6}$ are the numbers of Weyl spinors and
scalars, respectively, transforming in the bi-fundamental representations 
$\left(N_i,\overline{N}_j\right)$ \cite{LNV}. These numbers satisfy
\begin{equation}
3a_{ij}^{4}=3\overline{a}_{ij}^{4}=2a_{ij}^{6}.  \label{nu}
\end{equation}
As we shall see, this relation plays a crucial role in our analysis.
For now, we merely mention that (i) the solutions for $\beta _{i}=0$ are
related to the Dynkin diagrams of affine $\widehat{ADE}$
Kac-Moody algebras \cite{KMV,BFS,BS1,BS2}, 
and that (ii) these solutions may be extended to the
class of field theories based on finite and indefinite Lie algebras 
\cite{ABS1,ABS2} as well. This property reflects the fact that these classes of
field theories can be treated in a unified way.

\paragraph{Geometric engineering of $\mathcal{N}=2$ QFT$_{4}$:}

A powerful method to deal with the classification of these CFT$_4$s is
geometric engineering of $\mathcal{N}=2$ quiver QFT$_{4}$ developed in 
\cite{KMV}, see also \cite{BFS,BS1,BS2}. It relies on the description and
treatment of singular  complex  manifolds in algebraic geometry. Field-theoretical
quantities such as mass and coupling-constant moduli are encoded in
equations describing Calabi-Yau manifolds as K3 fibrations. Interesting
cases are given by the following singular complex surfaces:

\begin{equation}  \label{surface}
\begin{tabular}{|c|l|}
\hline
Singularity & Geometry of ALE space \\ \hline
$A_{k-1}$ & $\zeta_{1}^{2}+\zeta_{2}^{2}=\zeta^{k}, \qquad\ \ \ \ \ \  k\geq 2$
\\ \hline
$D_{k}$ & $\zeta_{1}^{2}+\zeta_{2}^{2}\zeta =\zeta^{k-1}, \qquad\ k\geq 3$
\\ \hline
$E_{6}$ & $\zeta_{1}^{2}+\zeta_{2}^{3}+\zeta ^{4}=0$ \\ \hline
$E_{7}$ & $\zeta_{1}^{2}+\zeta_{2}^{3}+\zeta_{2}\zeta ^{3}=0$ \\ \hline
$E_{8}$ & $\zeta_{1}^{2}+\zeta_{2}^{3}+\zeta ^{5}=0$ \\ \hline
\end{tabular}%
\end{equation}
\newline
Here $\zeta_1,\zeta_2,\zeta$ are complex variables. In general, one is
interested in the deformation of the singularities as they are associated
with a large class of $\mathcal{N}=2$ quiver QFT$_{4}$s. The singular
surfaces (\ref{surface}), which are associated with the cases where the
gauge symmetry of the $\mathcal{N}=2$ QFT$_{4}$ is unbroken, have affine
extensions classified by affine Lie algebras $\widehat{A}_{k}$, 
$\widehat{D}_{k}$ and $\widehat{E}_{s}$.

It is recalled that deformations of the singularities can be obtained in two
ways, either by K\"ahler deformations or by deforming the complex structure.
In the second case, the complex deformations are carried by complex moduli 
$a_{i}$, and the singular surfaces $F(\zeta_{1},\zeta_{2},\zeta)=0$ are
replaced by non-singular surfaces described by equations of the type
\begin{equation}  \label{deformation}
F(\zeta_{1},\zeta_{2},\zeta;\{a_i\}) =0.
\end{equation}
Following \cite{KKV,KMV}, the complex moduli are in general polynomials
depending on two extra complex variables, $\xi$ and $\upsilon$,
\begin{equation}  \label{fiber}
a_{i}\left( \xi ,\upsilon \right)
=\sum_{n=1}^{N_{i}}\sum_{m=1}^{M_{i}}c_{i_{n,m}}\xi ^{n}\upsilon^{m}.
\end{equation}
These extra variables
allow one to engineer gauge groups and include fundamental matter,
where $N_i$ and $M_i$ characterize the gauge groups and their associated
matter content.

As an illustration of (\ref{deformation}), we recall the complex deformation
of the ALE surface with $A_{k-1}$ singularity (\ref{surface})
\begin{equation}
z_1z_2=\zeta^{k+1}+\sum_{i=1}^{k}a_i\zeta^{k-i},  
\label{zza}
\end{equation}
where $z_1=\zeta_1+i\zeta_2$ and $z_2=\zeta_1-i\zeta_2$. One can then relate
the various differentials as in the following example:
\begin{equation}
dz_2=-\left(\zeta^{k}+\sum_{i=1}^{k}a_{i}\zeta^{k-i}\right) 
\frac{dz_1}{z_1^{2}}+\left(k\zeta^{k-1}+\sum_{i=1}^{k}(k-i)
a_{i}\zeta^{k-1-i}\right) \frac{d\zeta }{z_1} 
+\sum_{i=1}^{k}\zeta^{k-i}da_{i}.  
\label{da}
\end{equation}
The terms depending on the differentials $da_{i}$ are important in the
metric building of pp-wave orbifolds based on deformations of $ADE$
singularities. In this formulation, the vanishing of the one-loop beta
function of these $\mathcal{N}=2$ quiver QFT$_{4}$s is translated into a
well-known problem of the classification of Lie algebras \cite{KMV,ABS1,ABS2}. 
Based on these results, we shall show explicitly how one can get the general
classes of $\mathcal{N}=2$ quiver CFT$_4$s, in particular those involving
affine $\widehat{ADE}$ and finite $ADE$ geometries. Since mainly the first
class is well studied in the literature, we take this opportunity to present
some explicit results for the finite $ADE$ symmetries.

\subsubsection{Brane realizations}

\quad\ As in the case of $\mathcal{N}=4$ $U(N)$ CFT$_4$s living in the
world-volume of $N$ coincident D3-branes, $\mathcal{N}=2$ quiver CFT$_4$s
based on $\Gamma$ orbifolds involve closed strings. The new feature is that
the $\mathcal{N}=2$ models also involve open-string sectors, not present in
the $\mathcal{N}=4$ models. Indeed, brane engineering of $\mathcal{N}=2$
quiver CFT$_4$s is based on (partially) wrapped D5- and D7-branes on
two- and four-cycles, respectively, in addition to the familiar D3-branes.
As the initial gauge symmetry is broken by the deformation of the orbifold
singularity, open strings are stretched between some of the D-branes. On the
field-theory side, this corresponds to $\mathcal{N}=2$ CFT$_4$s with 
\textit{bi-fundamental} matter, which has no analogue in 
$\mathcal{N}=4$ CFT$_4$.

With the above tools at hand, we can write down the correspondence rule
between string states on pp waves on $AdS_{5}\times S^{5}/{\Gamma}$ where 
$\Gamma$ is a discrete subgroup of $SU(2)$, and gauge-invariant field
operators in the dual $\mathcal{N}=2$ CFT$_4$s. As we will show, the
resulting correspondence rule for $\mathcal{N}=2$ quiver gauge theories is
richer than the one indicated in (\ref{st2}-\ref{st5}).

\section{Type IIB superstrings on pp-wave orbifolds}

\quad\ In this section, we consider the problem of metric building of pp
waves on $AdS_5\times S^5/\Gamma$ orbifolds with $\Gamma\subset SU(2)$. 
As $\Gamma$ may be any of the discrete groups $\widetilde{A}_{k-1}$, 
$\widetilde{D}_{2k}$, $\widetilde{E}_{6}$, $\widetilde{E}_{7}$ and 
$\widetilde{E}_{8}$,
we will study each case separately and treat both the singular and deformed
geometries. For an explicit treatment, we will first consider the abelian
group orbifolds where $\Gamma=\widetilde{A}_{k-1}$. The results for the
non-abelian cases and the results for the affine geometries 
are deferred to Appendix A.

\subsection{Metric of abelian orbifolds}

\quad\ Following the previous section, we shall use the metric
\begin{equation}
ds^2 =-4dx^{+}dx^{-}+d\mathbf{x}^2-\mu ^{2}(\mathbf{x}^2+|z_1|^2
+|z_2|^2)(dx^{+})^2+|dz_1|^2+|dz_2|^2  \label{ppwavemetric}
\end{equation}
to describe the Penrose limit of $AdS_5\times S^5$. The string coupling 
$g_{s}\sim g_{YM}^{2}$ of the type IIB superstring theory is kept fixed in
this limit, while the curvature, $R$, of the ten-dimensional background is
related to the gauge group $U(N)$ of the dual four-dimensional SYM$_4$
through $R^4\sim\left(g_{YM}^{2}N\right) l_{s}^{4}$. For non-zero $\mu$, the
metric (\ref{ppwavemetric}) is invariant under the action of $SO(4)\otimes
SO(4)$ on the eight coordinates $(\mathbf{x,y})$ of $\mathbb{R}^{4}\times
\mathbb{R}^{4}$. The $\mu$-dependent term proportional to $(dx^{+})^{2}$
manifestly breaks the $SO(1,1)$ sub-group of the symmetry group $SO(1,9)$ of
the ten-dimensional space with $\mu=0$. For small values of $\mu$, the
metric (\ref{ppwavemetric}) thus corresponds to a deformation of the flat
space-time background $\mathbb{R}^{1,9}$. 
After replacing $\mathbb{R}^4\times\mathbb{R}^4$ by 
$\mathbb{R}^4\times\mathbb{C}^2$, the group $SO(4)\otimes
U(2)$ naturally plays the role of $SO(4)\otimes SO(4)$.

The space $AdS_5\times S^5$ is known to be an exactly solvable, maximally
supersymmetric geometry preserving the 32 space-time supercharges 
\cite{BFHP1,BFHP2}. This preserving property survives in the Penrose limit. As
discussed above, 16 supercharges may be reached by considering type IIB
superstrings on orbifolds like $AdS_{5}\times S^{5}/\Gamma$ in certain
Penrose limits. Here we focus on Penrose limits of the orbifolds 
$AdS_{5}\times S^{5}/\mathbb{Z}_{k}$, while the other geometries are
discussed in Appendix A. To work out the explicit form of the metric, let
us list a couple of observations. (i) Under orbifolding by $\mathbb{Z}_{k}$,
the original supersymmetry of $\mathcal{N}=4$ $U(N)$ SYM$_{4}$ is reduced by
half, whereas the gauge group is promoted to $U(kN)$. The first property
implies that the massless $\mathcal{N}=4$ on-shell gauge multiplet 
$\left(0^{6},\frac{1}{2}^{4},1\right)$ splits into $\mathcal{N}=2$ representations
according to
\begin{equation}
\left( 0^{6},\frac{1}{2}^{4},1\right) \rightarrow \left( 0^{2},
\frac{1}{2}^{2},1\right) \oplus \left( 0^{4},\frac{1}{2}^{2}\right) ,
\end{equation}
where $\left( 0^{2},\frac{1}{2}^{2},1\right)$ is an $\mathcal{N}=2$ vector
multiplet and $\left( 0^{4},\frac{1}{2}^{2}\right)$ is a hypermultiplet.
Under this decomposition, the initial $SO(6)\sim SU(4)$ R-symmetry of the 
$\mathcal{N}=4$ theory is broken down to $SO(2)\otimes SO(4)$ which in terms
of complex parameters corresponds to $U(1)\otimes SU(2)\otimes SU(2)$. This
group contains the usual $U(2)\sim U(1)\otimes SU(2)$ R-symmetry of 
$\mathcal{N}=2$ gauge theories in four dimensions. (ii) Under orbifolding,
the transverse light-cone space $\mathbb{R}^{4}\times \mathbb{R}^{4}$ is now
replaced by the orbifold 
$\mathbb{R}^{4}\times\left(\mathbb{R}^{4}/\mathbb{Z}_k\right)$.

The action of the orbifold group given in (\ref{gac}) has a fixed point at 
$z_1=z_2=0$. This means that quantum fields and string states at this point
have $k$ sectors. To determine the metric of this pp-wave background, one
should understand the geometry in the vicinity of the orbifold (fixed)
point. In this regard, we recall that the real four-dimensional orbifold 
$\mathbb{R}^{4}/\mathbb{Z}_{k}$ near the singular point $y^1=y^2=y^3=y^4=0$
looks like the ALE space with an $A_{k-1}$ singularity which can be embedded in
 three-dimensional complex plane.  This is a crucial point in
our construction, and it is important here to specify the nature of this
singularity which can be either elliptic, ordinary or indefinite 
\cite{ABS1,ABS2}. In the elliptic case, one has only closed string states while
for the other two geometries one needs open strings as well. On the
field-theory side, this corresponds to adding fundamental matter in order to
ensure the vanishing of the beta function. Put differently, one needs
complex moduli $a_{i}$ with non-zero differentials $da_{i}$ as in 
(\ref{deformation}). To show how the machinery works, we start by treating
orbifolds with an ordinary $A_{k-1}$ singularity and initially disregard the 
open strings. The rationale for starting with this example is that it
is sufficiently simple to handle while it nevertheless illustrates the
implementation of partially wrapped D-branes. Once we get the explicit form
of the pp-wave metric with $A_{k-1}$ geometry, open strings can be
introduced in conjunction with the complex moduli (\ref{fiber}).

It is recalled that an ALE space with an $A_{k-1}$ singularity is a complex
surface embedded in $\mathbb{C}^{3}$, and may be
defined by the algebraic equation
\begin{equation}  
\label{Asing}
zt=\zeta^k.
\end{equation}
The metric of this complex space is induced from 
$ds^{2}=|dz|^{2}+|dt|^{2}+|d\zeta|^2$ by substituting (\ref{Asing}). In terms
of the coordinates $z_1=z$ and $z_2=\frac{\zeta^{k}}{z}$, the differentials
are related as follows:
\begin{equation}
dz_2=k\frac{\zeta ^{k-1}}{z}d\zeta -\frac{\zeta ^{k}}{z^{2}}dz.  \label{dz2}
\end{equation}
It is noted that $\mathbb{Z}_{k}$ acts on these variables as $z\rightarrow
\alpha_{z}z$, $t\rightarrow\alpha_{t}t$ and $\zeta\rightarrow
\alpha_{\zeta}\zeta$ where $\alpha_{z}$, $\alpha _{t}$ and $\alpha_\zeta$
are $k$-th roots of unity. Invariance of (\ref{Asing}) requires $\alpha_{t}=%
\overline{\alpha}_{z}$ while $\alpha_{\zeta}$ can be any of the $k$-th roots
of unity. As in (\ref{gac}), we shall work with
\begin{equation}  \label{Zk}
z_1\rightarrow \omega z_1,\qquad z_2\rightarrow \overline{\omega} z_2,\qquad
\zeta \rightarrow \overline{\omega }\zeta,\qquad w=\exp\left(\frac{2\pi i}{k}%
\right),
\end{equation}
where we have set $\alpha_{\zeta}=\overline{\omega}$. This is a convenient
choice to be used below in the classification of the possible deformations.

Returning to the pp-wave orbifold (\ref{ppwavemetric}), we thus find that
its metric near the $A_{k-1}$ singularity reads
\begin{eqnarray}
ds^{2}|_{A_{k-1}} &=&-4dx^{+}dx^{-}+d\mathbf{x}^{2} -\mu ^{2}\left(\mathbf{x}%
^{2}+\left| z\right| ^{2} +\left| \frac{\zeta ^{k}}{z}\right|
^{2}\right)(dx^{+})^{2}  \notag \\
&&+\left( 1+\left| \frac{\zeta ^{k}}{z^{2}}\right| ^{2}\right) \left|
dz\right| ^{2}+k^{2}\frac{\left| \zeta \right|^{2(k-1)}}{\left| z\right|^{2}}%
\left| d\zeta \right| ^{2} -k\frac{|\zeta|^{2(k-1)}}{|z|^4}\left[ \zeta%
\overline{z}dzd\overline{\zeta } +z\overline{\zeta}d\zeta d\overline{z}%
\right].  \label{singmetric}
\end{eqnarray}
The singularity of the orbifold is resolved by replacing it by a complex
two-dimensional manifold representing the deformed geometry near the origin
of the $A_{k-1}$ ALE space \cite{HOV}. Since there are two mirror ways of
deforming the singularity, namely K\"ahler and complex deformations, we will
study both of them in the following and work out the corresponding
non-singular pp-wave metrics. For the time being, we note that type IIB
superstring theory in the Penrose limit of $AdS_{5}\times S^{5}/\mathbb{Z}_{k}$
involves $kN$ D5-branes (partially) wrapped on a system of shrunken
two-cycles. The world-volume variables of these wrapped D5-branes are given
by the local variables $\left\{x^{+},x^{-},x^{1},x^{2}\right\}$ as indicated
in the following table:

\begin{equation}  \label{tab}
\begin{tabular}{|l|c|c|c|c|c|c|c|c|}
\hline
Coordinate: & $x^{+}$ & $x^{-}$ & $x^{1}$ & $x^{2}$ & $x^{3}$ & $x^{4}$ & $%
z_1=z$ & $z_2=\zeta^k/z$ \\ \hline
Wrapped D5-branes: & $\surd $ & $\surd $ & $\surd $ & $\surd $ & $-$ & $-$ &
$-$ & $-$ \\ \hline
$\mathbb{Z}_k$ charge: & $0$ & $0$ & $0$ & $0$ & $0$ & $0$ & $1$ & $-1$ \\
\hline
\end{tabular}
\end{equation}

\subsection{PP waves on orbifolds}

\subsubsection{Orbifolds with K\"ahler deformations}

\quad\ The K\"{a}hler resolution of the singular $A_{k-1}$ geometry is
obtained by blowing up the singular point using $k-1$ intersecting
two-spheres, $\mathbb{CP}^{1}\sim S^{2}$, arranged as shown here
\be
    \mbox{
         \begin{picture}(20,30)(70,0)
        \unitlength=2cm
        \thicklines
    \put(0,0.2){\circle{.2}}
     \put(.1,0.2){\line(1,0){.5}}
     \put(.7,0.2){\circle{.2}}
     \put(.8,0.2){\line(1,0){.5}}
     \put(1.4,0.2){\circle{.2}}
     \put(1.6,0.2){$.\ .\ .\ .\ .\ .$}
     \put(2.5,0.2){\circle{.2}}
     \put(2.6,0.2){\line(1,0){.5}}
     \put(3.2,0.2){\circle{.2}}
     \put(-1.2,.15){$A_{k-1}:$}
  \end{picture}
}
\label{ordAk}
\ee
where the nodes represent $\mathbb{CP}^1$ curves, while their intersections
are represented by the links.
To describe the blow-up of the $A_{k-1}$ singularity, one may introduce $k$
complex two-dimensional planes $\mathcal{U}_{i}$ parameterized by 
$(u_{i},v_{i})$ \cite{HOV}. In this way, the coordinates of the original
singular manifold are realized as
\begin{eqnarray}
z_{1} &=&u_{i}^{i}v_{i}^{i-1},  \notag \\
z_{2} &=&u_{i}^{k-i}v_{i}^{k+1-i},  \notag \\
\zeta &=&u_{i}v_{i}.  \label{blow1}
\end{eqnarray}
The transition functions describing how the patches are glued together are
given by
\begin{equation}
v_{i}u_{i+1}=1,\ \ \ \ \ \ \ \ \ \ \ u_{i}v_{i}=u_{i+1}v_{i+1}.
\label{transition}
\end{equation}
Combined, the variables $\{u_{i},v_{i}\}$ describe a system of $k-1$
intersecting rational curves, $\mathcal{C}_{i}$, given by
\begin{eqnarray}
\mathcal{C}_{i} &=&\left\{ v_{i}u_{i+1}=1,\quad u_{i}=v_{i+1}=0\right\} ,\ \
\ \ \ \ \ \ i=1,\ldots ,k-1,  \notag \\
\mathcal{C}_{i}\cap \mathcal{C}_{j} &=&\varnothing ,\ \ \ \text{unless}\
j=i\pm 1,  \notag \\
\mathcal{C}_{i}\cap \mathcal{C}_{i+1} &=&\{u_{i+1}=v_{i+1}=0\}.  \label{cc}
\end{eqnarray}
The manifold characterized by these intersecting curves is smooth and mapped
isomorphically into the singular $A_{k-1}$ surface, except at the inverse
image of the singular point $(z_{1},z_{2},\zeta )=(0,0,0)$.

{}From the relations (\ref{transition}) and the transformations of $z_1$, 
$z_2$ and $\zeta$ in (\ref{Zk}), one sees that the orbifold group 
$\mathbb{Z}_{k}$ acts on the variables $u_{i}$ and $v_{i}$ as
\begin{equation}
\mathbb{Z}_k\ :\ \ \ \ \ \ u_{i}\mapsto \omega^iu_i, \qquad v_{i}\mapsto
\overline{\omega}^{i+1}v_{i}.
\label{Zuv}
\end{equation}

Based on (\ref{blow1}), one can write down the pp-wave metric of the Penrose
limit of type IIB superstrings on the blown-up singularity of $AdS_{5}\times
S^{5}/\mathbb{Z}_{k}$. On the $j$-th patch $\mathcal{U}_{j}$, the
differentials $dz_1$, $dz_2$ and $d\zeta$ may be expressed in terms of 
$du_{j}$ and $dv_{j}$ as
\begin{eqnarray}
dz_1 &=&ju_{j}^{j-1}v_{j}^{j-1}du_{j}+(j-1) u_{j}^{j}v_{j}^{j-2}dv_{j},
\notag \\
dz_2 &=&(k-j) u_{j}^{k-j-1}v_{j}^{k+1-j}du_{j}+(k+1-j)
u_{j}^{k-j}v_{j}^{k-j}dv_{j},  \notag \\
d\zeta &=&du_{j}v_{j}+u_{j}dv_{j}.  \label{dz1dz2}
\end{eqnarray}
The metric of the $j$-th patch $\mathcal{U}_{j}$ now follows directly from 
(\ref{ppwavemetric}) by substituting the relations (\ref{blow1}) and the
expressions (\ref{dz1dz2}) for $dz_1$ and $dz_2$. We find
\begin{eqnarray}
ds^2|_{A_{k-1},\mathcal{U}_j}&=&-4dx^{+}dx^{-} +d\mathbf{x}^2-\mu ^{2}
\left(\mathbf{x}^2+|u_j|^{2j}|v_j|^{2j-2}
+|u_j|^{2k-2j}|v_j|^{2k+2-2j}\right)(dx^{+})^2  \notag \\
&&+\left\{j^2|u_j|^{2j-2}|v_j|^{2j-2}
+(k-j)^2|u_j|^{2k-2j-2}|v_j|^{2k-2j+2}\right\}|du_j|^2  \notag \\
&&+\left\{(j-1)^2|u_j|^{2j}|v_j|^{2j-4}
+(k-j+1)^2|u_j|^{2k-2j}|v_j|^{2k-2j}\right\}|dv_j|^2  \notag \\
&&+\left\{j(j-1)|u_j|^{2j-2}|v_j|^{2j-4}
+(k-j)(k-j+1)|u_j|^{2k-2j-2}|v_j|^{2k-2j}\right\}  \notag \\
&&\ \ \ \ \times\left(\overline{u}_jv_jdu_jd\overline{v}_j 
+u_j\overline{v}_jd\overline{u}_jdv_j\right).  
\label{ds2Uj}
\end{eqnarray}
In the special case of a $\mathbb{Z}_{2}$ orbifold, we have two coordinate
patches, $\mathcal{U}_{1}$ and $\mathcal{U}_{2}$, to describe the blown-up
geometry. The metric of the resolved pp wave based on the coordinate patch 
$\mathcal{U}_{1}$ reads
\begin{eqnarray}
ds^2|_{A_1,\mathcal{U}_1}&=&-4dx^{+}dx^{-} +d\mathbf{x}^2-\mu ^{2}
\left(\mathbf{x}^2+|u_1|^2 +|u_1|^2|v_1|^4\right)(dx^{+})^2  \notag \\
&&+\left(1+|v_1|^4\right)|du_1|^2+4|u_1|^2|v_1|^2|dv_1|^2 \notag \\
&&+2|v_1|^2\left(\overline{u}_1v_1du_1d\overline{v}_1 
+u_1\overline{v}_1d\overline{u}_1dv_1\right)  \label{ds2U1}
\end{eqnarray}
while the metric based on the coordinate patch $\mathcal{U}_{2}$ reads
\begin{eqnarray}
ds^2|_{A_1,\mathcal{U}_2}&=&-4dx^{+}dx^{-} +d\mathbf{x}^2
-\mu^{2}\left(\mathbf{x}^2+|u_2|^4|v_2|^2 +|v_2|^2\right)(dx^{+})^2  \notag \\
&&+4|u_2|^2|v_2|^2|du_2|^2+\left(1+|u_2|^4\right)|dv_2|^2 \notag \\
&&+2|u_2|^2\left(\overline{u}_2v_2du_2d\overline{v}_2 
+u_2\overline{v}_2d\overline{u}_2dv_2\right).  
\label{ds2U2}
\end{eqnarray}
It is recalled that the transition functions between the two realizations 
(\ref{ds2U1}) and (\ref{ds2U2}) of the metric $ds^{2}$ of the pp wave are
given by (\ref{transition}).

On the blown-up manifold, the original $kN$ wrapped D5-branes are now
partitioned into $k-1$ subsets of $N_{i}$ D5-branes at the origins of the
rational curves, $\mathcal{U}_i$, and the original gauge group $U(kN)$
breaks down according to
\begin{equation}
U(kN)\ \rightarrow\ \left[\otimes _{i=1}^{k-1}U(N_i)\right],
\end{equation}
where
\begin{equation}
kN=\sum_{i=1}^{k-1}N_i.
\end{equation}

Just as $z_1$ and $z_2$, the new coordinates $\{u_i,v_i\}$ have
superpartners. They will be discussed in Section 4 when we study
supersymmetric blow-ups, and in the derivation of an extended BMN
correspondence for $ADE$ orbifolds.

\subsubsection{Orbifolds with complex deformations}

\quad\ In mirror geometry, the deformation of the $A_{k-1}$ singularity 
(\ref{Asing}) is obtained by introducing $k+1$ complex variables $\tau _{i}$
satisfying the following relation \cite{KKV,KMV}
\begin{equation}
\tau _{i}\tau _{i+2}=\tau _{i+1}^{2}.  \label{miro}
\end{equation}
These constraint  equations can also be put into a more convenient form as
\begin{equation}
\prod_{i=1}^{k+1}\tau _{i}^{\ell _{i}^{(a)}}=1,\qquad a=1,\ldots ,k-1
\end{equation}%
where the $\ell ^{(a)}$ are integer vectors given by
\begin{eqnarray}
\ell ^{(1)} &=&(1,-2,1,0,0,0,\ldots ,0),  \notag \\
\ell ^{(2)} &=&(0,1,-2,1,0,0,\ldots ,0),  \notag \\
&\vdots &  \notag \\
\ell ^{(k-1)} &=&(0,0,0,0,\ldots ,1,-2,1).  \label{mori}
\end{eqnarray}
These vectors are recognized as a simple extension of minus the Cartan
matrix $K_{ab}$ of the $A_{k-1}$ Lie algebra. As far as explicit
realizations of $\tau _{i}$ are concerned, (\ref{miro}) may be solved
naturally in terms of monomials of two independent complex variables 
$\eta$ and $\xi$ as follows
\begin{equation}
\tau _{i}=\eta ^{k+1-i}\xi ^{i-1}=\eta ^{k}\zeta ^{i-1},\ \ \ \ \ \ \
i=1,\ldots ,k+1,\ \ \ \ \ \ \ \quad \zeta \ =\ \frac{\xi }{\eta }.
\label{tau1}
\end{equation}
In toric geometry \cite{KMV}, the monomials $\zeta ^{i}$ are associated with
the compact curves $\mathcal{C}_{i}$ of (\ref{cc}), and the quiver graph for
the mirror geometry is the Dynkin diagram of the $A_{k-1}$ Lie algebra as
shown in (\ref{ordAk}). By varying the homogeneous factor, one may in
general fix one of the variables $\eta $ or $\xi $. In the particular case
where $\eta =1$, which corresponds to $\tau _{1}=1$, it follows from 
(\ref{tau1}) that $\tau _{i}=\xi ^{i-1}$. It is also noted that
\begin{equation}
\tau _{i}\rightarrow \omega ^{i-1}\tau _{i},\ \ \ \ \ \ \ \ \ i=1,\ldots ,k+1
\label{mzktr}
\end{equation}
provides a natural representation of the action of $\mathbb{Z}_{k}$ on the
variables $\tau _{i}$, cf. (\ref{miro}).

To get the pp-wave metric associated with the complex
deformation of $A_{k-1}$ singularity, one may consider the situation where
the singularity is partially solved (see (\ref{zza})) by taking
\begin{equation}
z_1z_2=\zeta^{k}+\sum_{i=1}^{k-j}a_{i}\zeta^{k-i}, \qquad 1\leq j\leq k
\end{equation}
where $z_1=\eta^k$ and $z_2=\tau_{j+1}$. In this case, the $N_{j}$ D5-branes
wrapping the spheres represented by  
$j$-th node $\mathcal{O}_j$ of the complex deformation of the 
$A_{k-1}$ singularity are parameterized as indicated in the following table:

\begin{equation}
\begin{tabular}{|l|l|l|l|l|l|l|l|c|}
\hline
Coordinate: & $x^{+}$ & $x^{-}$ & $x^{1}$ & $x^{2}$ & $x^{3}$ & $x^{4}$ & $%
z_1$ & $\tau _{j}$ \\ \hline
Wrapped D5-branes: & $\surd $ & $\surd $ & $\surd $ & $\surd $ & $-$ & $-$ &
$-$ & $-$ \\ \hline
$\mathbb{Z}_{k}$ charge: & $0$ & $0$ & $0$ & $0$ & $0$ & $0$ & $1$ & $j-1$
\\ \hline
\end{tabular}%
\end{equation}
\newline
Using these mirror complex variables, we have $|z_1|^{2}=|\eta|^{2k}$, $%
|\tau _{j}|^{2}=|\eta|^{2k}|\zeta|^{2j}$ and
\begin{eqnarray}
dz_1&=&k\eta ^{k-1}d\eta ,  \notag \\
dz_2&=&\zeta ^{j}dz_1+jz_1\zeta ^{j-1}d\zeta.
\end{eqnarray}
The metric of the pp wave on the deformed geometry that follows from (\ref%
{ppwavemetric}) by substituting the expressions of $\tau_j$ in terms of $z_1$
and $\zeta$ reads
\begin{eqnarray}
ds^{2}|_{A_{k-1},\mathcal{O}_{j}} &=&-4dx^{+}dx^{-}+d\mathbf{x}^2 -\mu
^{2}\left(\mathbf{x}^{2}+\left| z_1\right| ^{2}\left( 1+\left| \zeta \right|
^{2j}\right) \right)(dx^{+})^{2}  \notag \\
&&+\left(1+|\zeta|^{2j}\right)\left|dz_1\right|^2+j^2|z_1|^2|\zeta|^{2(j-1)}
|d\zeta|^2  \notag \\
&&+j|\zeta|^{2(j-1)}\left(\zeta\overline{z}_1dz_1d\overline{\zeta} +z_1%
\overline{\zeta }d\zeta d\overline{z}_1\right).  \label{Ak1}
\end{eqnarray}
Particular cases are easily described by substituting given values for $j$.

The above analysis for the $A_{k-1}$ singularity can be extended to the
other possible finite and affine singularities of the ALE space. These
extensions are addressed in Appendix A, and are based on a specification
of the K\"ahler and complex properties of the sets $\{\mathcal{O}_j\}$ used
to resolve the singularity.

\textbf{Open-string sector}: \quad\ A way to see how open strings can be
implemented is to consider the adjunction of fundamental matter in the dual
field theory. To that purpose recall that in the dual four-dimensional 
$\mathcal{N}=2$ field theory of the above type IIB superstring on the 
$AdS_{5}\times S^{5}/\mathbb{Z}_{k}$ orbifold, the $a_{i}$ moduli
parameterizing the complex deformation of the singularity are associated
with the vacuum expectation values of the scalars of the $\mathcal{N}=2$ 
$\left[ \otimes _{i=1}^{k-1}U(N_{i})\right] $ hypermultiplets 
$\left( 0^{4},\frac{1}{2}^{2},1\right) $ in the bi-fundamental representation of 
$U(N_{i})\otimes U(N_{i+1})$. To see how this follows from the original 
$\mathcal{N}=4$ $U(kN)$ SYM$_{4}$, it is enough to decompose the gauge
multiplet
\begin{equation}
\left( 0^{6},\frac{1}{2}^{4},1\right) \otimes (kN,\overline{kN}),
\label{promotion}
\end{equation}%
with $6(kN)^{2}+2(kN)^{2}=8(kN)^{2}$ bosonic degrees of freedom and as many
fermionic ones, on two representations of the orbifold group along the lines
of \cite{LNV}. Under deformation of the $A_{k-1}$ singularity, the orbifold
gauge group $U(kN)$ is generally broken down to $\left[ \otimes
_{i=1}^{k-1}U(N_{i})\right] $ with $kN=\sum_{i=1}^{k-1}N_{i}$ and 
$\mathcal{N}=4$ supersymmetry is broken down to $\mathcal{N}=2$. 
In this case, the previous $\mathcal{N}=4$ vector multiplet 
decomposes as $\mathcal{N}=2$ gauge multiplets 
$\left( 0^{2},\frac{1}{2}^{4},1\right) $ plus
hypermultiplets $\left( 0^{4},\frac{1}{2}^{2}\right) $. 
In the $\mathcal{N}=2$ quiver gauge theory where the $A_{k-1}$ 
singularity is completely resolved, the $U(kN)$ gauge-group multiplet 
(\ref{promotion}) decomposes into $k-1$ $\mathcal{N}=2$ on-shell 
gauge multiplets in the adjoint representations of $U(N_{i})$
\begin{equation}
\left( 0^{2},\frac{1}{2}^{2},1\right) \otimes \left( N_{i},\overline{N}_{i}\right),
\ \ \ \ \ \ \ \ \ i=1,\ldots ,k-1,
\end{equation}
in addition to $k-2$ hypermultiplets $\left( 0^{4},\frac{1}{2}^{2}\right) $
in the bi-fundamental representations of the $U(N_{i})\otimes U(N_{i+1})$
subgroups of the gauge group $[\otimes _{i=1}^{k-1}U(N_{i})]$ as
\begin{equation}
\left( 0^{4},\frac{1}{2}^{2}\right) \otimes \left( N_{i},\overline{N}_{i+1}\right),
\ \ \ \ \ \ \ \ \ i=1,\ldots ,k-2.
\end{equation}
Under deformation of the orbifold point, the initial group symmetry 
$SO(6)\otimes U(kN)$ breaks down to 
$SO(2)\otimes SU(2)\otimes \left[ \otimes_{i=1}^{k-1}U(N_{i})\right]$. 
In this way, the original six scalar fields $\phi _{1},\ldots ,\phi _{6}$ 
of (\ref{promotion}), transforming in the 
$\left( 6,N\otimes \overline{N}\right) $ representation of $SO(6)\otimes
U(kN) $, split as follows
\begin{equation}
\left( 6,kN\otimes k\overline{N}\right) \ \longrightarrow \ 
\left[ \oplus_{i=1}^{k-1}\left( 2,1,N_{i}\otimes \overline{N}_{i}\right) \right] 
\oplus \left[ \oplus_{i=1}^{k-2}\left(1,2,N_{i}\otimes\overline{N}_{i+1}
\right)\right] .  
\label{b}
\end{equation}
For the spin-one gauge field $A_{\mu }$ at the orbifold point, the
decomposition that follows after the complex deformation of $A_{k-1}$
singularity reads
\begin{equation}
\left( 1,kN\otimes k\overline{N}\right) \ \longrightarrow \ \left[ \oplus
_{i=1}^{k-1}\left( 1,1,N_{i}\otimes \overline{N}_{i}\right) \right] .
\label{c}
\end{equation}
Similarly for the gauginos where we have
\begin{equation}
\left( 4,kN\otimes \overline{kN}\right) \ \longrightarrow \ \left[ \oplus
_{i=1}^{k-1}\left( 1,2,N_{i}\otimes \overline{N}_{i}\right) \right] \oplus 
\left[ \oplus _{i=1}^{k-2}\left( \pm 1,1,N_{i}\otimes \overline{N}_{i+1}
\right) \right] .  \label{b1}
\end{equation}

Note that the first contributions to (\ref{b}) and (\ref{b1}) combined with 
(\ref{c}) make up the $\mathcal{N}=2$ gauge multiplets in the adjoint
representation of $U(N_{i})$, while the second contributions to (\ref{b})
and (\ref{b1}) give bi-fundamental matter. Note also that under this
orbifolding there is no fundamental matter that follows from the reduction
of $\mathcal{N}=4$ SYM$_{4}$. 
Finally, while $\mathcal{N}=4$ SYM$_{4}$ is
scale invariant, its reduction to $\mathcal{N}=2$ SYM$_{4}$ is not necessary
a conformal theory since the beta function (\ref{beta}) may not
vanish. They do, though, for
orbifolds associated with $\mathcal{N}=2$ quiver gauge theories based on
affine $\widehat{ADE}$ singularities, while the models based on finite
singularities are not conformal. This is illustrated by the example of
ordinary $A_{k-1}$ in which case $\beta _{1}$ and $\beta _{k-1}$
are non zero. To recover conformal invariance one needs the introduction of
fundamental matter by allowing the complex moduli to have extra dependencies
as in (\ref{fiber}). On the string-theory side this corresponds to allowing
open-string sectors stretching between wrapped D5-branes and wrapped
D7-branes. In Section 5, we describe such systems. In Section 6, we 
give the correspondence between type IIB superstring states on pp waves and
gauge-invariant operators in the above $\mathcal{N}=2$ CFT$_{4}$s,
including those involving fundamental matter.

\section{$\mathcal{N}=2$ superconformal theories in four dimensions}

\quad\ Due to their special features, supersymmetric conformal field
theories in various dimensions have been a subject of great interest over the
last couple of decades. A particular one of these scale-invariant theories
has recently been studied intensively.
It is the four-dimensional $\mathcal{N}=4$ 
superconformal field theory with both gauge and matter fields in the adjoint
representation of the gauge group \cite{M1}. A simple counting of the
degrees of freedom of $\mathcal{N}=4$ $SU(N)$ QFT$_{4}$ shows that the
following beta function vanishes identically
\begin{equation}
\beta =\frac{11}{3}C(G)-\frac{2}{3}T(R)-\frac{1}{6}T(S).
\end{equation}
In this equation, $C(G)$ is the Casimir of the gauge group, $T(R)$ is the
number of fundamental fermions, while $T(S)$ is the number of adjoint
scalars. For $SU(N)$ gauge symmetry, we have $C(G)=N$, $T(R)=4N$
and $T(S)=6N$, in which case the beta function indeed vanishes.
The $\mathcal{N}=4$ $SU(N)$ gauge theory is then a
critical theory which can be embedded in string theory compactifications
preserving 32 supercharges. It also plays a central role in the study of 
extensions of the BMN proposal. 

Besides this standard example, there are several
other four-dimensional conformal field theories, 
though with a lower number of conserved supercharges.
These models are essentially obtained from four-dimensional $\mathcal{N}=4$ 
$SU(N)$ gauge theory by appropriate deformations preserving a fraction of the
32 original supercharges. Examples are the four-dimensional $\mathcal{N}=2$ 
models being obtained as resolutions of orbifolds based on discrete
subgroups of $SU(2)$. In these examples the beta functions $\beta_{i}$ (\ref{beta})
may be written as
\begin{equation}
\beta_{i}=\frac{11}{3}\left(N_{i}-\frac{1}{4}\sum_{j}a_{ij}^{4}N_{j}\right),  
\label{bii}
\end{equation}
where we have used the relations (\ref{nu}). The vanishing conditions for
these beta functions can be brought to a form familiar from Lie algebra
theory, namely
\begin{equation}
\beta _{i}=\frac{11}{6}\sum_{j}K_{ij}N_{j}=0,  
\label{bii2}
\end{equation}
where the matrix $K_{ij}$ is given by
\begin{equation}
K_{ij}=2\delta _{ij}-\frac{1}{2}a_{ij}^{4}.
\end{equation}
This matrix is not necessarily linked directly to a standard Cartan matrix,
though affine Kac-Moody algebras do provide natural candidates,
as we then have simple identities of the form
\begin{equation}
a_{ij}^{4}=2\delta _{i,j-1}+2\delta _{i,j+1}.
\end{equation}

Situations based on finite or indefinite
Lie algebras have been considered in \cite{ABS1,ABS2}. 
They require adding fundamental matter to the bi-fundamental matter,
already present in the affine case, in order to satisfy the 
vanishing condition of the beta function. The engineering of $\mathcal{N}=2$ 
CFT$_{4}$s therefore depends on the matter content of the gauge theory, 
where the possible matter contents are linked to the
classification theorem of Kac-Moody algebras \cite{Kac,malika,iran,AABDS}. 
This theorem classifies the algebras according to three sectors, in particular,
namely finite ($q=1$), affine ($q=0$) and indefinite ($q=-1$) Lie algebras
\begin{equation}
\sum_{j}K_{ij}^{(q)}n_{j}=qm_{i},\ \ \ \ \qquad q=0,\pm 1,  
\label{clas}
\end{equation}
where $K_{ij}^{(q)}$ denotes the generalized Cartan matrix, while 
$n_{j}$ and $m_{j}$ are positive integers. One should therefore 
expect that there are three classes of $\mathcal{N}=2$ CFT$_4$s
\cite{ABS1,ABS2}. Here we will discuss the two classes associated to 
ordinary or affine Lie algebras. We leave the third class
based on indefinite Lie algebras to future investigations.

Following \cite{KMV,BFS}, one may distinguish between two classes of 
$\mathcal{N}=2$ superconformal $\left[\otimes_{i}SU(N_i)\right]$ quiver gauge 
theories in four dimensions according to whether or not the model contains
fundamental matter. That is, we have the following two classes of models: 
(i) Four-dimensional $\mathcal{N}=2$ SCFT with gauge group $G_{g}$ of the form
\begin{equation}
G_{g}=\left[\otimes_{i}SU(N_i)\right]  \label{Gg},
\end{equation}
with a special set of bi-fundamental matter $\left(N_{i},\overline{N}_{j}\right)$. 
(ii) Four-dimensional $\mathcal{N}=2$ SCFT with symmetry group
$G$ composed of a gauge group $G_{g}$ as in the first class but here
with an extra $G_{f}$ flavour symmetry, i.e.,
\begin{equation}
G=G_{g}\otimes G_{f}.
\label{GGG}
\end{equation}
In this case, there is a chain of bi-fundamental matter 
$\left(N_{i},\overline{N}_{j}\right)$
in addition to the fundamental matter $\left(N_{l}\right)$ transforming under 
the flavour group. The numbers $n_{f}$ and $n_{bif}$ of fundamental and
bi-fundamental matter, respectively, cannot be arbitrary because of 
the condition that the beta function must vanish.
Furthermore, $n_{f}$ and $n_{bif}$ depend on the type of the underlying 
orbifold. 

The four-dimensional $\mathcal{N}=2$ quiver gauge models
we will consider in the present study admits an elegant description in terms
of the Cartan matrices $K$ and $\widehat{K}$ of ordinary 
$ADE $ and affine $\widehat{ADE}$ Lie algebras, respectively, where
$K$ refers to $K^{(+)}$ while $\widehat{K}$ corresponds to $K^{(0)}$ in 
the notation of (\ref{clas}). If we suppose that the symmetry group $G$ of 
the quiver gauge theory is given by (\ref{GGG}) with gauge group
(\ref{Gg}) and flavour symmetry 
\begin{equation}
G_f=\left[\otimes_{l}SU(M_l)\right],
\end{equation}
then the conformal invariance conditions $\beta_i=0$ can be translated into
an algebraic constraint equation on the matter content of the model.
More precisely, we have
\begin{equation}  \label{betaaff}
\beta_i =0\qquad \Leftrightarrow \qquad \sum_{j} \widehat{K}_{ij}N_{j}=0
\end{equation}
in the case of bi-fundamental matter only, and
\begin{equation}  \label{betaord}
\beta_i =0\qquad \Leftrightarrow \qquad \sum_{j}K_{ij}N_{j}=M_{i},
\end{equation}
in the case where one has both bi-fundamental and fundamental matter. These
algebraic relations were first noted in the context of geometric
engineering of $\mathcal{N}=2$ QFT$_{4}$ embedded in type II
superstring theory \cite{KMV,BFS}. Here we want to exploit this important
observation to discuss the various types of $\mathcal{N}=2$ CFT$_4$s. Then
we use these results to study extensions of the BMN correspondence
between gauge-invariant operators of $\mathcal{N}=2$ CFT$_4$ and type IIB
superstring theory on pp waves with $ADE$ geometry. 
First, though, we shall discuss the solutions to the vanishing beta function 
conditions with initial focus on (\ref{betaaff}) followed by an analysis
of (\ref{betaord}).

Quiver CFT$_4$s based on indefinite Lie algebras have been studied in 
\cite{ABS1,ABS2}.

\subsection{$\mathcal{N}=2$ affine CFT$_4$s}

\quad\ We will discuss individually the various $\mathcal{N}=2$
quiver CFT$_4$s associated with the affine $\widehat{ADE}$ Kac-Moody
algebras. The $\mathcal{N}=2$ affine 
$\widehat{A}_{k}$ CFT$_4$s are analyzed in the following, 
whereas results for $\widehat{D}_{k}$ and 
$\widehat{E}_{s}$ are deferred to Appendix A.

In our context, the simplest example of an 
$\mathcal{N}=2$ superconformal field theory without
fundamental matter is based on the affine Lie algebra $\widehat{A}_{k}$. This
model has no flavour symmetry while the gauge group is 
$G_{g}=\left[\otimes_{i=0}^kU(N_i)\right]$ subject to $\sum_{i=0}^{k}N_{i}=(k+1)N$. 
One may also evaluate the $N_{i}$'s explicitly as 
conformal invariance imposes the condition
\begin{equation}
\left(
\begin{array}{ccccccc}
2 & -1 & 0 & 0 & \ldots & 0 & -1 \\
-1 & 2 & -1 & 0 & \ldots & 0 & 0 \\
0 & -1 & 2 & -1 & \ldots & 0 & 0 \\
\vdots & \ddots & \ddots & \ddots & \ddots & \ddots & \vdots \\
0 & 0 & \ddots & -1 & 2 & -1 & 0 \\
0 & 0 & \ldots & 0 & -1 & 2 & -1 \\
-1 & 0 & \ldots & 0 & 0 & -1 & 2
\end{array}
\right) \left(
\begin{array}{c}
N_{0} \\
N_{1} \\
N_{2} \\
\vdots \\
N_{k-2} \\
N_{k-1} \\
N_{k}
\end{array}
\right) =\left(
\begin{array}{c}
0 \\
0 \\
0 \\
\vdots \\
0 \\
0 \\
0
\end{array}
\right),
\end{equation}
corresponding to the following system of linear equations
\begin{eqnarray}
2N_{1}-N_{2}-N_{k} &=&0,  \label{aff1} \\
-N_{j-1}+2N_{j}-N_{j+1} &=&0,\quad \ \ \ \ \ 2\leq j\leq k-1  \label{aff2} \\
-N_{1}-N_{k-1}+2N_{k} &=&0.  \label{aff3}
\end{eqnarray}
These equations can be solved straightforwardly by
\begin{equation}
N_{i}=N,
\end{equation}
in which case the gauge group of the $\mathcal{N}=2$ SCFT$_{4}$ simply reads
\begin{equation}
G=\left[U(N)\right]^{k+1}.
\end{equation}

The set of bi-fundamental matter 
\begin{equation}
\{ \mathcal{H}_{i,\overline{j}}= \left(N_{i},\overline{N}_{j}\right) \ 
|\ j-i\equiv 1\ \text{mod}\ k+1\}
\end{equation} 
forms a one-dimensional quiver with $k+1$
edges where an edge like $\mathcal{H}_{i,\overline{i+1}}$ is associated 
with a hypermultipet in the bi-fundamental of 
$U(N_{i})\otimes U(N_{i+1})= U(N)\otimes U(N)$. 
This system is seen to be
described by the affine $\widehat{A}_{k}$ Dynkin diagram
\begin{equation}
    \mbox{
         \begin{picture}(20,90)(50,-20)
        \unitlength=2cm
        \thicklines
      \put(-1.2,.3){$\widehat{A}_k:$}
    \put(0,0){\circle{.2}}
     \put(.1,0){\line(1,0){.5}}
     \put(.7,0){\circle{.2}}
     \put(.9,0){$.\ .\ .\ .\ .\ .$}
     \put(1.8,0){\circle{.2}}
     \put(1.9,0){\line(1,0){.5}}
     \put(2.5,0){\circle{.2}}
     \put(0,.1){\line(2,1){1.15}}
     \put(1.25,.7){\circle{.2}}
     \put(2.5,.1){\line(-2,1){1.15}}
  \end{picture}
}
\label{affAk}
\end{equation} 
The complete set of hypermultiplets is thus given by the following chain
\begin{equation}
\mathcal{H}_{0,\overline{1}},\ \ \mathcal{H}_{1,\overline{2}},\  \ldots,\
\ \mathcal{H}_{k-1,\overline{k}},\ \ \mathcal{H}_{k,\overline{0}}.
\end{equation}

Each $\mathcal{N}=2$ hypermultiplet 
$\mathcal{H}_{i,\overline{i+1}}$ carries $4+4$ on-shell degrees 
of freedom:
\begin{eqnarray}
\mathcal{H}_{i,\overline{i+1}} &=&\left( 0^{4},\frac{1}{2}^{2}\right) ,
\notag \\
0^{4} &\sim &\mathcal{\phi }_{i,\overline{i+1}}^{\alpha },\quad \alpha =1,2,
\notag \\
\frac{1}{2}^{2} &\sim &\mathcal{\psi }_{i,\overline{i+1}}\oplus 
\overline{\mathcal{\chi }}_{i,\overline{i+1}}.
\end{eqnarray}
It may be decomposed into two $\mathcal{N}=1$ chiral multiplets, here
denoted as $Q_{i,\overline{i+1}}\sim \left( q_{i,\overline{i+1}},
\psi _{i,\overline{i+1}}\right) $ and $\widetilde{Q}_{i,\overline{j}}\sim \left(
\widetilde{q}_{i,\overline{i+1}},\widetilde{\psi }_{i,\overline{i+1}}\right)$.
Their charges under a $U(1)$ subgroup of the 
$SU(2)\subset SO(4)$ R-symmetry are $\frac{1}{2}$ and $-\frac{1}{2}$,
respectively. In terms of these multiplets, the matter content of the 
affine $\widehat{A}_{k}$ quiver CFT$_{4}$ reads
\begin{equation}
\left(
\begin{array}{c}
Q_{0,\overline{1}} \\
\widetilde{Q}_{0,\overline{1}}
\end{array}
\right) ,\ \ldots ,\left(
\begin{array}{c}
Q_{k,\overline{0}} \\
\widetilde{Q}_{k,\overline{0}}
\end{array}
\right) .
\end{equation}
The scalar fields $q_{i,\overline{j}}$ and $\widetilde{q}_{i,\overline{j}}$ of
the $Q_{i,\overline{j}}$ and $\widetilde{Q}_{i,\overline{j}}$ chiral
multiplets are associated with the complex deformations of the affine 
$\widehat{A}_{k}$ singularity of the pp waves.

\subsection{Ordinary CFT$_4$ models}

\quad\ Here we discuss how the above results can be extended to the
case where, in addition to bi-fundamental matter, we also have fundamental
matter carrying flavour charges. A class of such extensions is
comprised of the four-dimensional $\mathcal{N}=2$ conformal field
theories based on \textsl{finite} $ADE$ geometries. After a brief review of 
these models, we show how the vanishing of the corresponding beta function 
may be ensured.

It is recalled that for 
any Cartan matrix $K_{ij}$ of finite $ADE$ Lie algebras   
and a positive integer vector $N_j$, the linear combination $M_{i}$ with
\begin{equation}
M_{i}=\sum_{j}K_{ij}N_{j}
\label{MKN}
\end{equation}
is in general a positive integer vector. These positive integers 
$M_{i}$, which were null in affine models as in (\ref{betaaff}), are
interpreted as the number of fundamental matter of the supersymmetric 
$U(N_i)$ quiver gauge theory.

Conformal field theories
with fundamental matter require a non-trivial flavour symmetry $G_{f}$. 
This result is manifestly exhibited by the constraint equations (\ref{MKN}), 
where  the $N_{i}$ integers are the orders
of the gauge group factors $U(N_{i})\subset \left[ \otimes
_{i=1}^{k-1}U(N_{i})\right] $ and the $M_{i}$s are the orders of the 
$U(M_{i})$ flavour global symmetries. In the following table, we list the
basic content of an $\mathcal{N}=2$ ordinary $A_{k-1}$
CFT$_{4}$. The flavour symmetry is $G_{f}=\left[
\otimes _{i=1}^{k-1}U(M_{i})\right]$, while the 
gauge group $G_{g}$ is taken as a product of $U(N_{i})$ factors with the
condition $2N_{i}\geq N_{i-1}+N_{i+1}$ for $2\leq i\leq k-2$ and $2N_{1}\geq
N_{2}$ and $2N_{k-1}\geq N_{k-2}$.

\begin{equation}  \label{fem}
\begin{tabular}{|l|l|}
\hline
Gauge group $G_{g}$: & $\left[\otimes_{i=1}^{k-1}U(N_i)\right]$ \\ \hline
Flavour group $G_{f}$: & $U(2N_1-N_2) \otimes U(2N_{k-1}-N_{k-2})$ \\
& $\otimes \left[\otimes_{i=2}^{k-2}U(2N_{i}-N_{i-1}-N_{i+1})\right]$ \\
\hline
Bi-fundamental matter: & $\oplus _{i=1}^{k-2}\left(N_{i},\overline{N}%
_{i+1}\right) $ \\ \hline
Fundamental matter: & $\oplus _{i=1}^{k-1}(M_{i}N_{i})$ \\ \hline
\end{tabular}
\end{equation}
\newline
The scale invariance now fixes the number of fundamental matter $M_{i}$ as
follows: $M_{1}=2N_{1}-N_{2}$, $M_{k-1}=2N_{k-1}-N_{k-2}$, and 
$M_{i}=2N_{i}-N_{i-1}-N_{i+1}$ for $2\leq i\leq k-2$. These numbers satisfy
manifestly the natural constraint equation
\begin{equation}
N_{1}+N_{k-1}=M_{1}+\ldots+M_{k-1}.
\end{equation}
Actually this relation may be thought of as an equation classifying the
kinds of critical models for the finite $A_{k-1}$ category. The partitions
of the number $N_{1}+N_{k-1}$ give various kinds of finite $A_{k-1}$
critical models with fundamental matter. Note that for the special case of a
flavour symmetry group of type $U(M_2)\otimes U(M_{k-2})$ with $M_{2}N_{2}$ 
and $M_{k-2}N_{k-2}$ fundamental matter, the constraint equations for conformal
invariance reads
\begin{equation}
M_{i}=0,\qquad \text{for \ }i\neq2,k-2.
\end{equation}
In the simplest case of this kind in which $M_{2}=M_{k-2}=N$, 
it follows that the constraint equation for scale invariance is solved by
\begin{eqnarray}
N_1&=&N,  \notag \\
N_i&=&2N,\ \ \ \qquad 2\leq i\leq k-2,  \notag \\
N_{k-1}&=&N.
\end{eqnarray}

\section{String and brane interpretations}

\quad\ Before studying the extension of the BMN correspondence between the
various $\mathcal{N}=2$ CFT$_4$s and superstrings in the Penrose limit of 
$AdS_{5}\times S^{5}/\Gamma $ orbifolds, let us first give their
interpretations  in terms of D-branes living on $ADE$ geometries.

\subsection{String-theory analysis}

\quad\ In the Penrose limit of $AdS_{5}\times S^{5}/{\Gamma}$ taken along
the great circle of the $S^{5}$ fixed by the orbifold, the initial 32
supercharges are reduced by half. In this case, type IIB superstring theory
on a pp-wave background with $ADE$ geometry may be 
described by 8 worldsheet
scalars $X^{I}$ and eight pairs of worldsheet Majorana fermions $(\vartheta
_{1}^{\mathrm{A}},\vartheta _{2}^{\mathrm{A}})$. All of these fields are
free but massive. In type IIB superstring theory, $\vartheta _{1}$ and $\vartheta _{2}$
have the same chirality, while the masses of the $X^{I}$ scalars and 
the $\vartheta_{i}^{\mathrm{A}}$ fermions, being equal by worldsheet 
supersymmetry, are effectively given by the RR 5-form field strength $\mu$.

Moreover, the description of the Penrose limit of $AdS_{5}\times S^{5}$ is  
based on $\mathbb{R}^{4}\mathbb{\times R}^{4}$ rather than $\mathbb{R}^{8}$.
This is manifested in the breaking of the eight-dimensional string light-cone 
$SO(8)$ symmetry rotating the eight
worldsheet scalars $X^{I}$ and their $(\vartheta _{1}^{\mathrm{A}},\vartheta
_{2}^{\mathrm{A}})$ superpartners, as it is broken down to the $SO(4)_{1}\otimes
SO(4)_{2}$ homogeneous group of the $\mathbb{R}^{4}\times\mathbb{R}^{4}$
space of coordinates $(x^{i},y^{j})$. The eight scalars $X^{I}$
and their $(\vartheta _{1}^{\mathrm{A}},\vartheta _{2}^{\mathrm{A}})$
superpartners therefore decompose as
\begin{eqnarray}
X^{I} &\sim &8_v=(4_v,1)+(1,4_v)  \notag \\
\vartheta ^{\mathrm{A}} &\sim &16_s=(4_s,4_s) ; \qquad \Gamma _{11}\vartheta
^{\mathrm{A}}=0,  \label{fer}
\end{eqnarray}
where we have set $\vartheta^{\mathrm{A}}\equiv (\vartheta _{1}^{\mathrm{A}}
+i\vartheta _{2}^{\mathrm{A}})/\sqrt{2}$, and where the numbers appearing in
the above relations stand for the dimensions of the representations. In these
relations, the representation $(4_v,1) \mathbf{+}(1,4_v)$ refer to 
$(x^{i},y^{j})$, and, due to the homomorphism $SO(4)=SU(2)\otimes SU(2)$,
each $SO(4)$ spinor representation $4_s$ can be read as 
$(2,2)$. The second relation of (\ref{fer}) is solved as 
$\vartheta ^{\mathrm{A}}\sim (\chi _{+}^{\alpha }, \xi _{-}^{\dot{\alpha}})$. 
Here $\alpha $ and $\dot{\alpha}$ are spinor indices of 
$SU(2)\otimes SU(2)$ while the sub-indices carried by 
$(\chi _{+}^{\alpha },\xi _{-}^{\dot{\alpha}})$
refer to the chiralities with respect to $SO(4)$. These group multiplets
can be further decomposed by working in the complex space 
$\mathbb{R}^{4}\mathbb{\times C}^{2}$ instead of 
$\mathbb{R}^{4}\mathbb{\times R}^{4}$. As
such the $SO(4)_{2}$ symmetry group rotating the four $y^{j}$s is broken
down to $U(2) $ and so the $SO(4)$ four vector is now viewed as a $U(2)$ two
spinor. Similarly the $SO(4)$ bi-spinor $(2,2)$ is now reduced to $(2,+1)
+(2,-1)$. A further breaking of $U(2)$ down to $U(1)\otimes U(1)$ leads to
the following splitting of the worldsheet fields 
\begin{eqnarray}
X^{I} &\rightarrow &(x^{i},y^{j})\rightarrow (x^{i},z^{1},z^{2})\,,  \notag
\\
\vartheta ^{\mathrm{A}} &\rightarrow &(\chi _{+}^{\alpha }, 
\xi_{-}^{\dot{\alpha}})\rightarrow (\lambda ^{\alpha },\psi ,\overline{\chi })  
\label{sp}
\end{eqnarray}
where $\chi_{+}^{\alpha }\equiv \lambda ^{\alpha }$ and 
$\xi _{-}^{\dot{\alpha}}\equiv\xi ^{\dot{\alpha}}$. It is noted that
the spinor components $\xi ^{\dot{1}}$ and $\overline{\xi }^{\,\dot{2}}$
transform in the same manner under the abelian orbifold group $\mathbb{Z}_{k}$.
It is therefore
convenient to combine $\xi ^{\dot{1}},$ $\overline{\xi }^{\,\dot{2}}$ and 
$\overline{\xi }^{\,\dot{1}},\xi ^{\dot{2}}$ into a Dirac spinor as 
$\left( \psi ,\overline{\chi }\right) $. This change of variables has been
used in \cite{KPRT} to study type IIB superstrings propagating on pp waves
over $\mathbb{R}^{4}/\mathbb{Z}_{k}$. We will also use it here.

\subsubsection{Ordinary $A_{k-1}$ geometry}

\quad\ Under blowing up of the $\mathbb{Z}_{k}$ orbifold singularity, the
initial orbifold point is replaced by the covering of open sets 
$\mathcal{U}_{i}$, $1\leq i\leq k$. The bosonic variables near 
the origin of $\mathbb{C}^{3}$ parameterized by $(\mathbf{x},z_1,z_2)$ 
with $z_1=z,$ $z_2=\frac{\zeta ^{k+1}}{z}$ and their superpartners 
$(\mathbf{\lambda }^{\alpha},\psi ,\overline{\chi})$ thus get promoted to the 
sets $(\mathbf{x}_{i},u_{i},v_{i})$ and $(\mathbf{\lambda}_{i}^{\alpha },
\psi _{i},\overline{\chi }_{i})$, respectively. Here the variables $u_{i}$ 
and $v_{i}$ are as in (\ref{blow1},\ref{transition}) and transform as in
(\ref{Zuv}) whereas $\mathbf{x}_{i}\longrightarrow\mathbf{x}_{i}$
under $\mathbb{Z}_{k}$. The fields transform as
\begin{eqnarray}
\mathbf{\lambda}_{i}^{\alpha }&\longrightarrow& \mathbf{\lambda}_{i}^{\alpha},  
\notag \\
\left( \psi _{i},\overline{\chi }_{i}\right)&\longrightarrow&
\left(\omega^i\psi _i,\overline{\omega}^{i+1}\overline{\chi}_i\right).
\label{zt}
\end{eqnarray}
In the mirror description, the analogue to the set 
$\left( \mathbf{x}_{i},u_{i},v_{i}\right)$ is given by 
$\left( \mathbf{x}_{i},\tau_{i}\right)$, while the analogue to
its superpartner 
$(\mathbf{\lambda }_{i}^{\alpha },\psi_{i},\overline{\chi }_{i})$ is 
$(\mathbf{\lambda }_{i}^{\alpha },\vartheta_{i})$. Here 
$\tau _{i}$ are the mirror complex variables given by (\ref{miro}) and 
$\vartheta _{i}$ are their superpartners. Their transformations under the
orbifold group are given by $\tau _{i}\longrightarrow \omega ^{i-1}\tau_{i}$, cf.
(\ref{mzktr}), with a similar change for their fermionic partners.
The different phases $\omega ^{n}$ define the various sectors 
of the worldsheet fields sensitive to the $\mathbb{Z}_{k}$ orbifolding. Put
differently, the fields get monodromies under 
$\mathbb{Z}_{k}$ transformations, resulting in the following twisted
string sectors
\begin{eqnarray}
\tau_{q}(\sigma +2\pi \alpha ^{\prime }p^{+},\tau )
&=&\omega_{q}^{q-1}\tau(\sigma ,\tau )\,,  \notag \\
\vartheta _{q}(\sigma +2\pi \alpha ^{\prime }p^{+},\tau )
&=&\omega^{q-1}\vartheta _{q}(\sigma ,\tau ).
\end{eqnarray}
The other worldsheet fields $\mathbf{x}_{i}$ and 
$\mathbf{\lambda }_{i}^{\alpha }$ remain periodic as usual
\begin{eqnarray}
\mathbf{x}_{q}(\sigma +2\pi \alpha ^{\prime }p^{+},\tau ) 
&=&\mathbf{x}_{q}(\sigma ,\tau )\,,  \notag \\
\mathbf{\lambda }_{q}^{\alpha }(\sigma +2\pi \alpha ^{\prime }p^{+},\tau )
&=&\mathbf{\lambda }_{q}^{\alpha }(\sigma ,\tau ).
\end{eqnarray}
The modes of these fields are given by

\begin{equation}  \label{amodes}
\begin{tabular}{|l|l|l|l|l|}
\hline
Field: & $\mathbf{x}_{q}^{i}$ & $\tau _{q}$ & $\mathbf{\lambda}_{q}^{\alpha}$
& $\vartheta _{q}$ \\ \hline
Mode: & $a_{n}^{i}$ & $\tau _{n(q)}$ & $\mathbf{\lambda}_{n}^{\alpha }$ & 
$\vartheta _{n(q)}$ \\ \hline
\end{tabular}
\end{equation}
\newline
where $n(q)=n+\frac{n}{k}$, $n\in\mathbb{Z}$. This analysis extends
naturally to the other orbifolds. The basic points are indicated here below 
in the cases of affine $\widehat{A}_k$ and affine $\widehat{D}_k$ geometries.

\subsubsection{Affine $\widehat{A}_k$ geometry}

\quad\ In mirror affine $\widehat{A}_k$ geometry, there are $k+1$ sets of
world-sheet variables $\left(\mathbf{x}_{i},\mathbf{\lambda }_{i}^{\alpha
}\right)$ and $\left(\tau _{i},\vartheta_{i}\right) $, $i=0,\ldots,k$,
associated with the $k+1$ nodes of the $\widehat{A}_k$ Dynkin diagram. The
monodromy conditions of the $A_{k}$ geometry in the $q$-th twisted string
sector, $q=0,\ldots ,k$, are given by
\begin{eqnarray}
\mathbf{x}_{q}(\sigma +2\pi \alpha ^{\prime }p^{+},\tau) 
&=&\mathbf{x}_{q}(\sigma ,\tau),  \notag \\
\mathbf{\lambda }_{q}^{\alpha }(\sigma +2\pi \alpha ^{\prime }p^{+},\tau )
&=&\mathbf{\lambda }_{q}^{\alpha }(\sigma ,\tau),  \notag \\
\tau_{q}(\sigma +2\pi \alpha ^{\prime }p^{+},\tau) &=&\omega ^{q}\tau(\sigma
,\tau),  \notag \\
\vartheta _{q}(\sigma +2\pi \alpha ^{\prime }p^{+},\tau )
&=&\omega^{q}\vartheta _{q}(\sigma ,\tau ),
\end{eqnarray}
while their oscillator modes are as in the table (\ref{amodes}).

\subsubsection{Affine $\widehat{D}_k$ geometry}

\quad\ In the affine $\widehat{D}_k$ geometry, we have $k+1$ sets of 
world-sheet variables $\left( \mathbf{x}_{i},\mathbf{\lambda }_{i}^{\alpha}\right)$
and $\left( \tau _{i},\vartheta_{i}\right) $, $i=1,\ldots,k+1$, associated
with the $k+1$ nodes of the affine $\widehat{D}_k$ Dynkin diagram. Since the
dihedral group of $\widehat{D}_k$ geometry contains $\mathbb{Z}_{k-2}$ as an
abelian subsymmetry, the monodromy conditions of the $\widehat{D}_k$
geometry read
\begin{eqnarray}
\tau_{q}(\sigma +2\pi \alpha ^{\prime }p^{+},\tau ) &=&\alpha
_{q}\tau_{q}(\sigma ,\tau ),\ \ \ \ \ \ \ \ \ \quad q=1,2,  \notag \\
\tau_{q}(\sigma +2\pi \alpha ^{\prime }p^{+},\tau)
&=&\omega^{q-3}\tau_{q}(\sigma ,\tau ),\ \ \ \ \ \ \quad 3\leq q\leq k-1,
\notag \\
\tau_{q}(\sigma +2\pi \alpha ^{\prime }p^{+},\tau ) &=&\beta
_{q}\tau_{q}(\sigma ,\tau ),\ \ \ \ \ \ \ \ \ \quad q=k,k+1
\end{eqnarray}
and similarly for their fermionic partners. The modes are

\begin{equation}
\begin{tabular}{|l|l|l|l|l|l|}
\hline
Field: & $\tau _{1}$ & $\tau _{2}$ & $\ldots$ & $\tau _{k}$ & $\tau _{k+1}$
\\ \hline
Mode: & $\tau _{n\left( \alpha _{1}\right) }$ & $\tau _{n(\alpha_2)}$ & 
$\ldots$ & $\tau _{n(\alpha _k)}$ & $\tau_{n(\alpha _{k+1})}$ \\ \hline
\end{tabular}
\end{equation}
\newline
In the special case where all $\alpha _{q}$s are equal to $1$, one recovers
the usual periodic boundary conditions. 

Having described the essentials on the string-theory side, we now turn to 
the field-theory interpretation of these models.

\subsection{Field-theory analysis}

\quad\ The AdS/CFT correspondence connects the spectrum of type IIB
superstring theory
on $AdS_{5}\times S^{5}$ with $\mathcal{N}=4$ $U(N)$ SYM$_{4}$
(CFT$_4$) operators on the boundary $\partial(AdS_5)$. It is also at the
basis for other induced correspondences involving theories with a lower
number of supercharges. Constructions based on taking Penrose limits and
those using orbifolding by discrete groups $\Gamma$ are concrete examples.
In the previous subsection, we have studied the string-theory side. Here we
want to explore the field-theory side. Both of them are needed for working
out the extension of the BMN correspondence which will be considered in 
the next section.

We start by noting that in type IIB brane language, the engineering of $U(kN)
$ SYM$_{4}$ theories can be achieved in different ways. The latter live in
the world volume of D-branes which, according to the number of preserved
supercharges and criticality, may involve various systems of 
branes. In particular, we have the following D-brane configurations:

(\textbf{i}) $kN$ parallel D3-branes filling the four-dimensional 
space-time and located
at the origin of the transverse $\mathbb{R}^{6}$ space as one usually does
in brane engineering of $\mathcal{N}=4$ $U(kN)$ SYM$_{4}$.

(\textbf{ii}) $kN$ parallel D5-branes wrapping vanishing two-cycles
and filling the four-dimensional space-time. These branes are located at the
origin of the transverse $\mathbb{R}^{4}$ space but the resulting $U(kN)$
gauge theory has $\mathcal{N}=2$ supersymmetry. Blowing up the vanishing
spheres breaks the $U(kN)$ gauge symmetry down to 
$\left[\otimes_{i=1}U(N_i)\right]$, as already discussed. 
This is the usual system involved in the engineering of $\mathcal{N}=2$ 
quiver QFT$_4$s based on \textit{finite} Lie algebras.

(\textbf{iii}) One may also have systems involving both D3- and D5-branes.
The D3-branes are located at the origin of the transverse $\mathbb{R}^{6}$
space, whereas the D5-branes are wrapped on vanishing two-cycles located at
the origin of the transverse space $\mathbb{R}^{4}\subset \mathbb{R}^{6}$.
This is the kind of brane system one has in the brane engineering
of $\mathcal{N}=2$ quiver CFT$_4$s classified by affine Kac-Moody algebras.
In quiver Dynkin diagram representation, the affine node is associated with
D3-branes and the others with D5-branes.

(\textbf{iv}) Parallel D7-branes wrapped on four-cycles and located at a
point on $\mathbb{R}^{2}\subset \mathbb{R}^{4}\subset \mathbb{R}^{6}$. In the
case where the four-cycles are complex surfaces of the Hirzebruch type, that is,
spheres fibered over spheres, the resulting gauge theory has $\mathcal{N}=1$
supersymmetry. However, here we will consider the D7-branes wrapped on
four-cycles with large volumes. They are needed in the brane engineering of 
$\mathcal{N}=2$ CFT$_4$s based on either finite or indefinite Lie algebras.

\subsubsection{Affine $\widehat{A}_k$ model}

\quad\ As noted before this kind of $\mathcal{N}=2$ quiver CFT$_4$ involves
D3-branes and open strings stretching between them. The total number of
D-branes is $kN$ and may be partitioned into $k+1$ factors. 
$\mathcal{N}=2$ conformal invariance therefore
requires a $U(N)^{k+1}$ gauge symmetry
with $k+1$ bi-fundamentals interpreted as open strings stretching between
them. A picture of the repartition of these branes is as in the following
loop
\begin{equation}
N\cdot\mathrm{D5}+\ldots+N\cdot\mathrm{D5}+N\cdot\mathrm{D5}
+N\cdot\mathrm{\mathbf{D3}} +N\cdot\mathrm{D5}+N\cdot\mathrm{D5}
+\ldots+N\cdot\mathrm{D5},
\end{equation}
where the two sets of $N$ D5-branes at the ends are identified. From this
D-brane configuration, one sees that we have two kinds of open strings: those
stretching between D3-D5 and open strings between D5-D5. In geometric
engineering, this quiver is represented by (\ref{affAk}) which is merely the 
Dynkin diagram of affine $\widehat{A}_{k}$. Each node, represented by a 
$N_{i}\times N_{i}$ matrix projector $\pi_{i}$ which when restricted to the proper
subspace is just the $N\times N$ identity matrix, corresponds to wrapping $N$
D5-branes ($N$ D3-branes for affine node) over two-spheres. On this node
thus lives an $\mathcal{N}=2$ $U(N)$ gauge model. Links between the nodes
describe bi-fundamental matter. The explicit field content of this 
$\mathcal{N}=2$ CFT$_4$ is described presently.

\paragraph{Four-dimensional $\mathcal{N}=2$ vector 
multiplet $\left(\mathcal{C}_{i},\mathcal{\protect\lambda}_i^\protect
\protect\alpha,\mathcal{A}_\protect\protect\mu^i\right)$:}

The gauge multiplet of the $\mathcal{N}=2$ model, 
$\left(\mathcal{C},\mathcal{\lambda}^{\alpha},\mathcal{A}_{\mu}\right)$, 
is reducible and splits as $\oplus _{i=1}^{k+1}\left(\mathcal{C}_{i},
\mathcal{\lambda}_{i}^{\alpha }, \mathcal{A}_{\mu }^{i}\right) $. 
This property can be also
derived from orbifold group theory analysis. Since the elements $g$ of the
orbifold group with an affine $\widehat{A}_{k}$ singularity are generated
by $\Pi=$diag$\left( 1,\omega ,\ldots,\omega ^{k}\right)$, transformations
of $\left( \mathcal{C},\mathcal{\lambda }^{\alpha },\mathcal{A}_{\mu
}\right)$ reads
\begin{equation}
g\mathcal{A}_{\mu}g^{-1}=\mathcal{A}_{\mu},\qquad g\mathcal{\lambda}^{\alpha
}g^{-1}=\mathcal{\lambda}^{\alpha},\qquad g\mathcal{C}g^{-1}=\mathcal{C},
\label{gt}
\end{equation}
leaving the gauge-multiplet invariant. Moreover, as $g\Pi =\Pi g$, the
solution to these equations may be expressed in terms of the projectors 
$\pi_i=|{i}\rangle\langle {i}|$:
\begin{equation}
V_{0}=\sum_{j=1}^{k+1}V_{j}\pi _{j}.
\end{equation}
Here $V_{0}$ stands for the massless 
$\mathcal{C},\mathcal{\lambda}_{a}^{\alpha },\mathcal{A}_{\mu }$ fields and 
where each component $V_{j}$
of the expansion is a $N\times N$ hermitian matrix. The first $k$ $V_{j}$'s
describe the gauge degrees of freedom in the $j$-th subset of $N$ D5-branes
while the last one describes the gauge fields on the $N$ D3-branes
associated with the affine node.

\paragraph{Four-dimensional $\mathcal{N}=2$ hypermultiplets 
$\mathcal{H}_{i,i+1}=\left(\mathcal{\protect\phi }_{i,i+1}^{\protect\alpha }, 
\mathcal{\protect\psi }_{i,i+1},\overline{\protect\chi }_{i,i+1}\right)$:}

Together with the $\mathcal{N}=2$ gauge multiplets described above, there
are also massless $k$ hypermultiplets $\mathcal{H}_{i,i+1}$ describing the
transverse positions of the $k+1$ subsets of the $N$ D-branes in the pp-wave
geometry. These hypermultiplets form collectively a matrix $\mathcal{H}$
whose component fields $\mathcal{\phi }^{\alpha }$, $\mathcal{\psi}$ and 
$\overline{\chi}$ transform under the orbifold group as
\begin{equation}
g\mathcal{\phi }^{\alpha }g^{-1}=\omega \mathcal{\phi }^{\alpha }, \qquad 
g\mathcal{\psi }g^{-1}=\omega \mathcal{\psi }, \qquad g\overline{\chi }
g^{-1}=\overline{\omega }\overline{\chi }.  \label{bif}
\end{equation}
The solution to these constraint equations reads
\begin{equation}
\mathcal{H}=\sum_{q=1}^{k+1}\mathcal{H}_{i,i+1}a_{i}^{+},
\end{equation}
where $\mathcal{H}$ transforms in the $\left(N_{i},\overline{N}_{i+1}\right)$
bi-fundamental representation of $U(N_i)\otimes U(N_{i+1})$, 
and where $a_{j}^{\pm}$ are the usual creation and annihilation operators
\begin{eqnarray}
a_{i}^{+} &=&|{i+1}\rangle\langle {i}|,\ \ \ \ \ \ \ \quad a_{j}^{-}
=|{i}\rangle\langle {i+1}|,  \notag \\
a_{i}^{-}a_{i}^{+} &=&\pi _i=|{i}\rangle\langle {i}|,  \notag \\
a_{i}^{+}\pi _{i} &=&\pi _{i+1}a_{i}^{+},\ \ \ \ \ \ \ \ \ \quad
a_{i}^{+}\pi _{i}=\pi_{i+1}a_{i}^{+}.
\end{eqnarray}
This analysis can be extended naturally to the other $\mathcal{N}=2$ affine 
$\widehat{DE}$ CFT$_4$s. As an illustration, the example based on the series 
$\widehat{D}_{k}$ is discussed in the following.

\subsubsection{Affine $\widehat{D}_{k}$ CFT$_4$}

\quad\ If we leave aside geometry, the field-theory properties of 
$\widehat{D}_{k}$ CFT$_4$s are quite similar to those we have described for
the $\widehat{A}_{k}$ affine case. Here we have $N$ D3-branes and $(2k-5)N$
D5-branes with open strings stretching between them. The D-brane
configuration we have in this case is
\begin{equation}
N\cdot\mathrm{D3}+N\cdot\mathrm{D5}+2(k-4)N\cdot\mathrm{D5}+N\cdot\mathrm{D5}
+N\cdot\mathrm{D5}.
\end{equation}
So we have $2N^{2}$ open strings between D3-D5 and open strings stretching
between the remaining D5-D5 pairs. Some of these open strings transform in
the bi-fundamental representations $\left(N,\overline{N}\right)$, some in 
$\left(2N,2\overline{N}\right)$, some in $\left(2N,\overline{N}\right)$,
while some transform in $\left(N,2\overline{N}\right)$. The transformation
of the gauge multiplet is essentially the same as in (\ref{gt}), while
transformations of hypermultiplets in the bi-fundamental representations involve

\begin{equation}
\mathcal{\phi }^{\alpha }\ =\ \left(
\begin{array}{ccccccccc}
0 & 0 & \mathcal{\phi }_{1,3}^{\alpha } & 0 &  &  & \ldots &  & 0 \\
0 & 0 & \mathcal{\phi }_{2,3}^{\alpha } & 0 &  &  &  &  & 0 \\
0 & 0 & 0 & \mathcal{\phi }_{3,4}^{\alpha } & 0 &  &  &  & 0 \\
0 & 0 & 0 & 0 & \mathcal{\phi }_{4,5}^{\alpha } & 0 &  &  & 0 \\
\vdots & \vdots & \vdots & \vdots & \ddots & \ddots & \ddots & \vdots &
\vdots \\
0 &  &  &  &  & 0 & \mathcal{\phi }_{k-2,k-1}^{\alpha } & 0 & 0 \\
0 &  &  &  &  & 0 & 0 & \mathcal{\phi}_{k-1,k}^\alpha & 
\mathcal{\phi}_{k-1,k+1}^{\alpha } \\
0 &  &  &  &  & 0 & 0 & 0 & 0 \\
0 &  &  & \ldots &  & 0 & 0 & 0 & 0
\end{array}
\right)
\end{equation}
and likewise for their fermionic partners. 

Similar results may also be
derived for the exceptional affine series. These solutions
can be obtained directly from the corresponding Dynkin diagrams. 

Now we turn
to $\mathcal{N}=2$ CFT$_4$s with fundamental matter. We first consider 
$\mathcal{N}=2$ CFT$_4$s based on ordinary $A_{k-1}$, after which 
we give the results for the other geometries.

\subsubsection{Ordinary $A_{k-1}$ model}

\quad\ Contrary to the affine case, there are no D3-branes in the ordinary case. 
The $\mathcal{N}=2$ quiver CFT$_4$ based on finite $A_{k-1}$ geometry,
in particular, lives on
the world-volume of wrapped D5-branes on spheres of the deformed $A_{k-1}$
singularity; but has in addition D7-branes needed to ensure conformal
invariance. This is a gauge theory having in addition to the gauge group 
$\left[\otimes_{i=1}^{k-1}U_{g}(N_i)\right]$, a 
$\left[\otimes_{i=1}^{k-1}U_{f}(M_i)\right]$ flavour invariance dictated by the
D7-branes wrapped on four-cycles. The latter are supposed to have volumes big
enough for the corresponding gauge symmetry to become a global
gauge symmetry. The brane system describing the $\mathcal{N}=2$ 
quiver CFT$_4$s we are interested in here is indicated in the following table

\begin{equation}
\begin{tabular}{|l|l|l|l|l|l|l|l|l|l|l|}
\hline
Coordinate: & $x^{+}$ & $x^{-}$ & $x^{1}$ & $x^{2}$ & $x^3$ & $x^4$ & $y^1$
& $y^2$ & $y^3$ & $y^4$ \\ \hline
$kN$ D5-branes: & $\surd $ & $\surd $ & $\surd $ & $\surd $ & $-$ & $-$ & 
$\surd $ & $\surd $ & $-$ & $-$ \\ \hline
$kM$ D7-branes: & $\surd $ & $\surd $ & $\surd $ & $\surd $ & $\surd $ & 
$\surd $ & $\surd $ & $\surd $ & $-$ & $-$ \\ \hline
\end{tabular}
\end{equation}
\newline
The coordinates $x^{+},x^{-},x^{1}$ and $x^{2}$ parameterize the world-volume of
 the D5-branes wrapped on the compact spheres of the deformed
singularity of $\mathbb{R}^{4}/\mathbb{Z}_{k}$. The latter is parameterized
by the transverse coordinates $(y^1,y^2,y^3,y^4)$. The other two transverse
coordinates $x^3$ and $x^4$, which are interpreted as the scalar partners of
the gauge fields on the D5-branes, describe the motion of the wrapped D-branes in
the direction transverse to the orbifold $\mathbb{R}^{4}/\mathbb{Z}_{k}$. We
also have D5-D5 and D5-D7 systems in the bi-fundamental and
fundamental representations, respectively, of the gauge groups. 
These models can be engineered as follows.

\begin{itemize}
\item A set of $kN$ D5-branes wrapped on two-cycles in the deformed $A_{k-1}$
singularity of the orbifold point in the transverse space. These branes are
located at the origin of $\mathbb{R}^{4}\subset \mathbb{R}^{6}$ and are
partitioned as follows
\begin{equation}  \label{D5}
N_{1}\cdot\mathrm{D5}+N_{2}\cdot\mathrm{D5}+\ldots+N_{k-1}\cdot\mathrm{D5}
\end{equation}
with $kN=N_{1}+N_{2}+\ldots+N_{k-1}$. Strings emanating and ending on the
same subset of $N_{i}$ D5-branes are associated with the $A_{\mu}^{i}$ gauge
fields of the supersymmetric multiplets.

\item Between these wrapped D5-branes, there are $k-1$ open stings
stretching and transforming in the bi-fundamentals of $U(N_i)\otimes
U(N_{i+1})$. In geometric engineering of QFT$_{4}$, one may think about the
subset of $N_i$ D5-branes as corresponding to a D5-branes wrapped $N_{i}$
folds on the $i$-th two-sphere of the $A_{k-1}$ geometry (\ref{ordAk}).
The resulting quantum field theory based on (\ref{D5}) is a gauge theory
with a gauge group $\left[\otimes_{i=1}^{k-1}U(N_i)\right]$ and admits 
$\mathcal{N}=2$ supersymmetry without full conformal invariance. To recover
scale invariance one has to add fundamental matter. This is achieved as
follows.

\item A set of $kM$ D7-branes wrapped on four-cycles with large volumes.
These D7-branes are located at the origin of 
$\mathbb{R}^{2}/\mathbb{Z}_{k}\subset \mathbb{R}^{4}/\mathbb{Z}_{k} 
\subset \mathbb{R}^{6}/\mathbb{Z}_{k}$ and are partitioned as follows
\begin{equation}
M_{1}\cdot\mathrm{D7}+M_{2}\cdot\mathrm{D7}+\ldots+M_{k-1}\cdot\mathrm{D7}
\end{equation}
where $kM=M_1+M_2+\ldots+M_{k-1}$ and where the $M_i$s are as in (\ref{fem}).

\item On each subset of $N_{i}$ wrapped D5-branes emanate $M_{i}$ open
strings in the fundamental representation of $U(N_i)$ and ending on the 
$M_{i}$ D7-branes. These open strings are needed to ensure conformal
invariance.
\end{itemize}

\paragraph{Massless four-dimensional $\mathcal{N}=2$ vector multiplets, 
$\mathcal{V}=\left(\mathcal{C}, \mathcal{\protect\lambda}_{a}^{\protect\alpha},
\mathcal{A}_{\protect\mu }\right)$:}

Under transformations $g$ generated by the element $\Pi=$diag$\left( 1,\omega
,\ldots,\omega ^{k-1}\right) $ of the orbifold group $\mathbb{Z}_{k}$,
the $\mathcal{N}=2$ massless gauge multiplet is invariant, that is,
\begin{equation}
g\mathcal{A}_{\mu }g^{-1}=\mathcal{A}_{\mu }, \qquad g\mathcal{\lambda}^{\alpha}
g^{-1}=\mathcal{\lambda }^{\alpha }, \qquad g\mathcal{C}g^{-1}=\mathcal{C}.
\end{equation}
The solution of these equations is $V=\sum_{j=1}^{k-1}V_{j}\pi _{j}$, where
each component $V_{j}$ of the expansion is a $N_{j}\times N_{j}$ hermitian
matrix. Along with these massless representations, there are also some
massive $\mathcal{N}=2$ gauge multiplets $W_{ij}=\left( \mathcal{C}_{ij},
\mathcal{\lambda }_{ij}^{\alpha },\mathcal{A}_{\mu }^{ij}\right)$ associated
with the broken generator of the original $U(N_1+N_k)$ group down to 
$\left[\otimes_{i}U(N_i)\right]$. These states break the conformal invariance,
and will not be discussed here.

\paragraph{Four-dimensional $\mathcal{N}=2$ bi-fundamental matter, 
$\mathcal{H}_{i,i+1}=\left( \mathcal{\protect\phi }_{i,i+1}^{\protect\alpha},
\mathcal{\protect\psi }_{i,i+1},\overline{\protect\chi }_{i,i+1}\right)$:}

Along with the $V_{i,j}$ gauge multiplets described above, there are also
massless hypermultiplets $\mathcal{H}_{i,i\pm 1}$ in the bi-fundamental 
representations of the $U(N_i)\otimes U(N_{i+1})$ gauge subgroups of 
$\left[\otimes_{j=1}^{k-1}SU(N_j)\right]$ describing the positions of the sets of 
$N_i$ D3-branes, $i=1,\ldots,k$, in the pp-wave background
\begin{equation}
\mathcal{H}_{1,\overline{2}},\mathcal{H}_{2,\overline{3}},\ldots,\ 
\mathcal{H}_{k-1,\overline{k}}.
\end{equation}
Each hypermultiplet $\mathcal{H}_{i,\overline{j}}$ contains two 
$\mathcal{N}=1 $ chiral multiplets, which we denote as $q_{i,\overline{j}}$ 
and $\widetilde{q}_{i,\overline{j}}$, and whose charges under the $U_{R}(1)$
subsymmetry are $\frac{1}{2}$ and $-\frac{1}{2}$, respectively. These matter
fields transform under the orbifold group as in (\ref{bif}).

\paragraph{Four-dimensional $\mathcal{N}=2$ fundamental matter, 
$\mathcal{H}_{i}=\left( \mathcal{\protect\phi }_{i}^{\protect\alpha }, 
\mathcal{\protect\psi }_{i},\overline{\protect\chi }_{i}\right)$:}

The fundamental matter 
$\left\{\mathcal{H}_{i}^{a_{i}}\equiv\left(N_{i},M_{a}\right)\right\}$, denoted also as 
$\left( q_{i}^{a_{i}},\widetilde{q}_{i}^{a_{i}}\right)$ in terms of $\mathcal{N}=1$ chiral
multiplets, transforms in the $(N_i,M_i)$ representation of the 
$U(N_i)\otimes U(M_i)$ subsymmetries of the $A_{k-1}$ CFT$_4$. These
multiplets carry no orbifold charge. \\

This analysis can be extended naturally to the $\mathcal{N}=2$ CFT$_4$ based
on finite or indefinite Lie algebras.

\section{Operator/String-state correspondence in $\mathcal{N}=2$ CFT$_4$}

\quad\ With all the above tools at hand, we are now ready to write down the
extension of the BMN correspondence between type IIB superstring theory
on pp-waves
with $ADE$ orbifold geometries and $\mathcal{N}=2$ CFT$_4$s. A path to
obtain this extension is to use the following:
(\textbf{a}) the original BMN proposal for 
$\mathcal{N}=4$ $U(|\Gamma|N)$ SYM$_{4}$ and type IIB superstring 
theory on a Penrose limit of $AdS_{5}\times S^{5}$;
(\textbf{b}) reduction of $\mathcal{N}=4$ 
$U(|\Gamma|N)$ SYM$_{4}$ down to $\mathcal{N}=2$ quiver gauge CFT$_4$s
according to the picture (\ref{pic}); and (\textbf{c}) $ADE$ orbifolds of
the Penrose limit of $AdS_{5}\times S^{5}$ studied in Section 3. In what
follows, we first build gauge-invariant field operators, then we give the
corresponding string states. We conclude this section by commenting
on certain specific states of the various classes of $\mathcal{N}=2$ CFT$_4$s.

\subsection{Gauge-invariant field operators}

\quad\ Following \cite{BMN}, there is a correspondence between type IIB
superstring states on pp-wave geometries with quantized energy $E=\Delta
-J=n $, with $n$ a positive integer, and gauge-invariant states of 
$\mathcal{N}=4$ $U(N)$ SYM$_{4}$. This relation applies to orbifolds as well, 
except that now $n$ is a positive \textit{half-integer} since the scalars in the
matter hypermultiplets carry positive half-integer charges $J$.

On the field-theory side, orbifolding $\mathcal{N}=4$ $U(N)$ 
SYM$_{4}$ by a discrete group $\Gamma $ promotes the gauge group 
to $U(|\Gamma|N)$. However, deformations of the orbifold singularity breaks the
$\mathcal{N}=4$ $U(|\Gamma|N)$ SYM$_{4}$ down to $\mathcal{N}=2$ 
quiver gauge CFT$_4$s with gauge group $\left[\otimes_iU(N_i)\right]$. 
The initial $\mathcal{N}=4$ $U(|\Gamma|N)$ vector multiplet 
$\left(0^{6},1,\frac{1}{2}^{4}\right) \otimes\left[U(|\Gamma|N)\right]_{\mathrm{adj}}$, 
where $\left[U(N_i)\right]_{\mathrm{adj}} =N_{i}\otimes \overline{N}_{i}$, decomposes
into two kinds of representations: (a) $\mathcal{N}=2$ vector multiplets 
$\left(0^{2},1,\frac{1}{2}^{2}\right)_i\otimes
\left[U(N_i)\right]_{\mathrm{adj}}$ which in terms of the component fields read
\begin{equation}
\sum_{i}\left( \mathcal{C}_{i},A_{\mu }^{i},\lambda _{1,2}^{i}\right)
\otimes \left[U(N_i)\right]_{\mathrm{adj}} ,
\end{equation}
where $\mathcal{C}_{i}$ is the usual complex scalar gauge field with 
superpartners $A_{\mu}^{i}$ and $\lambda _{1,2}^{i}$; (b)
hypermultiplets in bi-fundamental representations of the gauge group with
field content given by
\begin{equation}
\left[ \sum_{i,j}\left(0^{4},\frac{1}{2}^{2}\right)_{ij}\otimes(N_i\otimes
\overline{N}_j)\right] \oplus\left[ \sum_{i}M_{i}\left( 0^{4},
\frac{1}{2}^{2}\right) _{i}\otimes N_{i}\right] ,  \label{fund}
\end{equation}
with the second term describing fundamental matter. It goes beyond the residue
one gets from the reduction $\mathcal{N}=4$ $U(|\Gamma|N)$ SYM$_{4}$ 
and has been introduced here to recover conformal invariance for the case of
singularities classified by finite $ADE$ Lie algebras.

In $\mathcal{N}=1$ formalism, the $\mathcal{N}=2$ vector multiplet 
$\left(0^{2},1,\frac{1}{2}^{2}\right) _{i}$ is represented by a 
$\mathcal{N}=1 $ gauge multiplet $\left( \frac{1}{2},1\right) $ and a chiral multiplet 
$\left( 0^{2},\frac{1}{2}\right) \equiv C_{i}$ in the adjoint representation of 
$U(N_i)$. We shall think about these adjoint supermultiplets as representing
the nodes of a quiver diagram as
\begin{equation}
\ldots\quad O_{N_{i-1}}^{\overline{N}_{i-1}}\qquad 
O_{N_{i}}^{\overline{N}_{i}}\qquad O_{N_{i+1}}^{\overline{N}_{i+1}} \quad\ldots,
\end{equation}
Similarly, hypermultiplets $\left( 0^{4},\frac{1}{2}^{2}\right) _{ij}$ in
bi-fundamental representations of $U(N_i)\otimes U(N_i)$ decompose into two
kinds of chiral multiplets, $\left( 0^{2},\frac{1}{2}\right) _{ij}$ and 
$\left( 0^{2},\frac{1}{2}\right) _{ji}$. The two chiral superfields, which we
denote as $\Phi _{ij}$ and $\Phi _{ji}$ with $i<j$, transform as upper
components of two $SU_{R}(2)$ iso-doublets and have a charge $J=\frac{1}{2}$.
They are in the $\left( N_{i},\overline{N}_{j}\right) $ and $\left(
\overline{N}_{j}, N_{i}\right)$ bi-fundamental representations of 
$U(N_i)\otimes U(N_i)$, respectively, and may be thought of as the links
between nodes
\begin{equation}
\ldots\leftrightarrows \quad O_{N_{i-1}}^{\overline{N}_{i-1}} \quad
\rightleftarrows \quad O_{N_{i}}^{\overline{N}_{i}} \quad \rightleftarrows
\quad O_{N_{i+1}}^{\overline{N}_{i+1}} \quad \rightleftarrows \ldots.
\end{equation}
The arrow $\longrightarrow$ from $i$ to $j$ refers to $\Phi _{ij}$ and the
opposite $\Phi _{ji}$ one. Besides $U(N_i)$, fundamental hypermultiplets 
$\left( 0^{4},\frac{1}{2}^{2}\right) _{i}$ of (\ref{fund}) transform moreover
under the flavour group $U(M_i)$, and are decomposed in two chiral
multiplets $Q_{i}^{\pm }$ with charge $J_{i}=\frac{1}{2}$. The $Q_{i}^{-}$
and $Q_{i}^{+}$ multiplets transform as $\left(N_i,\overline{M}_i\right)$
and $\left(\overline{N}_i,M_i\right)$ of $U(N_i)\otimes U_{f}(M_i)$. The
scalar fields associated to the $\Phi _{ij}$ and $Q_{i}^{\pm }$ multiplets
will be denoted as $\phi _{ij}$ and $q_{i}^{\pm }$.

Using these fields, we can build gauge-invariant field operators involving a
single trace. The simplest ones are given by
\begin{equation}
\mathbb{O}_{i}^{(0)}=\text{Tr}\left[ \mathcal{C}_{i}^{J}\right] ;\qquad
\mathbb{O}_{j}^{(1)}= \text{Tr}\left(\sum_{i}\phi _{ji}\mathcal{C}_{i}^{J_{i}}\phi
_{ij} \mathcal{C}_{j}^{J_{j}}\right).
\end{equation}
The first field operator has $\Delta -J=0$, while the second one has $\Delta
-J=1$. Note that introducing the $\Pi _{i}$ projectors on the $i$-th set of 
$N_{i}$ D-branes, one can put the previous operators in compact
form as
\begin{equation}
\mathbb{O}_{0}=\sum_{i}\Pi _{i}\mathbb{O}_{i}^{(0)}, \qquad 
\mathbb{O}_{1}=\sum_{i}\Pi _{i}\mathbb{O}_{i}^{(1)},  \label{0}
\end{equation}
with $\Delta-J=0$ and $\Delta-J=1$, respectively. Along with these
gauge-invariant operators, one also may have gauge-invariant field operators
that are not singlets with respect to the flavour symmetry. This concerns operator
involving fundamental matter such as the operators
\begin{equation}
\mathbb{M}_{\alpha _{i}\overline{\beta }_{i}}^{(1)} 
=q_{\alpha i}^{+}\mathcal{C}_{i}^{J_{i}}q_{i\overline{\beta}}^{-}; \qquad 
\mathbb{M}_{\alpha_{i}\overline{\beta }_{i}}^{(2)} 
=q_{\alpha i}^{+}\mathcal{C}_{i}^{J_{i}}\phi _{ij} 
\mathcal{C}_{j}^{J_{j}}\phi _{ji}q_{i\overline{\beta }}^{-},
\end{equation}
with $\Delta-J=1$ and $\Delta-J=2$ transforming in the adjoint
representation of the flavour symmetry $U(M_i)$. These relations are
generalized straightforwardly.

\subsection{Correspondence}

\quad\ On the string-theory side, we have both closed and open-string
states. The open sector is associated with fundamental matter. Before giving
these states, recall that orbifolding by $\Gamma$ induces twisted string
sectors with $\left| \Gamma \right| -1$ closed-string ground states 
$|{0,p^{+}}\rangle_{i}^{lc}$. The remaining state corresponds to the untwisted
case, though we shall treat them collectively here below. Since these vacua have
zero energy, it is natural to identify them with the following chiral
operators
\begin{equation}
\frac{1}{N_{i}^{J/2}\sqrt{J_{i}}}\mathbb{O}_{i}^{(0)}\longleftrightarrow 
|{0,p^{+}}\rangle_{i}^{lc};\quad i=1,\ldots,|\Gamma|.
\end{equation}
As on the field-theory side, here also one may use the $\Pi_{i}$ projectors
on the $i$-th block of D-branes to reformulate this correspondence in a
compact form. We thus have
\begin{equation}
|{0,p^{+}}\rangle_{lc}=\sum_{i}|{0,p^{+}}\rangle_{i}^{lc}\Pi _{i}, \qquad 
|{0,p^{+}}\rangle_{i}^{lc}=\Pi _{i}|{0,p^{+}}\rangle_{lc},
\end{equation}
resulting in the correspondence
\begin{equation}
\mathbb{O}_{0}\longleftrightarrow |{0,p^{+}}\rangle_{lc}
\end{equation}
where $\mathbb{O}_{0}$ is as in (\ref{0}). This relation formally resembles
the original BMN proposal for the vacuum in type IIB superstring theory on
pp waves of $AdS_{5}\times S^{5}$. One can therefore mimic their approach
to write down the
excited closed-string states. Note that one may also write the light-cone
gauge hamiltonian $H_{lc}$ using eigenvalues as
\begin{equation}
H_{lc}=\sum_i(\Delta_i-J_i) \Pi _{i},\ \ \ \ \ \ \ \Delta_i-J_i>0.
\end{equation}
Each string oscillator corresponds to the insertion of a $\Delta -J=1$ 
gauge-invariant field operator, summing over all positions with an independent
phase, according to the rule
\begin{eqnarray}
a_{n_{i}}^{\dagger \mu } &\rightarrow &D_{\mu }\mathcal{C}_{i},\ \ \ \
\qquad \mu =1,\ldots,4,\qquad i=1,2,\ldots,  \notag \\
b_{_{ij}}^{\dagger } &\rightarrow &\phi _{ij},\ \ \ \ \ \ \ \qquad i<j,
\notag \\
\widetilde{b}_{_{ij}}^{\dagger}&\rightarrow&\phi _{ji},\ \ \ \ \ \ \ \qquad
i<j,  \notag \\
\lambda ^{\dagger \alpha } &\rightarrow &\lambda _{J=1/2}^{\alpha },  \notag
\\
\psi _{ij}&\rightarrow&\psi _{J=1/2},
\end{eqnarray}
where $D_{\mu}$ is the gauge-covariant derivative. For states with $\Delta
_{i}-J_{i}=1$, for instance, we have the following two examples of correspondence 
for $\mathcal{N}=2$ CFT$_4$ closed-string states:
\begin{eqnarray}
\frac{1}{N_{i}^{J_{i}/2}\sqrt{J_{i}}} 
\text{Tr}\left[ \mathcal{C}_{i}^{J_{i}}D_{\mu } \mathcal{C}_{i}\right] 
&\longleftrightarrow &
a_{i}^{\mu \dagger }\Pi_{i}|0,p^{+}\rangle _{lc}  \notag \\
\frac{1}{N_{i}^{J_{i}/2}\sqrt{J_{i}}} 
\text{Tr}\left[ \mathcal{C}_{i}^{J_{i}} \lambda_{i}^{\alpha }\right] 
&\longleftrightarrow & \lambda
_{0}^{\alpha +}\Pi_{i}|0,p^{+}\rangle _{lc}
\end{eqnarray}
where $\lambda _{0}^{\alpha +}$ is a fermionic zero mode with $J=1/2$. For
states with $\Delta _{i}-J_{i}=2$, we have, for example,
\begin{equation}
\frac{1}{N_{i}^{J_{i}/2}\sqrt{J_{i}}}\sum_{l_{i}=1}^{J_{i}} 
\text{Tr}\left[\mathcal{C}_{i}^{l_{i}}\left(D_{\mu}\mathcal{C}_i\right)
\mathcal{C}_{i}^{(J_i-l_i)} D_{\nu }\mathcal{C}_{i}\right] 
\exp\left(\frac{2i\pi l_{i}}{J_{i}}\right) \longleftrightarrow 
a_{n_{i}}^{\mu +}a_{-n_{i}}^{\nu +}\Pi
_{i}|0,p^{+}\rangle _{lc}
\end{equation}
where $n_{i}$ is a positive integer for all $i$. 

Together with these closed-string
states, there are states involving open strings stretching between D-branes.
These correspond to the insertion of matter fields $\phi_{ij}$ having
no $J_{ij}$ charge. For states with 
$\Delta-J=2 $ or $3$, we have among other examples the following typical
gauge-invariant field operators
\begin{equation}
\phi _{ji}\mathcal{C}_{i}^{J_{i}}\phi _{ij}\mathcal{C}_{j}^{J_{j}} \ \ \ \ \
\mathrm{and}\ \ \ \ \ \phi _{ji}\mathcal{C}_{i}^{J_{i}}\phi _{ik}
\mathcal{C}_{k}^{J_{k}} \phi _{kj}\mathcal{C}_{j}^{J_{j}}.
\end{equation}
They correspond to the states
\begin{equation}
\widetilde{b}_{n_{ij}}^{+}b_{-n_{ij}}^{+}\Pi _{j}|0,p^{+}\rangle_{lc} \ \ \
\ \ \mathrm{and}\ \ \ \ \ 
\widetilde{b}_{n_{ji}}^+b_{-n_{ik}}^+b_{-n_{kj}}^+\Pi_j|0,p^+\rangle _{lc},
\end{equation}
respectively, where $n_{ij}=n-(j-i)/|\Gamma|$ and 
$n_{ij}+n_{jk}+\ldots+n_{li}=0$. It is noted that the first closed-string state 
is built out of
two open strings stretching between the $i$-th and $j$-th D-branes and
constituting a loop emanating and ending on the $j$-th D-brane. The second
state involves three open strings forming a loop. Note also that on the
field-theory side, open-string states stretching between the $i$-th and 
$k$-th wrapped D5-branes are associated with 
$\mathcal{C}_{i}^{J_{i}}\phi _{ik}\mathcal{C}_{k}^{J_{k}}$. 
The composition of two open strings may give either an open string as for 
$\mathcal{C}_{j}^{J_{j}}\phi _{ji}\mathcal{C}_{i}^{J_{i}}\phi _{ik}
\mathcal{C} _{k}^{J_{k}}$ or a closed string state as for the gauge-invariant 
operator $\phi _{ji}\mathcal{C}_{i}^{J_{i}}\phi _{ij}\mathcal{C}_{j}^{J_{j}}$. 
In this D-brane engineering of QFT, bi-fundamental matter corresponds
to open strings stretching between D3-D5 and D5-D5
branes. However, one may still build gauge-invariant field operators that
are dual to open strings stretching between D7-branes. This requires the
introduction of fundamental matter as in the following examples
\begin{equation}
q_{\alpha i}\mathcal{C}_{i}^{J_{i}}\widetilde{q}_{i\overline{\alpha }} \ \ \
\ \ \mathrm{and}\ \ \ \ \ q_{\alpha i}\mathcal{C}_{i}^{J_{i}}\phi _{ij}
\mathcal{C}_{j}^{J_{j}}\phi _{ji}\widetilde{q}_{i\overline{\alpha }}.
\end{equation}
These gauge-invariant field states involve two quarks in the fundamental
representation of the flavour symmetry $U(M_i)$, and have a light-cone energy
$E_{lc}$ equal to 2 and 3, respectively.

\subsection{Specific states}

\quad\ Here we want to describe some particular gauge-invariant field
operators that are specific to the various $\mathcal{N}=2$ $ADE$ quiver 
CFT$_4$s. These states concern both closed and open strings, and are related to
the topology of $ADE$ Dynkin diagrams.

\subsubsection{$\mathcal{N}=2$ $A_{k-1}$ quiver CFT$_4$s}

To fix the ideas, we consider here the example of $\mathcal{N}=2$ quiver
gauge SYM$_{4}$ based on an ordinary $A_{k-1}$ singularity. This is in
general not a conformal $\mathcal{N}=2$ QFT$_4$ but may be 
converted into one by
adding fundamental matter. In this case the group symmetry of the CFT is of
the form $G=G_{g}\otimes G_{f}$. A simple example corresponds to a gauge
group $G_{g}=U(N)^{k-1}$ and $G_{f}=U(M_1) \otimes U(M_{k-1})$ given by 
$U(N)^2$. This is then a supersymmetric conformal model with 
$M_{1}+M_{k-1}=N+N$ fundamental matter respectively in the $N_1$ and $N_{k-1}$
representations of the $U(N_1)\otimes U(N_{k-1})$ gauge symmetry, which in
the present example is also equal to $U(N)^2$. A gauge-invariant field
operator of this model, having light-cone energy $k+1$ with two quarks at
the ends, is shown here below
\begin{equation}
\mathbb{M}_{\alpha ,\overline{\alpha }}= \text{Tr}\left(q_{\alpha 1}^{-}\phi
_{01}\mathcal{C}_{1}^{J_{1}} \phi_{12}\mathcal{C}_{2}^{J_{2}}\ldots\phi
_{i-1i}\mathcal{C}_{i}^{J_{i}} \phi_{ii+1}\ldots
\mathcal{C}_{k-1}^{J_k}q_{k-1\overline{\alpha}}^+\right),
\end{equation}
where the trace is over the gauge group. The introduction of the two quarks
ensures gauge invariance. This field operator transforms in the
bi-fundamental representation of the flavour 
group $U(N_1)\otimes U(N_{k-1})$, and it
corresponds to an open string stretching between the sets of $M_{1}$ and 
$M_{k-1}$ D7-branes.

\subsubsection{$\mathcal{N}=2$ affine $\widehat{A}_k$ quiver CFT$_4$s}

An $\mathcal{N}=2$ quiver gauge theory based on affine $\widehat{A}_{k}$ is
a CFT$_4$ without the need of introducing fundamental matter. As affine 
$\widehat{A}_{k}$ has a Dynkin diagram represented by a loop (\ref{affAk}),
the latter may be viewed as a closed string of energy $k+1$ built out of
the open strings 
stretching between the D-branes. Adjunction of fundamental matter in this 
class of CFTs is possible, but leads to interacting open strings.

Let us discuss hereafter the case without fundamental matter. In this case
affine $\widehat{A}_{k}$ $\mathcal{N}=2$ quiver CFT$_{4}$s has gauge
group $U(N)^{k+1}$ and bi-fundamental matter. Closed-string state of
light-cone energy $k+1$ involves $k-1$ D5-D5 open strings and $2$ D3-D5 open
strings. In the large-$N$ limit, the corresponding gauge-invariant field operator 
reads
\begin{equation}
\text{Tr}\left( \phi _{01}\mathcal{C}_{1}^{J_{1}}\phi _{12}
\mathcal{C}_{2}^{J_{2}}\ldots \phi _{i-1i}\mathcal{C}_{i}^{J_{i}}\phi _{ii+1}\ldots
\mathcal{C}_{k}^{J_{k}}\phi _{k0}\mathcal{C}_{0}^{J_{0}}\right) .
\label{efa}
\end{equation}
Here $\phi _{k0}$ and $\phi _{01}$ represent open strings stretching
between D5-D3 branes, while the remaining fields
$\phi_{ii+1}$ correspond to open strings
stretching between D5-D5 branes. As above, the $J_{i}$s are all of them of
order $\sqrt{N}$. 

Adding fundamental matter while keeping scale
invariance requires more general group symmetries. In geometric engineering,
such a CFT is described by diagrams with open topologies involving
trivalent vertices as in \cite{KMV}. On the string-theory side, these
topologies may be interpreted as describing interacting open strings with
quarks as external legs.

\subsubsection{$\mathcal{N}=2$ affine $\widehat{DE}$ quiver CFT$_4$s}

One may also build the analogue of the previous gauge-invariant field
operators for the other classes of $\mathcal{N}=2$ CFT$_4$s. For affine 
$\widehat{D}_{k}$ series with gauge group $G_{g}=U(N)^4\otimes U(2N)^{k-3}$,
the situation is interesting as it involves two three-vertices as indicated in
the associated Dynkin diagram:
\be
    \mbox{
         \begin{picture}(20,110)(60,-50)
        \unitlength=2cm
        \thicklines
      \put(-1.8,-.05){$\widehat{D}_k:$}
    \put(0,0){\circle{.2}}
     \put(.1,0){\line(1,0){.5}}
     \put(.7,0){\circle{.2}}
     \put(.8,0){\line(1,0){.5}}
     \put(1.4,0){\circle{.2}}
     \put(1.6,0){$.\ .\ .\ .\ .\ .$}
     \put(2.5,0){\circle{.2}}
     \put(2.6,0){\line(1,0){.5}}
     \put(3.2,0){\circle{.2}}
     \put(3.2,.1){\line(1,1){.35}}
     \put(3.64,.5){\circle{.2}}
     \put(3.2,-.1){\line(1,-1){.35}}
     \put(3.64,-.5){\circle{.2}} 
     \put(0,.1){\line(-1,1){.35}}
     \put(-.44,.5){\circle{.2}}
     \put(0,-.1){\line(-1,-1){.35}}
     \put(-.44,-.5){\circle{.2}}
  \end{picture}
}
\label{affDk}
\ee
A direct way to obtain gauge-invariant field operators is to use
the following representation: (a) Vertices with two legs are associated with
the adjoint representation $N_i\otimes \overline{N}_i$ of the 
corresponding gauge group $U(N_i) $, and are denoted as

\begin{equation}
\begin{tabular}{l}
\ldots$\quad\longleftarrow\quad N_{i-1}\quad\longrightarrow\quad 
\overline{N}_{i}\quad \longleftarrow \quad N_{i+1}\quad \longrightarrow
\quad\ldots$ \\
\ldots$\quad \longrightarrow\quad\overline{N}_{i-1}\quad\longleftarrow\quad
N_{i}\quad \longrightarrow \quad \overline{N}_{i+1}\quad \longleftarrow
\quad \ldots$
\end{tabular}
\end{equation}
\newline
where an arrow from $N_{i}$ to $\overline{N}_{i+1}$ refers to bi-fundamental
matter $\phi _{i,i+1}$, while an arrow from $N_{i+1}$ to $\overline{N}_{i}$
refers to $\phi _{i+1,i}$. (b) The three-vertices, cf. (\ref{affDk}), are
represented by

\begin{equation}
\begin{tabular}{l}
\ldots $\quad \longleftarrow \quad N_{k-3}\quad \longrightarrow \quad $ \\
\ldots $\quad \longrightarrow \quad \overline{N}_{k-3}\quad \longleftarrow
\quad $
\end{tabular}
\
\begin{tabular}{lll}
& $\longleftarrow $ & $N_{k}$ \\
& $\longleftarrow $ & $N_{k-1}$ \\
$\overline{N}_{k-2}\quad $ &  &  \\
$N_{k-2}\quad $ &  &  \\
& $\longrightarrow $ & $\overline{N}_{k-1}$ \\
& $\longrightarrow $ & $\overline{N}_{k}$
\end{tabular}
\end{equation}
\newline
The use of arrows is quite similar to the situation for two-vertices. 
Using this representation, one can write down
the analogue of (\ref{efa}). We find the following gauge-invariant field
operator for affine $\widehat{D}_{k}$ CFT$_{4}$s,
\begin{eqnarray}
&&\text{Tr}\left( \left[ \mathcal{C}_{0}^{J_{0}}\phi _{02}
\mathcal{C}_{2}^{J_{2}}+\mathcal{C}_{1}^{J_{1}}\phi _{12}
\mathcal{C}_{2}^{J_{2}^{\prime}}\right] \times \phi _{23}
\mathcal{C}_{3}^{J_{3}}\phi _{34}\ldots \phi_{i-1i}
\mathcal{C}_{i}^{J_{i}}\phi _{ii+1}\ldots \phi _{k-3,k-2}\right.
\notag \\
&&\ \ \ \ \ \ \ \left. \times \left[ \mathcal{C}_{k-2}^{J_{k-2}}\phi
_{k-2,k-1}\mathcal{C}_{k-1}^{J_{k-1}}+\mathcal{C}_{k-2}^{J_{k-2}^{\prime
}}\phi _{k-2,k}\mathcal{C}_{k}^{J_{k}}\right] \right)
\end{eqnarray}%
Open-string configurations may be also considered here and one has just to
add fundamental matter by keeping gauge invariance. The results we derived
here can be extend naturally to the remaining classes of $\mathcal{N}=2$ 
CFT$_{4}$s \cite{ABS1,ABS2}.

\section{Conclusion}

\quad\ In this paper, we have studied the extension of the BMN proposal to
type IIB superstring theory on pp-wave orbifolds preserving 16 supercharges.
In particular,  we have considered a large class of models describing 
certain limits of type IIB superstrings propagating on pp-wave with ADE
geometries. Based initially on closed strings only, 
we have developed an analysis for the two classes involving
ordinary $ADE$ singularities and affine 
$\widehat{ADE}$ elliptic geometries.
These models are dual to $\mathcal{N}=2$ CFT$_{4}$ quiver 
theories classified respectively by finite $ADE$ Lie algebras and affine 
$\widehat{ADE}$ Kac-Moody algebras. Concerning the finite $ADE$ quiver models,
we have shown that conformal invaraince requires the incorporation of 
fundamental matter being linked to open strings. Our main results in this paper
may be summarized as follows.  First, we have 
reviewed type IIB superstring theory on pp-wave orbifolds, in particular for orbifold
groups $\Gamma \subset SU(2)$, itself contained in the R-symmetry $SU(4)$ of
$\mathcal{N}=4$ SYM$_{4}$. Then we have given the explicit form of the
corresponding pp wave metrics for both ordinary and affine geometries. After
that we have studied the various classes of $\mathcal{N}=2$ CFT$_{4}$s
classified by finite and ordinary $ADE$ Dynkin diagrams of Lie algebras.
Finally, we have derived the appropriate extension of a BMN proposal for these
models. In particular, we have given the correspondence rule
between leading closed-string states and gauge-invariant operators in the 
$\mathcal{N}=2$ superconformal quiver theories. 

Our work opens up for further studies. 
One interesting problem is to complete this analysis
by considering  $\mathcal{N}=2$ CFT$_{4}$ quiver models classified by
indefinite Lie algebras \cite{ABS1,ABS2,malika,iran,AABDS}. In particular, 
it is intersting to write down the pp-wave geometries corresponding to such models. 
A natural  question concers the  extension of this work to the case of pp-wave
orbifolds with eight or four supercharges. We hope to report elsewhere on these
open questions.

\begin{acknowledgement}
AB is very grateful to  A. Sebbar for discussion, hospitality  and  scientific helps. 
JR thanks C. Cummins for discussions.
This work is supported in part by the program Protars III, CNRST D12/25, Rabat.
\end{acknowledgement} 

\appendix

\section{Non-abelian and affine orbifolds}

\quad\ So far we have worked out explicit examples of pp waves with abelian
orbifold geometry. Here we study their non-abelian analogues by first
considering the ordinary orbifold geometries and subsequently their
affine counterparts.

\subsection{Non-abelian orbifolds}

\quad\ Since there are two basic classes of $ADE$ singularities that one
encounters in the study of type II superstrings on singular Calabi-Yau
manifolds, namely those ordinary singularities associated with ordinary $ADE$
Lie algebras and those affine ones associated with the $\widehat{ADE}$
affine Kac-Moody extensions, we will divide our discussion on pp-wave
backgrounds into two 
parts.\footnote{For an extension to geometries involving indefinite 
Lie algebras, see \cite{ABS1,ABS2}.} 
The first one deals with pp-wave geometries in connection
with ordinary $ADE$ Lie algebras and the second part concerns the geometries
in connection with the affine $\widehat{ADE}$ Lie algebras.

\subsubsection{Ordinary $DE$ pp waves}

\quad\ The $DE$ singularities of the ALE space leading to four-dimensional 
$\mathcal{N}=2$ models are described by the equations appearing in the table 
(\ref{surface}). The analysis of the blow up of
these singularities goes along the same lines as in the abelian
case $A_{k-1}$, except now the non-abelian property of the orbifold
groups makes the analysis a bit tedious. 
We will give some details regarding both the blow-ups and the
mirrors of the ordinary $D_{k}$ and $E_{6}$ geometries. The results for 
$E_{7}$ and $E_{8}$ geometries follow from a similar analysis and will
be omitted.

\paragraph{$D_{k}$ pp-wave geometry:}

Starting from (\ref{ppwavemetric}) and using the equation defining $D_{k}$
singularity, which allow to express the $z_1$ variable in terms of $z_2$ and
$\zeta$ as
\begin{equation}
z_1=\sqrt{\zeta ^{k-1}-z_2^{2}\zeta },
\end{equation}
and the differential $dz_1$ in terms of $dz_2$ and $d\zeta$ as
\begin{equation}
dz_1=\frac{1}{2\left(\zeta^{k-1}-z_2^{2}\zeta\right)^{\frac{1}{2}}} \left[
2z_2\zeta dz_2+\left( (k-1)\zeta ^{k-2}-z_2^{2}\zeta \right) d\zeta \right],
\end{equation}
one can work out the metric of the pp-wave background near the orbifold point
with $D_{k}$ singularity. The result is given by
\begin{eqnarray}
ds^{2}|_{D_{k}} &=&-4dx^{+}dx^{-}-\mu^2\left(\mathbf{x}^{2}+
\left|\zeta^{k-1}-z_2^{2}\zeta\right|^2+\left|z_2\right|^2\right)
(dx^{+})^{2}+d\mathbf{x}^{2}  \notag \\
&&+\left( 1+\left| \frac{2z_2\zeta }{2\left( \zeta ^{k-1}-z_2^{2}\zeta
\right) ^{\frac{1}{2}}}\right|^2\right) \left| dz_2\right| ^{2}
+\left| \frac{(k-1)\zeta ^{k-2}-z_2^{2}\zeta }{2\left(\zeta ^{k-1}
-z_2^{2}\zeta \right)^{\frac{1}{2}}}\right| ^{2}\left| d\zeta \right| ^{2}  \notag \\
&&+\frac{2z_2\zeta }{2\left| \zeta ^{k-1}-z_2^{2}\zeta \right| }
\left( (k-1)\overline{\zeta }^{k-2}-\overline{z}_{2}^{2}\overline{\zeta }\right) 
dz_2d\overline{\zeta }  \notag \\
&&+\frac{2\overline{z}_{2}\overline{\zeta }}{2\left|\zeta
^{k-1}-z_2^{2}\zeta \right| }\left( (k-1)\zeta ^{k-2}-z_2^{2}\zeta \right) 
d\overline{z}_{2}d\zeta,  \label{metricD}
\end{eqnarray}
which is a singular metric at the origin $z_2=\zeta =0$. As usual, this can
be resolved either by K\"ahler or complex deformations. We will consider
both cases to which we shall refer hereafter as blown-up $D_{k}$ pp-wave and
mirror $D_{k}$ pp-wave geometries, respectively.

\paragraph{Blown-up $D_{k}$ pp-wave geometry:}

In the blow up of the ordinary $D_{k}$ singularity, one introduces a 
basis of $k$ complex three-dimensional open sets 
$\mathcal{U}_{i}=\left\{(u_{i},v_{i},w_{i}), 1\leq i\leq k\right\}$ glued together
as \cite{HOV}
\begin{eqnarray}
u_{i} &=&u_{i+1}v_{i+1}w_{i+1},  \notag \\
v_{i} &=&u_{i+1}, \qquad\ \ \ \ \ 1\leq i\leq k-4,  \notag \\
w_{i}v_{i+1} &=&w_{k-3}v_{k-2}=w_{k-2}v_{k-1}=w_{k}v_{k-1}=1,
\end{eqnarray}
supplemented by
\begin{eqnarray}
u_{k-3} &=&u_{k-2}v_{k-2}^{2}w_{k-2}  \notag \\
v_{k-3}&=&u_{k-2}v_{k-2}  \notag \\
u_{k-2}v_{k} &=&v_{k-1}w_{k-1}v_{k}=1  \notag \\
v_{k-2} &=&u_{k-2}=u_{k}v_{k}.
\end{eqnarray}
In terms of these variables, the $(z_1,z_2,\zeta)$ coordinates are
iteratively realized on the $k-2$ open sets $\mathcal{U}_{1},\ldots,$ 
$\mathcal{U}_{k-3}$ and $\mathcal{U}_{k-1}$ as follows
\begin{eqnarray}
z_1 &=&u_{2j-1}^{j}w_{2j-1}^{j-1}=u_{2j}^{j}v_{2j}w_{2j}^{j},  \notag \\
z_2 &=&u_{2j-1}^{j-1}v_{2j-1}w_{2j-1}^{j-1}=u_{2j}^{j}w_{2j}^{j-1},  \notag
\\
\zeta &=&u_{2j-1}w_{2j-1}=u_{2j}w_{2j},  \label{link1}
\end{eqnarray}
or equivalently by substituting the expressions of $\zeta $ in terms of the 
$u_{j}$ and $v_{j}$'s
\begin{eqnarray}  \label{link2}
z_1 &=&u_{2j-1}\zeta ^{j-1}=v_{2j}\zeta ^{j},  \notag \\
z_2 &=&\zeta ^{j-1}v_{2j-1}=u_{2j}\zeta ^{j-1}.
\end{eqnarray}
Note in passing that using the relation $w_{i}v_{i+1}=1$, and the identity 
$v_{i+1}\zeta =u_{i}$ following from the third relation of (\ref{link1}), one
can show that on the $\mathcal{U}_{2j-1}$ patch, for instance, the equation
of the singularity is given by
\begin{equation}
u_{2j-1}^{2}+v_{2j-1}^{2}\zeta =\zeta ^{k-1-2(j-1)}
\end{equation}
which is nothing but the defining equation of a $D_{k-(2j-1)}$ singularity.
On the remaining two other open sets $\mathcal{U}_{k-2}$ and 
$\mathcal{U}_{k} $, we have $\zeta =u_{k-2}v_{k-2}w_{k-2}=u_{k}w_{k}$ 
and the following projections for $z_2$
\begin{equation}
z_2=\zeta ^{\frac{k-4}{2}}u_{k-2}v_{k-2}=u_{k}\zeta ^{\frac{k-4}{2}},
\end{equation}
for $k$ even integer, and
\begin{equation}
z_2=v_{k-2}\zeta ^{\left[ \frac{k}{2}\right] -1}=u_{k}v_{k}\zeta ^{\left[
\frac{k}{2}\right] -1}
\end{equation}
for $k$ odd. Note that in terms of the $\{(u_{i},v_{i},w_{i})\}$ system of
variables of the $\mathcal{U}_{i}$ patches, the first $k-2$ intersecting 
$\mathcal{C}_{i}$ curves are described by the following equations
\begin{eqnarray}
\mathcal{C}_{i} &=&\left\{ w_{i}v_{i+1}=1,\quad u_{i}=v_{i}=0,\quad
u_{i+1}=w_{i+1}=0\right\}  \notag \\
\mathcal{C}_{i-1}\cap \mathcal{C}_{i} &=&\left\{
u_{i}=v_{i}=w_{i}=0\right\}, \quad 1\leq k-2
\end{eqnarray}
while $\mathcal{C}_{k-1}$ and $\mathcal{C}_{k}$ are given by
\begin{eqnarray}
\mathcal{C}_{k-1} &=&\left\{u_{k-2}v_{k-1}=1,\quad
v_{k-2}=w_{k-2}-1=0\right\} ,  \notag \\
\mathcal{C}_{k} &=&\left\{ u_{k-2}v_{k}=1,\quad v_{k-2}=w_{k-2}+1=0\right\} ,
\notag \\
\mathcal{C}_{k-2}\cap \mathcal{C}_{k-1}
&=&\left\{u_{n-2}=v_{n-2}=w_{n-2}-1=0\right\} ,  \notag \\
\mathcal{C}_{k-2}\cap \mathcal{C}_{k}
&=&\left\{u_{n-2}=v_{n-2}=w_{n-2}+1=0\right\} .
\end{eqnarray}
On the blown-up $D_k$ geometry, the $kN$ D3-branes are partitioned into $k$
subsets of $N_i$ D3-branes wrapping the $\mathcal{C}_{i}$ curves, and the
initial gauge group $U(kN)$ breaks down to $[\otimes _{i=1}^{k}U(N_i)]$.
Using (\ref{link2}), one can write down the pp-geometry for the Penrose
limit of type IIB superstring theory on the blown-up $D_k$ singularity. The pp-wave
metrics on the $\mathcal{U}_{2j-1}$ and $\mathcal{U}_{2j}$ patches one
gets are derived from (\ref{ppwavemetric}) by substituting $z_1$ and $z_2$
as in (\ref{metricD})\ and the abelian differentials
\begin{eqnarray}
dz_1 &=&\zeta ^{j-1}du_{2j-1}+(j-1)\zeta ^{j-2}u_{2j-1}d\zeta  \notag \\
dz_2 &=&\zeta ^{j-1}dv_{2j-1}+(j-1)\zeta ^{j-2}v_{2j-1}d\zeta
\end{eqnarray}
for $\mathcal{U}_{2j-1}$ and
\begin{eqnarray}
dz_1 &=&\zeta ^{j}dv_{2j}+j\zeta ^{j-1}v_{2j}d\zeta ,  \notag \\
dz_2 &=&\zeta ^{j}du_{2j}+j\zeta ^{j-1}u_{2j}d\zeta .
\end{eqnarray}
on the $\mathcal{U}_{2j}$ patch. In this way, the pp-wave metric on 
$\mathcal{U}_{2j}$ is given by
\begin{eqnarray}
ds^{2}|_{D_k,\mathcal{U}_{2j}}&=&-4dx^{+}dx^{-}+d\mathbf{x}^{2}  \notag \\
&&-\mu ^{2}\left(\mathbf{x}^{2}+\left|\zeta^{j-1}\right|^2\left(\left|
v_{2j}\zeta \right|^2+\left| u_{2j}\right|^2\right)+\left|
z_2\right|^2\right)(dx^{+})^2  \notag \\
&&+\left| \zeta ^{j}\right|^2\left(\left| du_{2j}\right|^2+\left|
dv_{2j}\right| ^{2}\right)+j^2\left|\zeta^{j-1}\right|^2
\left(\left|v_{2j}\right|^2+\left|u_{2j}\right|^2\right)
\left|d\zeta\right|^2  \notag \\
&&+j\left| \zeta ^{j-1}\right| ^{2}\left( \zeta \overline{u}_{2j}du_{2j} 
d\overline{\zeta }_{2j}+\overline{\zeta }u_{2j}d\zeta _{2j} 
d\overline{u}_{2j}\right)  \notag \\
&&+j\left| \zeta ^{j-1}\right| ^{2}\left( \zeta \overline{v}_{2j}dv_{2j} 
d\overline{\zeta }_{2j}+\overline{\zeta }v_{2j}d\zeta _{2j} 
d\overline{v}_{2j}\right).
\end{eqnarray}
A quite similar result applies to the patch $\mathcal{U}_{2j-1}$.

\paragraph{Mirror $D_{k}$ pp-wave geometry:}

Mimicking the situation for $A_{k}$, the mirror of the blow up of the $D_{k}$
singularity is obtained by introducing a system of $k+4$ complex variables 
$\tau _{i}$ satisfying the following $k$ holomorphic constraint equations
\begin{equation}
\prod_{i=1}^{k}\tau _{i}^{\ell _{i}^{(a)}}=1,
\end{equation}
in addition to two extra constraints. The $\ell _{i}^{(a)}$'s appearing in these
equations are integers and are basically given by the Cartan matrix of the 
$D_{k}$ finite Lie algebra. These constraint equations resemble 
(\ref{miro}) concerning the case $A_{k-1}$, 
and are naturally solved in terms of monomials of three
independent complex variables $\eta $, $\xi $ and $\sigma $ in one-to-one
correspondence with the nodes of the $D_{k}$ Dynkin diagram
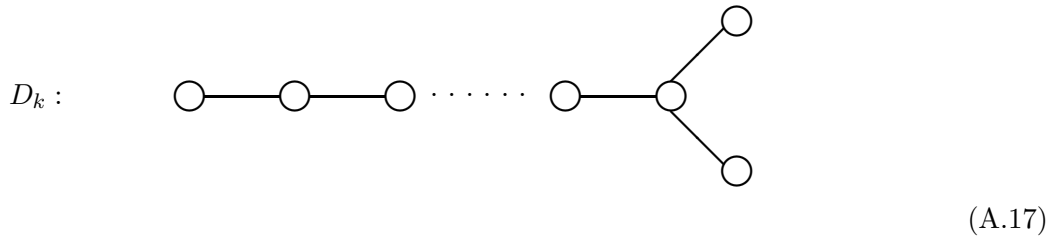
\be
    \mbox{
         \begin{picture}(20,100)(90,-50)
        \unitlength=2cm
        \thicklines
      \put(-1.2,-.05){$D_k:$}
    \put(0,0){\circle{.2}}
     \put(.1,0){\line(1,0){.5}}
     \put(.7,0){\circle{.2}}
     \put(.8,0){\line(1,0){.5}}
     \put(1.4,0){\circle{.2}}
     \put(1.6,0){$.\ .\ .\ .\ .\ .$}
     \put(2.5,0){\circle{.2}}
     \put(2.6,0){\line(1,0){.5}}
     \put(3.2,0){\circle{.2}}
     \put(3.2,.1){\line(1,1){.35}}
     \put(3.64,.5){\circle{.2}}
     \put(3.2,-.1){\line(1,-1){.35}}
     \put(3.64,-.5){\circle{.2}}
  \end{picture}
}
\label{ordDk}
\ee
To get the corresponding pp-wave 
geometry, one may follow the same steps we
took in the case of $A_{k-1} $. 
However, since the Dynkin diagram of $D_{k}$ has a
trivalent vertex exactly like its affine extension, one must resolve to
methods regarding this kind of geometry. This so-called
trivalent geometry is conveniently formulated in terms of
elliptic fibrations over the complex plane making the study of the $D_k$
pp-wave mirror geometry and its affine extension feasible. 
Details are provided below.

\subsubsection{$E_{6}$ pp-wave geometry}

Using the equation defining the $E_{6}$ singularity (\ref{surface}), we can
eliminate variable $z_{1}$ in terms of $z_{2}$ and $\zeta $:
\begin{equation}
z_{1}=\sqrt{\zeta ^{4}+z_{2}^{3}},
\end{equation}
The differential $dz_{1}$ may subsequently be expressed
as a function of $dz_{2}$ and $d\zeta$:
\begin{equation}
dz_{1}=\frac{1}{2\sqrt{\zeta ^{4}+z_{2}^{3}}}\left( 3z_{2}^{2}dz_{2}+4\zeta
^{3}d\zeta \right).
\end{equation}
The metric of a pp-wave background near the orbifold point with $E_{6}$
singularity then reads
\begin{eqnarray}
ds^{2}|_{E_{6}} &=&-4dx^{+}dx^{-}+d\mathbf{x}^{2}-\mu ^{2}
\left( \mathbf{x}^{2}+\left\vert \zeta ^{4}+z_{2}^{3}\right\vert 
+\left\vert z_{2}\right\vert^{2}\right) (dx^{+})^{2}  \notag \\
&&+\left( 1+\frac{9\left\vert z_{2}\right\vert ^{4}}{2\left\vert \zeta
^{4}+z_{2}^{3}\right\vert }\right) \left\vert dz_{2}\right\vert ^{2}
+\frac{4\left\vert \zeta \right\vert ^{6}}{\left\vert \zeta
^{4}+z_{2}^{3}\right\vert }\left\vert d\zeta \right\vert ^{2}  \notag \\
&&+\frac{6}{\left\vert \zeta ^{4}+z_{2}^{3}\right\vert }
\left( z_{2}^{2}\overline{\zeta }^{3}dz_{2}d\overline{\zeta }
+\overline{z}_{2}^{2}\zeta ^{3}d\overline{z}_{2}d\zeta \right) .  
\label{e6b}
\end{eqnarray}%
This is a degenerate metric at the orbifold point $\zeta =z_{2}=0$,
but as before, such a degeneracy may be lifted by resolving the singularity
either by blowing it up or by complex deformations. In the blow up of the 
$E_{6}$ singularity, one introduces a topological basis of five open sets 
$\mathcal{U}_{i}$ parameterized by the holomorphic coordinates 
$(u_{i},v_{i},w_{i})$ and glued together as
\begin{equation}
v_{1}u_{2}=w_{1}u_{3}=w_{3}u_{5}=w_{2}v_{3}=w_{4}v_{5}=1.
\end{equation}%
On these open sets, $\mathcal{U}_{i}$, the relations for 
$\left(z_{1},z_{2},\zeta \right) $ read
\begin{eqnarray}
z_{1}
&=&u_{1}v_{1}=u_{2}v_{2}^{6}w_{2}=u_{3}v_{3}^{4}w_{3}^{6}
=u_{4}v_{4}^{2}w_{4}^{4}=u_{5}w_{5}^{2},
\notag \\
z_{2}
&=&v_{1}=v_{2}^{2}w_{2}=v_{3}^{3}w_{3}^{3}=v_{4}^{2}w_{4}^{3}=v_{5}w_{5}^{2},
\notag \\
\zeta
&=&v_{1}w_{1}=v_{2}^{3}w_{2}=v_{3}^{3}w_{3}^{3}=v_{4}^{4}w_{4}^{2}=w_{5},
\end{eqnarray}
together with
\begin{eqnarray}
u_{1} &=&-\sqrt{v_{1}+v_{1}^{2}w_{1}^{4}},  \notag \\
u_{2} &=&-\sqrt{v_{2}+w_{2}^{2}},  \notag \\
u_{3} &=&-\sqrt{1+v_{3}},  \notag \\
u_{4} &=&-\sqrt{1+v_{4}^{2}w_{4}^{2}},  \notag \\
u_{5} &=&-\sqrt{1+v_{5}^{3}w_{5}^{2}}.
\end{eqnarray}
{}From these relations, one can compute the explicit expressions of the 
$dz_{1}$, $dz_{2}$ and $d\zeta $ differentials in terms of $dv_{i}$ and 
$dw_{i}$. Indeed, putting back these into (\ref{e6b}), one gets the $E_{6}$
pp-wave metric on the $\mathcal{U}_{i}$ open sets. The mirror geometry of
the $E_{6}$ singularity may be also derived by following the same lines we
have presented above, now based on the Dynkin diagram
\be
    \mbox{
         \begin{picture}(20,80)(60,-20)
        \unitlength=2cm
        \thicklines
      \put(-1.2,-.05){$E_6:$}
    \put(0,0){\circle{.2}}
     \put(.1,0){\line(1,0){.5}}
     \put(.7,0){\circle{.2}}
     \put(.8,0){\line(1,0){.5}}
     \put(1.4,0){\circle{.2}}
    \put(1.5,0){\line(1,0){.5}}
    \put(2.1,0){\circle{.2}}
    \put(2.2,0){\line(1,0){.5}}
    \put(2.8,0){\circle{.2}}
   \put(1.4,.1){\line(0,1){.5}}
   \put(1.4,.7){\circle{.2}}
  \end{picture}
}
\label{ordE6}
\ee
As for $D_{k}$, it is more convenient
to use here the elliptic fibrations over the complex plane.

\subsection{Affine $\widehat{ADE}$ pp waves}

\quad\ Here we continue studying the Penrose limits of the
$AdS_{5}\times S^{5}/\Gamma $ orbifolds with affine 
$\widehat{ADE}$ singularities. The deformations of these 
singularities are represented by their Dynkin diagrams (\ref{affAk}),
(\ref{affDk}) and
\be
    \mbox{
         \begin{picture}(20,350)(50,-20)
        \unitlength=2cm
        \thicklines
    \put(0,4){\circle{.2}}
     \put(.1,4){\line(1,0){.5}}
     \put(.7,4){\circle{.2}}
     \put(.8,4){\line(1,0){.5}}
     \put(1.4,4){\circle{.2}}
    \put(1.5,4){\line(1,0){.5}}
    \put(2.1,4){\circle{.2}}
    \put(2.2,4){\line(1,0){.5}}
    \put(2.8,4){\circle{.2}}
   \put(1.4,4.1){\line(0,1){.5}}
   \put(1.4,4.7){\circle{.2}}
   \put(1.4,4.8){\line(0,1){.5}}
   \put(1.4,5.4){\circle{.2}}
   \put(-.7,2){\circle{.2}}
   \put(-.6,2){\line(1,0){.5}}
    \put(0,2){\circle{.2}}
     \put(.1,2){\line(1,0){.5}}
     \put(.7,2){\circle{.2}}
     \put(.8,2){\line(1,0){.5}}
     \put(1.4,2){\circle{.2}}
    \put(1.5,2){\line(1,0){.5}}
    \put(2.1,2){\circle{.2}}
    \put(2.2,2){\line(1,0){.5}}
    \put(2.8,2){\circle{.2}}
   \put(1.4,2.1){\line(0,1){.5}}
   \put(1.4,2.7){\circle{.2}}
  \put(-1.4,0){\circle{.2}}
  \put(-1.3,0){\line(1,0){.5}}
  \put(-.7,0){\circle{.2}}
   \put(-.6,0){\line(1,0){.5}}
    \put(0,0){\circle{.2}}
     \put(.1,0){\line(1,0){.5}}
     \put(.7,0){\circle{.2}}
     \put(.8,0){\line(1,0){.5}}
     \put(1.4,0){\circle{.2}}
    \put(1.5,0){\line(1,0){.5}}
    \put(2.1,0){\circle{.2}}
    \put(2.2,0){\line(1,0){.5}}
    \put(2.8,0){\circle{.2}}
   \put(1.4,.1){\line(0,1){.5}}
   \put(1.4,.7){\circle{.2}}
   \put(-2.3,.3){$\widehat{E}_8:$}
   \put(-2.3,2.3){$\widehat{E}_7:$}
   \put(-2.3,4.7){$\widehat{E}_6:$}
  \end{picture}
}
\label{affE}
\ee

In the cases of affine $\widehat{A}_{k-1}$ and $\widehat{D}_{k}$, the
standard defining equations of the singularities read as follows 

\begin{equation}  \label{tablaff}
\begin{tabular}{|l|l|}
\hline
Affine singularity & Geometry near the orbifold point \\ \hline
\multicolumn{1}{|c|}{$\widehat{A}_{k-1}$} & $z_1^{2}+z_2^{3}+z_2^{2} =\zeta
^{k}, \qquad\ \ \ \ \ k\geq 2$ \\ \hline
\multicolumn{1}{|c|}{$\widehat{D}_{k}$} & $z_1^{2}+z_2^{3}+z_2^{2}\zeta
=\zeta ^{k-1}, \qquad k\geq 3$ \\ \hline
\end{tabular}
\end{equation}
\newline
They differ from the ordinary ones by the extra $z_2^{3}$ terms. For the
affine $\widehat{E}_{6}$, $\widehat{E}_{7}$ and $\widehat{E}_{8}$
geometries, the defining equations of singularities are conveniently
described by help of trivalent geometry based on elliptic fibrations over
the complex plane. In this language \cite{KMV}, the
mirrors of the blow ups of these exceptional affine singularities read

\begin{equation}  \label{affe}
\begin{tabular}{|l|l|}
\hline
Singularity & Mirror geometry \\ \hline
$\widehat{E}_{6}$ & $\left( z_1^{3}+z_2^{3}+z_{3}^{3}+z_1z_2z_{3}\right)
+av\left( z_1^{2}+z_2^{2}+z_{3}^{2}\right) $ \\
& $+bv^{2}\left( z_1+z_2+z_{3}\right) +cv^{3}$ \\ \hline
$\widehat{E}_{7}$ & $\left( z_1^{2}+z_2^{4}+z_{3}^{4}+az_1z_2z_{3}\right) $
\\
& $+\sum_{i=1}^{4}v^{i}\left( a_{i}z_2^{4-i}+b_{i}z_{3}^{4-i}\right)
+\sum_{i=1}^{2}c_{i}v^{i}z_1^{2-i}$ \\ \hline
$\widehat{E}_{8}$ & $\left( z_1^{2}+z_2^{3}+z_{3}^{6}+az_1z_2z_{3}\right) $
\\
& $+\sum_{i=1}^{6}a_{i}v^{i}z_{3}^{6-i}+
\sum_{i=1}^{3}b_{i}v^{i}z_2^{3-i}+\sum_{i=1}^2c_iv^iz_1^{2-i}$ \\ \hline
\end{tabular}
\end{equation}
\newline
where $a_{i}$, $b_{i}$ and $c_{i}$ are complex deformations of the
singularities. The monomials describing the mirror geometry are associated
with the nodes of the Dynkin diagrams of Lie algebras. Singular surfaces are
recovered by setting $a_{i}=b_{i}=c_{i}=0$ and describe elliptic K3
surfaces. Following \cite{KMV}, it is more interesting to think about affine
$\widehat{A}_{k-1}$ and $\widehat{D}_{k}$ singularities of (\ref{tablaff})
in the same manner as in (\ref{affe}). In this elliptic parameterization,
affine $\widehat{A}_{k-1}$ and $\widehat{D}_{k}$ singularities are realized
in terms of homogeneous polynomials as follows

\begin{equation}  \label{addad}
\begin{tabular}{|l|l|}
\hline
Elliptic singularity & Polynomial realization \\ \hline
$\widehat{A}_{2n}$ & $v\left( z_1^{2}+z_2^{3}
+\zeta^{6}+az_1z_2\zeta\right)+\zeta ^{2n+1}+z_1z_2^{n-1}$ \\ \hline
$\widehat{A}_{2n-1}$ & $v\left( z_1^{2}+z_2^{3} +\zeta^{6}+az_1z_2\zeta
\right) +\zeta ^{2n}+z_2^{n}$ \\ \hline
$\widehat{D}_{2n}$ & $\left( z_1^{2}+z_2^{3} +\zeta^{6}+az_1z_2\zeta \right)
+v^{2}\zeta ^{4n}$ \\
& $+v\left( z_1\zeta ^{2n}+\zeta ^{2n+3}+\zeta ^{4}z_1z_2^{n-2}
+\zeta^{3}z_2^{n}\right)$ \\ \hline
$\widehat{D}_{2n-1}$ & $\left( z_1^{2}+z_2^{3} +\zeta^{6}+az_1z_2\zeta
\right) +v^{2}\zeta ^{4n-2}$ \\
& $+v\left( z_1\zeta ^{2n-1}+\zeta ^{2n+2}+\zeta ^{3}z_1z_2^{n-2}
+\zeta^{4}z_2^{n}\right) $ \\ \hline
\end{tabular}%
\end{equation}
\newline
These equations have in general non-abelian discrete symmetries with
commutative cyclic subgroups. The latters are generated by phases 
$\omega_{i}$ and act on the coordinates $(z_1,z_2,\zeta ,v)$ as
\begin{equation}
(z_1,z_2,\zeta ,v)\rightarrow(\omega_{i}^{r_{1}}z_1,\omega_i^{r_{2}}z_2,
\omega_i^{r_{3}}\zeta ,\omega_i^{r_{4}}v).
\end{equation}
In this way, we have the following result

\begin{equation}
\begin{tabular}{|l|l|l|l|l|l|}
\hline
Singularity & $r_{1}$ & $r_{2}$ & $r_{3}$ & $r_{4}$ & $\omega _{i}$ \\ \hline
$\widehat{A}_{2n}$ & $3$ & $2$ & $1$ & $5-2n$ & 
$\exp\left(\frac{2\pi i}{2n+1}\right)$ \\ \hline
$\widehat{A}_{2n-1}$ & $3$ & $2$ & $1$ & $6-2n$ & 
$\exp\left(\frac{2\pi i}{2n}\right)$ \\ \hline
$\widehat{D}_{2n}$ & $3$ & $2$ & $1$ & $3-2n$ & 
$\exp\left(\frac{2\pi i}{4n}\right)$ \\ \hline
$\widehat{D}_{2n-1}$ & $3$ & $2$ & $1$ & $4-2n$ & 
$\exp\left(\frac{2\pi i}{4n-2}\right)$ \\ \hline
\end{tabular}.
\end{equation}
\newline
Degenerate metric building of affine $\widehat{ADE}$ pp-wave backgrounds
follows the outline in the case of the finite $A_k$ pp
wave studied above. 
Resolutions of these singularities lift this degeneracy and, as noted
before, this can be done either by K\"ahler or complex deformations. In what
follows, we will focus on the 
complex deformations of affine $\widehat{ADE}$. They are
classified in \cite{KMV}, see also \cite{BFS,BS1,BS2}.

\subsubsection{Mirror affine $\widehat{A}_{k}$ pp-wave metrics}

\quad\ Starting from (\ref{ppwavemetric}) and using the equation defining the
$\widehat{A}_{k}$ singularity in (\ref{addad}) for the case where $k$ is an
odd integer for instance, i.e., $k=2n-1$, one can solve the relation
\begin{equation}
z_1^{2}+az_1z_2\zeta +\left( z_2^{3}+\zeta ^{6}+ \frac{z_2^{n}
+\zeta ^{2n}}{v}\right) =0,
\end{equation}
and express $z_1$ in terms of the homogeneous variables $z_2$, $\zeta $ and 
$v$ of $\mathbb{WP}_{(3,2,1)}^2$ as
\begin{equation}  \label{az1}
2z_1=az_2\zeta \pm \sqrt{\left( az_2\zeta \right) ^{2} -4\left(z_2^{3}+\zeta
^{6}+\frac{z_2^{n}+\zeta ^{2n}}{v}\right) }.
\end{equation}
Likewise, the differential $dz_1$ reads in terms of $dz_2$ and $d\zeta $ as
follows
\begin{equation}
dz_1 =-\frac{nz_2^{n-1}+v\left( 3z_2+az_1\zeta \right)}{v(2z_1+az_2\zeta)}
dz_2 -\frac{2n\zeta^{n}+v\left(6\zeta^{5}+az_1z_2\right)}{v(2z_1+az_2\zeta) }d\zeta
+\frac{z_2^{n}+\zeta^{2n}}{v^2(2z_1+az_2\zeta)}dv.
\end{equation}
The metric of the pp-wave background near the orbifold point with affine 
$\widehat{A}_{k}$ singularity is then given by
\begin{eqnarray}
ds^{2}|_{\widehat{A}_{k}} &=&-4dx^{+}dx^{-}+d\mathbf{x}^2 
-\mu ^{2}\left(\mathbf{x}^{2}+\left|z_1\right| ^{2}+\left| z_2\right| ^{2}\right)
(dx^{+})^{2}  \notag \\
&&+\left( 1+\frac{\left| nz_2^{n-1}+v\left( 3z_2+az_1\zeta \right) \right|
^{2}}{\left| v\left( 2z_1+az_2\zeta \right) \right| ^{2}}\right) \left|
dz_2\right| ^{2}  \notag \\
&&+\frac{\left| 2n\zeta^n+v\left( 6\zeta ^{5}+az_1z_2\right) 
\right| ^{2}}{\left| v\left(2z_1+az_2\zeta \right) \right| ^{2}}\left| d\zeta \right| ^{2}
+\frac{\left|z_2^{n}+\zeta ^{2n}\right| ^{2}}{\left| v^{2}\left(
2z_1+az_2\zeta\right)\right|^2}\left|dv\right|^2  \notag \\
&&+\frac{1}{\left|v\left(2z_1+az_2\zeta \right)\right|^2} 
\{\left(nz_2^{n-1}+v(3z_2+az_1\zeta)\right)  \notag \\
&&\ \ \ \ \ \times\left(2n\overline{\zeta}^n
+\overline{v} \left(6\overline{\zeta}^5+a\overline{z}_1\overline{z}_2\right)\right) 
dz_2d\overline{\zeta }+h.c. \}  \notag \\
&&+\frac{1}{\left| v\left( 2z_1+az_2\zeta \right) \right| ^{2}}\left[ \left(
nz_2^{n-1}+v\left( 3z_2+az_1\zeta \right) \right) \frac{\overline{z}_{2}^{n}
+\overline{\zeta }^{2n}}{\overline{v}}dz_2d \overline{v}+h.c.\right]  \notag
\\
&&+\frac{1}{\left| v\left( 2z_1+az_2\zeta \right) \right| ^{2}}
\left[ \frac{z_2^{n}+\zeta ^{2n}}{v}\left( 2n\overline{\zeta }^{n} 
+\overline{v}\left(6\overline{\zeta}^5+a\overline{z}_1\overline{z}_2\right) \right) 
dvd\overline{\zeta}+h.c.\right],  \label{affasing}
\end{eqnarray}
where $z_1$ is as in (\ref{az1}), while $h.c.$ refers to the hermitian
conjugate. This pp-wave metric is degenerate at $z_1=z_2=\zeta =0$. However,
this degeneracy can be lifted by complex resolution by deforming 
(\ref{addad}) as follows
\begin{eqnarray}
\left[\widehat{A}_{k}\right]_{\text{complex def.}}:&&v\left(
z_1^{2}+z_2^{3}+\zeta ^{6}+az_1z_2\zeta \right) +\zeta
^{k+1}+\sum_{i=1}a_{i}z_2^{2i-1}\zeta^{k-2i-2}  \notag \\
&&+z_1\sum_{i=1}b_{i}z_2^{2i}\zeta ^{k-2i-1} 
+\left\{\begin{array}{lll}
z_1z_2^{\frac{k-2}{2}},\ \ \ \ \ \  & k\ \text{even} & z_2^{\frac{k+1}{2}},
\\
&  &  \\
z_2^{\frac{k+1}{2}}, & k\ \text{odd} &
\end{array}
\right.
\end{eqnarray}
where $a_{i}$ and $b_{i}$ are complex moduli. With this relation at hand,
one can go ahead and solve $z_1$ as a function of $z_2$ and $\zeta$ as in 
(\ref{az1}) but this time with non-zero $a_{i}$ and $b_{i}$, i.e.,
\begin{equation}
z_1( z_2,\zeta ,a_i,b_i).
\end{equation}
This extends the analysis we have developed for the case where 
$a_{i}=b_{i}=0 $. The relations are straightforwardly obtained and
similar to those we found before, albeit a bit lengthy, and since
they are not used explicitly below, they are omitted. In what follows we give a
general procedure to deal with the complex deformed affine geometries.

Before coming to the other kinds of pp-wave geometries, note that under the
complex deformation the degeneracy of the orbifold point with an affine 
$\widehat{A}_{k+1}$ singularity, embodied by the monomial $\zeta^{k+1}$, is
now lifted. The $(k+1)\times (k+1)$ matrix $\Pi$ of the characters
associated with the $\left\{\zeta^{k+1}, \tau _{i}, 0\leq i\leq k-1\right\}$
monomials of the $\widehat{A}_{k+1}$ geometry reads
\begin{equation}
\Pi =\mathrm{diag}(1,\omega ^{2},\omega ^{4},\ldots,\omega ^{2k-4},\omega
^{2k-2})
\end{equation}
where $\omega =\exp\left(\frac{2\pi i}{k+1}\right)$. These characters may be
derived from the properties of the deformation monomials $\tau _{i}$, which
read for $k$ even integer, $k=2n$, as
\begin{eqnarray}
\tau_{2i} &=&z_2^{i}\zeta ^{2n-2i+1},\quad\ \ \ \ \ \ \ i=0,\ldots,n  \notag
\\
\tau_{2i+1} &=&z_1z_2^{i-1}\zeta ^{2n-2i},\quad\ \ \ \ \ i=1,\ldots,n
\end{eqnarray}
and for $k=2n+1$
\begin{eqnarray}
\tau_{2i} &=&z_2^{i}\zeta ^{2n-2i+2},\ \ \ \ \ \ \ \ \ i=0,\ldots,n+1  \notag
\\
\tau_{2i+1} &=&z_1z_2^{i-1}\zeta ^{2n-2i+1},\ \ \ \ \ \ \ i=1,\ldots,n.
\end{eqnarray}
These monomials satisfy equations of the type (\ref{miro}), 
and are related to each other as
\begin{eqnarray}
\tau_{2i} &=&\frac{\zeta ^{2}}{z_2}\tau _{2i+2},\quad \tau _{2i+1}=\frac{
\zeta^{2}}{z_2}\tau _{2i+3},  \notag \\
\tau_{2i} &=&\frac{z_2\zeta }{z_1}\tau _{2i+1}.
\end{eqnarray}
In group-theory language, this link is described by an automorphism
generated by a $(k+1)\times(k+1)$ matrix $Q$ which in the present case reads
\begin{equation}
Q=\omega^2\left(
\begin{array}{cccccc}
0 & 1 &  &  &  &  \\
0 & 0 & 1 &  &  &  \\
& 0 & . & . &  &  \\
&  & . & . & . &  \\
&  &  & . & 0 & 1 \\
1 &  &  &  & 0 & 0
\end{array}
\right) .
\end{equation}
As the ring of monomials form is 
closed\footnote{In the ordinary $SU(k)$ geometry, the corresponding quiver 
diagram is not closed and so the periodicity condition $Q^{k+1}=Id$ of affine 
case is replaced by a nilpotency relation namely $Q^{k+1}=0$.}, the $\Pi $ and 
$Q$ matrices obviously obey $\Pi^{k+1}=Q^{k+1}=Id$, the $(k+1)\times(k+1)$
identity matrix. For the example of affine $\widehat{A}_{6}$ geometry, the 
$\Pi$ and $Q$ matrices read 

\begin{equation}
\Pi \left(\widehat{A}_{6}\right) =\left(
\begin{array}{ccccccc}
1 &  &  &  &  &  &  \\
& e^{i\frac{2\pi }{7}} &  &  &  &  &  \\
&  & e^{i\frac{4\pi }{7}} &  &  &  &  \\
&  &  & e^{i\frac{6\pi }{7}} &  &  &  \\
&  &  &  & e^{i\frac{8\pi }{7}} &  &  \\
&  &  &  &  & e^{i\frac{10\pi }{7}} &  \\
&  &  &  &  &  & e^{i\frac{12\pi }{7}}
\end{array}
\right) ,
\end{equation}
and

\begin{equation}
Q\left( \widehat{A}_{6}\right) =e^{i\frac{4\pi }{7}}\left(
\begin{array}{ccccccc}
0 & 1 &  &  &  &  &  \\
& 0 & 1 &  &  &  &  \\
&  & 0 & 1 &  &  &  \\
&  &  & 0 & 1 &  &  \\
&  &  &  & 0 & 1 &  \\
&  &  &  &  & 0 & 1 \\
1 &  &  &  &  &  & 0
\end{array}
\right)
\end{equation}
We will address these matrices again below.

\subsubsection{Affine $\widehat{D}_{k}$ pp-wave metrics}

\quad\ In the case where the orbifold point has an affine $\widehat{D}_{k}$
singularity, one may also determine the metric of the resulting pp-wave
background as follows. Since $z_1$ variable in (\ref{addad}) is at most
quadratic, it is more convenient to put the defining equations for the
affine $\widehat{D}_{k}$ singularity as follows
\begin{equation}  \label{quad}
z_1^{2}+2z_1f+g=0
\end{equation}
where the holomorphic complex functions $f$ and $g$ are given by
\begin{eqnarray}
f(z_2,\zeta ,v)&=&\frac{a}{2}z_2\zeta +v\left( \zeta ^{k}+\zeta ^{4-\epsilon
}z_2^{-2+\frac{k+\epsilon }{2}}\right),  \notag \\
g(z_2,\zeta ,v)&=&z_2^{3}+\zeta ^{6}+v\left( \zeta ^{k+3}+\zeta ^{3+\epsilon
}z_2^{\frac{k-\epsilon }{2}}\right).  \label{Dsing}
\end{eqnarray}
Using (\ref{quad}) $z_1$ can be solved in terms of the other variables and
the solutions are as usual given by
\begin{equation}
z_1=f\pm \sqrt{f^{2}-g}.
\end{equation}
In this way, one can then express $dz_1$ in terms of $dz_2$, $dv$ and 
$d\zeta $ (the holomorphic differentials of the homogeneous variables $z_2$, 
$v$ and $\zeta$) as follows
\begin{eqnarray}
dz_1 &=&\left(1\pm\frac{f}{\sqrt{f^{2}-g}}\right) df\ 
\mp\frac{1}{2\sqrt{f^{2}-g}}dg,  \notag \\
d\chi&=&\partial_{z_2}\chi dz_2+\partial_{\zeta}\chi d\zeta +\partial
_{v}\chi dv,
\end{eqnarray}
where $\chi$ stands for the functions $f$ and $g$. The metric of pp-wave
background near the orbifold point with affine $\widehat{D}_k$ singularity
preserving 16 supercharges is then given by
\begin{eqnarray}
ds^2|_{\widehat{D}_{k}} &=&-4dx^{+}dx^{-}+d\mathbf{x}^2 
-\mu^{2}\left(\mathbf{x}^{2}+\left| f\pm \sqrt{f^{2}-g}\right|^2 +\left|
z_2\right|^2\right)(dx^{+})^{2}  \notag \\
&&+\left[ 1+\left| \left( 1\pm \frac{f}{\sqrt{f^{2}-g}}\right)
\partial_{z_2}f\ \mp\frac{1}{2\sqrt{f^{2}-4g}} 
\partial_{z_2}g\right|^2\right] \left| dz_2\right|^2  \notag \\
&&+\left| \left(1\pm \frac{f}{\sqrt{f^{2}-g}}\right) \partial_{\zeta}f\ 
\mp\frac{1}{2\sqrt{f^{2}-g}} \partial_{\zeta}g\right| ^{2}\left|d\zeta \right|^2
\notag \\
&&+\left| \left( 1\pm \frac{f}{\sqrt{f^{2}-g}}\right) \partial_{v}f\ 
\mp\frac{1}{2\sqrt{f^{2}-g}} \partial_{v}g\right| ^{2}\left| dv\right|^2  \notag
\\
&&+\left[ \left( 1\pm \frac{f}{\sqrt{f^{2}-g}}\right) \partial _{z_2}f\ \mp
\frac{1}{2\sqrt{f^{2}-g}}\partial_{z_2}g\right]  \notag \\
&&\ \ \ \ \ \times\left[ \overline{\left( 1\pm \frac{f}{\sqrt{f^{2}-g}}\right) 
\partial_{\zeta }f\ \mp \frac{1}{2\sqrt{f^{2}-g}}\partial_{\zeta }g}\right] 
dz_2d\overline{\zeta }+h.c.  \notag \\
&&+\left[ \left( 1\pm \frac{f}{\sqrt{f^{2}-4g}}\right) \partial _{z_2}f\ \mp
\frac{1}{2\sqrt{f^{2}-g}}\partial_{z_2}g\right]  \notag \\
&&\ \ \ \ \ \times\left[ \overline{\left( 1\pm \frac{f}{\sqrt{f^{2}-4g}}\right) 
\partial_{v}f\text{ }\mp \frac{1}{2\sqrt{f^{2}-g}}\partial _{v}g}\right] 
dz_2d\overline{v}+h.c.  \notag \\
&&+\left[\left(1\pm \frac{f}{\sqrt{f^{2}-g}}\right) \partial_{v}f\ 
\mp \frac{1}{2\sqrt{f^{2}-g}}\partial _{v}g\right]  \notag \\
&&\ \ \ \ \ \times\left[ \overline{\left( 1\pm \frac{f}{\sqrt{f^{2}-g}}\right) 
\partial_{\zeta }f\ \mp\frac{1}{2\sqrt{f^{2}-g}}\partial _{\zeta }g}\right] 
dvd\overline{\zeta }+h.c..  \label{Dmetric}
\end{eqnarray}
As expected, this metric is degenerate at $z_2=\zeta =0$. Its
resolution can be obtained by performing deformations. Like for the 
$\widehat{A}_{k}$ case, complex deformations transform (\ref{Dsing}) to
\begin{eqnarray}
f(z_2,\zeta ,v)&=&\frac{a}{2}z_2\zeta+v\left(b_{1}\zeta^{k}
+c_{1}\zeta^{4-\epsilon}z_2^{-2+\frac{k+\epsilon }{2}}\right),  \notag \\
g(z_2,\zeta ,v)&=&z_2^{3}+\zeta ^{6}+v\left(b_{2}\zeta^{k+3}
+c_2\zeta^{3+\epsilon }z_2^{\frac{k-\epsilon }{2}}\right)
+v^{2}\sum_{i=1}^{k-1}a_{i}z_2^{i-1}\zeta ^{2(k+1-i)}.
\end{eqnarray}
In this case, the matrices $\Pi$ and $Q$ can be obtained from the properties
of the deformation monomials $\tau _i$ of the mirror singularity of 
$\widehat{D}_{k}$. Since these $\tau _{i}$'s satisfy the mirror geometry
constraint equations
\begin{equation}
\prod_{i}\tau_i^{\ell_{i}^{(a)}}=1,
\end{equation}
with $\ell_{i}^{(a)}$ essentially be given by minus the Cartan matrix 
of $\widehat{D}_{k}$. The matrices $\Pi $ and $Q$ read

\begin{equation}  \label{dcharac1}
\Pi \left( \widehat{D}_{k}\right) =\mathrm{diag}\left( \alpha _{1},\alpha
_{2},1,\omega ,\omega ^{2},\ldots,\omega ^{k-4},\alpha _{k},\alpha
_{k+1}\right) ,
\end{equation}
and

\begin{equation}  \label{dcharac2}
Q\left( \widehat{D}_{k}\right) =\left(
\begin{array}{cccccccccc}
0 & 0 & \alpha _{2} &  &  &  &  &  &  &  \\
& 0 & \alpha _{1} &  &  &  &  &  &  &  \\
&  & 0 & \omega &  &  &  &  &  &  \\
&  &  & 0 & \omega &  &  &  &  &  \\
&  &  &  & . & . &  &  &  &  \\
&  &  &  &  & . & . &  &  &  \\
&  &  &  &  &  & 0 & \omega &  &  \\
&  &  &  &  &  &  & 0 & \alpha _{k+1} & \alpha _{k} \\
&  &  &  &  &  &  &  & 0 & 0 \\
&  &  &  &  &  &  &  &  & 0
\end{array}
\right).
\end{equation}
The phases $\alpha _{i}$ appearing in these relations are constrained as
\begin{equation}
\alpha_q^2=1,\ \ \ \ \ \ \alpha_1\alpha_2=1,\ \ \ \ \ \
\alpha_k\alpha_{k+1}=1,\ \ \ \ \ \ \omega^{k-4}=1,
\end{equation}
together with either $\alpha_1=\alpha_k^{\ast}$ and 
$\alpha_{k+1}=\alpha_2^{\ast}$, or $\alpha_1=\alpha_{k+1}^{\ast}$ and 
$\alpha_k=\alpha_2^{\ast}$.

\subsubsection{Example}

\quad\ For the example of $\widehat{D}_{6}$ pp-wave geometries, the $\Pi$
and $Q$ matrices read as follows
\begin{equation}
\Pi \left( \widehat{D}_{6}\right) =\left(
\begin{array}{ccccccc}
\alpha _{1} &  &  &  &  &  &  \\
& \alpha _{2} &  &  &  &  &  \\
&  & 1 &  &  &  &  \\
&  &  & -1 &  &  &  \\
&  &  &  & 1 &  &  \\
&  &  &  &  & \alpha _{6} &  \\
&  &  &  &  &  & \alpha _{7}
\end{array}
\right)
\end{equation}
and
\begin{equation}
Q\left( \widehat{D}_{6}\right) =\left(
\begin{array}{ccccccc}
0 & 0 & \alpha _{2} &  &  &  &  \\
& 0 & \alpha _{1} &  &  &  &  \\
&  & 0 & -1 &  &  &  \\
&  &  & 0 & -1 &  &  \\
&  &  &  & 0 & \alpha _{7} & \alpha _{6} \\
&  &  &  &  & 0 & 0 \\
&  &  &  &  &  & 0
\end{array}
\right).
\end{equation}
Using the same approach as in (\ref{quad}), and following the same lines we
have used for the case of affine $\widehat{D}_{k}$, one can also work out
the pp-wave geometries and the matrices $\Pi $ and $Q$ for the exceptional
Lie algebras.

\subsection{$\mathcal{N}=2$ affine $\widehat{DE}$ CFT$_4$}

\subsubsection{$\mathcal{N}=2$ affine $\widehat{D}_{k}$ CFT$_4$}

\quad\ These $\mathcal{N}=2$ CFT$_4$ models are obtained by solving 
(\ref{betaaff}) for affine $\widehat{D}_{k}$ singularity (with $k>4$). Indeed, we
have the following
\begin{equation}
\left(
\begin{array}{cccccccc}
2 & 0 & -1 & 0 & 0 & 0 & 0 & 0 \\
0 & 2 & -1 & 0 & 0 & 0 & 0 & 0 \\
-1 & -1 & 2 & -1 & 0 & 0 & 0 & 0 \\
0 & 0 & -1 & 2 & -1 & 0 & 0 & 0 \\
\vdots & \vdots & \vdots & \vdots & \vdots & \vdots & \vdots & \vdots \\
0 & 0 & 0 & 0 & -1 & 2 & -1 & -1 \\
0 & 0 & 0 & 0 & 0 & -1 & 2 & 0 \\
0 & 0 & 0 & 0 & 0 & -1 & 0 & 2
\end{array}
\right) \left(
\begin{array}{c}
N_{1} \\
N_{2} \\
N_{3} \\
N_{4} \\
\vdots \\
N_{k-2} \\
N_{k-1} \\
N_{k}
\end{array}
\right)\ =\ \left(
\begin{array}{c}
0 \\
0 \\
0 \\
0 \\
\vdots  \\
0  \\
0\\
0
\end{array}
\right),
\end{equation}
which is equivalent to
\begin{eqnarray}
2N_{1}-N_{3} &=&0,\ \ \ \ \ \quad 2N_{2}-N_{3}=0,  \label{afd1} \\
-N_{1}-N_{2}+2N_{3}-N_{4} &=&0,  \label{afd2} \\
-N_{i-1}+2N_{i}-N_{i+1} &=&0,\quad\ \ \ \ \ \ 3\leq i\leq k-3  \label{afd3}
\\
-N_{k-3}+2N_{k-2}-N_{k-1}-N_{k} &=&0,  \label{afd4} \\
-N_{k-2}+2N_{k-1} &=&0,\ \ \ \ \ \quad -N_{k-2}+2N_{k}=0.  \label{afd5}
\end{eqnarray}
These equations are easily solved. Indeed, (\ref{afd1}) tells us that 
$N_{1}=N_{2}=N$ and $N_{3}=2N$, while (\ref{afd2}) and (\ref{afd3}) reveal
that $N_{3}=N_{4}=\ldots=N_{k-2}=2N$. The two last relations require that 
$N_{k-1}=N_{k}=N$. Therefore the resulting gauge group, $G$, of the 
$\mathcal{N}=2$ SCFT$_4$ reads
\begin{equation}
G_{\widehat{D}_{k}}=U(N)^{4}\otimes U(2N)^{k-3}.  \label{ggrd}
\end{equation}
Bi-fundamental matter is engineered by the links of the affine 
$\widehat{D}_{k}$ Dynkin diagram as shown in Figure (\ref{ordE6}). 
The scalar matter content of this quiver theory is
\begin{equation}
\left(
\begin{array}{c}
Q_{1,\overline{3}} \\
\widetilde{Q}_{1,\overline{3}}
\end{array}
\right) ,\left(
\begin{array}{c}
Q_{2,\overline{3}} \\
\widetilde{Q}_{2,\overline{3}}
\end{array}
\right) ,\left(
\begin{array}{c}
Q_{3,\overline{4}} \\
\widetilde{Q}_{3,\overline{4}}
\end{array}
\right) ,\ldots,\left(
\begin{array}{c}
Q_{k-3,\overline{(k-2)}} \\
\widetilde{Q}_{k-3,\overline{(k-2)}}
\end{array}
\right) ,\left(
\begin{array}{c}
Q_{k-2,\overline{(k-1)}} \\
\widetilde{Q}_{k-2,\overline{(k-1)}}
\end{array}
\right) ,\left(
\begin{array}{c}
Q_{k-2,\overline{k}} \\
\widetilde{Q}_{k-2,\overline{k}}
\end{array}
\right) .  \label{bifd}
\end{equation}
Like in the previous example, the $Q_{i,\overline{j}}$ and 
$\widetilde{Q}_{i,\overline{j}}$ fields are just complex moduli of the 
deformation of affine $\widehat{D}_{k}$ singularity of the pp-wave geometry.

\subsubsection{$\mathcal{N}=2$ affine $\widehat{E}_{s}$ CFT$_4$}

\quad\ Similarly, one can show that the gauge symmetries of the remaining
$\mathcal{N}=2$ CFT$_4$ models associated with $\widehat{E}_{s}$ 
($s=6,7,8$) orbifolds are given by
\begin{eqnarray}
G_{\widehat{E}_{6}} &=&U(N)^3\otimes U(2N)^3\otimes U(3N)  \notag \\
G_{\widehat{E}_{7}} &=& U(N)^2\otimes U(2N)^{3}\otimes U(3N)^{2}\otimes U(4N)
\notag \\
G_{\widehat{E}_{8}} &=&U(N) \otimes U(2N)^{2}\otimes U(3N)^{2}\otimes
U(4N)^{2}\otimes U(5N)\otimes U(6N).
\end{eqnarray}
Bi-fundamental matter is associated with the links of the affine 
$\widehat{E}_{s}$ Dynkin diagrams (\ref{affE}).

\section{$ADE$ CFT$_{4}$s and toric geometry}

\subsection{Interpretation of conformal invariance in toric geometry}

\quad\ The vanishing condition for the beta function dictated by the requirement
of conformal invariance of the $\mathcal{N}=2$ quiver gauge theory 
translates into a nice Lie algebraic condition \cite{ABS1,ABS2}. 
This condition appears in the
standard classification theorem of Kac-Moody 
algebras. For the special class of $\mathcal{N}=2$ affine $\widehat{A}_{k}$
CFT$_4$ with gauge group $G=\left[\otimes_{i=0}^{k}U(N_i)\right]$, the
above-mentioned condition reads
\begin{equation}  \label{aconfcond}
N_{i-1}-2N_{i}+N_{i+1}=0,\qquad 1\leq i\leq k,
\end{equation}
where $N_{i}$ is a positive integer for all $i$. 
For affine $\widehat{DE}$, the 
equations are slightly more complicated. In the case of $\widehat{D}_k$, for example,
we still have conditions of the type (\ref{aconfcond}), but also
\begin{equation}
N_{k-3}-2N_{k-2}+N_{k-1}+N_{k}=0.
\end{equation}
According to the classification theorem of
Lie algebras, the integers $N_{j}$ are proportional to the Dynkin weights 
${w}_{i}$, that is, $N_{i}=Nw_{i}$ where $N$ is a fixed positive integer. 
To indicate the ideas, we will focus on affine $\widehat{A}_{k}$ 
CFT$_4 $s and give the results for the general case. To that purpose note that
the above relations have a remarkable interpretation in toric geometry 
\cite{KMV,BFS,BS1,BS2}. They may be thought of as a toric geometry relation
of the form
\begin{equation}  \label{ver}
\sum_{j}\ell_{j}^{(a)}\mathbf{v}_{j}=0,
\end{equation}
where the $\mathbf{v}_{j}$ vertices are given by
\begin{equation}
\mathbf{v}_{j}=(1,N_j,n_j).
\end{equation}
As before, the $N_{j}$s are gauge-group orders, while the $n_{j}$s are
some integers with no major interest here and will therefore be ignored. 
The $\ell_{j}^{(a)}$s are given by minus the Cartan matrix, 
i.e., $\ell_{j}^{(a)}=\delta _{j+1}^{a}-2\delta _{j}^{a}+\delta _{j-1}^{a}$. 
The $\ell_{j}^{(a)}$s are integers satisfying the following toric geometry property
\begin{equation}  \label{conscond}
\sum_{j}\ell_{j}^{(a)}=0.
\end{equation}
This constraint, which defines the local Calabi-Yau condition in
toric geometry, is already contained in the toric equation (\ref{ver}) as
the projection on the first component of the vertices $\mathbf{v}$. In toric
geometry, (\ref{ver}) gives the links between the vertices.
For the resolved affine $\widehat{A}_{k}$ singularity, for instance, a given
vertex $\mathbf{v}_{a}$ is schematically represented by the usual 
$\widehat{A}_{k}$ Dynkin diagram
\begin{equation}
\ldots\bullet _{\mathbf{v}_{a-1}}\longleftrightarrow 
\bullet _{\mathbf{v}_{a}}\longleftrightarrow \bullet _{\mathbf{v}_{a+1}}
\end{equation}
Observe in passing that in this $\widehat{A}_{k}$ case, the non-zero 
$\mathbf{\ell}^{(a)}$ describing the links between the $\mathbf{v}_{j}$ 
vertices take the form $(1,-2,1)$. The $\mathcal{N}=2$ conformal 
invariance of affine $\widehat{A}_{k}$ CFT$_4$ is thereby encoded in the 
affine $\widehat{A}_{k}$ Dynkin
diagram. The total $\sum_{j}p_{j}^{a}$ should vanish and the same for 
$\sum_{j}p_{j}^{a}\mathbf{v}_{j}$ which include $\sum_{j}p_{j}^{a}N_{j}$ as a
projection on the second vector basis. 

This analysis may be extended to
the case of affine $\widehat{DE}$ CFT, but here one should allow also
vertices with more than two links. This kind of toric geometry
representation has been studied in details in \cite{KMV,BFS}. The geometry
is given by the following typical vertex
\begin{eqnarray}
1 &\searrow &\qquad \swarrow 1  \notag \\
&&\bullet \text{ }-2  \label{4vert} \\
1 &\nearrow &\text{ }\quad \nwarrow -1  \notag
\end{eqnarray}
with a vector charge
\begin{equation}  \label{4charg}
(1,-2,1,1,-1)
\end{equation}
satisfying the local Calabi-Yau condition (\ref{conscond}). In terms of this
vertex, the conformal condition may be thought of as a particular situation of
the following general relation
\begin{equation}  \label{nconfeq}
N_{i-1}-2N_{i}+N_{i+1}+(J_i-L_i)=0,
\end{equation}
which can also be rewritten as
\begin{equation}  \label{gen}
K_{ij}N_{j}=J_i-L_i.
\end{equation}
Before going ahead, we note the two following features regarding the above
relation: (i) As far as supersymmetric affine $\widehat{DE}$ CFT$_4$s are
concerned, one can interpret the constraint equations required by
conformal invariance as 
toric geometry equations involving (\ref{4vert}-\ref{4charg}) as for
the affine $\widehat{A}_{k}$ CFT$_4$s. In the case of affine 
$\widehat{D}_{k}$ CFT$_4$s, the corresponding 
constraint equations, in particular (\ref{afd2}) and (\ref{afd4}) associated 
with the two trivalent vertices $\mathbf{v}_{3}$ and $\mathbf{v}_{k-2}$ 
of the $\widehat{D}_{k}$ Dynkin diagram
\begin{eqnarray}
-N_{1}-N_{2}+2N_{3}-N_{4} &=&0,  \notag \\
-N_{k-3}+2N_{k-2}-N_{k-1}-N_{k} &=&0,
\end{eqnarray}
merely correspond to a special configuration of (\ref{nconfeq}) where we have
\begin{eqnarray}
&&J_{1} =N_{1}, \qquad \ \ \ \ \ \ J_{k}=N_{k},  \notag \\
&&L_{1} =L_{k}=0,\qquad\ 1\leq i\leq k.
\end{eqnarray}
For the other bivalent vertices, we simply have $J_{i}=L_{i}=0$ for $2\leq
i\leq k-1$. 

In fact this picture is a particular case of a more general
representation extending beyond the $\widehat{DE}$ Dynkin diagrams as
it concerns the Dynkin diagrams of general Lie algebras. In the general case,
the numbers $J_{i}$
and $L_{i}$ are interpreted as associated with fundamental matter and so one
ends up with a larger class of $\mathcal{N}=2$ gauge theories with fundamental
matter. For example, the affine $\widehat{D}_{k}$ CFT$_4$ quiver gauge theory
with gauge group (\ref{ggrd}) can, under the assumption that the YM gauge
couplings $g_1$ and $g_k$ associated with the first and $k$-th node of the 
$\widehat{D}_k$ Dynkin diagram tend to zero, be thought of as an ordinary 
$A_{k-2}$ CFT$_4$ quiver gauge theory with group symmetry 
$G_{g}\otimes G_{f}$ given by
\begin{eqnarray}
G_{g} &=&U(N)^2\otimes U(2N)^{k-3}  \notag \\
G_{f} &=&U(N)^2.  \label{agr}
\end{eqnarray}
It follows that (\ref{gen}) actually
describes a more general class of $\mathcal{N}=2$ conformal quiver
gauge theories. The number $J_{i}$ is now just the number of fundamental
matter one engineers on the nodes based on Dynkin diagrams of finite-dimensional
Lie algebras. For conformal field theories with both $J_{i}$ and $L_{i}$
fundamental matter, we refer to \cite{ABS1,ABS2}. In what follows, we give the
results for the ordinary Lie algebras.

\subsection{$\mathcal{N}=2$ $D_{k}$ CFT$_{4}$}

\quad\ To get the category of $\mathcal{N}=2$ $D_{k}$ CFT$_4$ based on
finite $D_{k}$ Dynkin diagrams, one may mimic our approach to
the supersymmetric $A_{k-1}$ CFT$_4$. Taking the quiver group symmetry
as $G_{g}\otimes G_{f}$ with gauge group
$G_{g}=\left[\otimes_{i=1}^{k}U(N_i)\right]$ and flavour group 
$G_{f}=\left[\otimes_{i=1}^{k}U(M_{i})\right]$, where each group factor 
$U(N_i)\otimes U(M_{i})$ is engineered over the $i$-th node of the finite 
$D_{k}$ Dynkin diagram as in (\ref{4vert}-\ref{4charg}), then scale 
invariance requires the following conditions to be satisfied
\begin{eqnarray}
M_{1} &=&2N_{1}-N_{2},  \notag \\
M_{i} &=&2N_{i}-(N_{i-1}+N_{i+1}),\ \ \ \ \qquad 2\leq i\leq k-2  \notag \\
M_{k} &=&2N_{k}-N_{k-2},
\end{eqnarray}
where $N_i,M_i\geq1$. The group symmetry and the matter content for the 
$D_{k}$ category of $\mathcal{N}=2$ CFT$_4$s are collected in the following
table:

\begin{equation}
\begin{tabular}{|l|l|}
\hline
Gauge group $G_g$: & $\left[\otimes_{i=1}^{k}U(N_i)\right]$ \\ \hline
Flavour group $G_f$: & $U(2N_1-N_2)\otimes U(2N_{k}-N_{k-2})$ \\
& $\otimes \left[\otimes_{i=2}^{k-2}U(2N_{i}-N_{i-1}-N_{i+1})\right]$ \\
\hline
Bi-fundamental matter: & $\oplus _{i=1}^{k-2}\left(N_i,
\overline{N}_{i+1}\right) \oplus \left(N_{k-2},\overline{N}_{k}\right) $ \\ \hline
Fundamental matter: & $\oplus _{i=1}^{k}(M_iN_i)$ \\ \hline
\end{tabular}
\end{equation}
\newline
One may recover other $\mathcal{N}=2$ conformal models by
considering the vanishing limits of some of the gauge-coupling constants.

\subsection{$\mathcal{N}=2$ $E_{6}$ CFT$_{4}$}

\quad\ For a group symmetry $G_{g}\otimes G_{f}$, with quiver gauge group 
$G=\left[\otimes_{i=1}^{6}U(N_{i})\right]$ and flavour group 
$G_{f}=\left[\otimes_{i=1}^{6}U(M_{i})\right]$, scale invariance 
requires the following relations to hold
\begin{eqnarray}
M_1&=&2N_1-N_2,  \notag \\
M_2&=&2N_2-(N_3+N_1),  \notag \\
M_3&=&2N_3-(N_2+N_4+N_6),  \notag \\
M_4&=&2N_4-\left(N_3+N_5\right),  \notag \\
M_5&=&2N_5-N_4,  \notag \\
M_6&=&2N_6-N_3,
\end{eqnarray}
where $N_i,M_i\geq1$. The symmetries in $\mathcal{N}=2$ $E_{6}$ CFT$_4$s are
as follows

\begin{equation}
\begin{tabular}{|l|l|}
\hline
Gauge group $G_{g}$: & $\left[ \otimes _{i=1}^{6}U(N_{i})\right] $ \\ \hline
Flavour group $G_{f}$: & $U(2N_{1}-N_{2})\otimes U(2N_{2}-N_{3}-N_{1})$ \\
& $\otimes \ U(2N_{3}-N_{2}-N_{4}-N_{6})\otimes U(2N_{4}-N_{3}-N_{5})$ \\
& $\otimes \ U(2N_{5}-N_{4})\otimes U(2N_{6}-N_{3})$ \\ \hline
Bi-fundamental matter: & $\left( N_{1},\overline{N}_{2}\right) \oplus 
\left(N_{2},\overline{N}_{3}\right) \oplus \left( N_{3},\overline{N}_{4}\right)$
\\
& $\oplus \left( N_{3},\overline{N}_{6}\right) \oplus \left( N_{4},
\overline{N}_{5}\right) $ \\ \hline
Fundamental matter: & $\oplus _{i=1,i\neq 3}^{6}\left( M_{i}N_{i}\right)$
\\ \hline
\end{tabular}
\end{equation}
\newline
This result is easily extended to the other exceptional
groups, $E_7$ and $E_8$.


\begin{thebibliography}{99}

\bibitem{M1} J. Maldacena, ``The large N limit of conformal field theories
and supergravity'', Adv. Theor. Phys. \textbf{2} (1998) 231, hep-th/9711200.

\bibitem{GKP} S.S. Gubser, I.R. Klebanov, A.M. Ployakov, ``Gauge theory
correlators from non-critical string theory'', Phys. Lett. \textbf{B 428}
(1998) 105, hep-th/9802109.

\bibitem{W1} E. Witten, ``Anti de Sitter space and holography'', Adv. Theor.
Math. Phys. \textbf{2} (1998) 253, hep-th/9802150.

\bibitem{Pet} J.L. Petersen, ``Introduction to the Maldacena conjecture on
AdS/CFT'', Int. J. Mod. Phys. \textbf{A 14} (1999) 3597, hep-th/9902131.

\bibitem{KW} I.R. Klebanov, E. Witten, ``AdS/CFT Correspondence and Symmetry
Breaking'', Nucl. Phys. \textbf{B 556} (1999) 89, hep-th/9905104.

\bibitem{AGMOO} O. Aharony, S.S Gubser, J. Maldacena, H. Oogori, Y. Oz,
``Large N field theories, string theory and gravity'', Phys. Rept. \textbf{%
323}, 183 (2000), hep-th/9905111.

\bibitem{Ve} P. Di Vecchia, ``Large N gauge theories and AdS/CFT
correspondence'', hep-th/9908148.

\bibitem{K} I.R. Klebanov, ``TASI Lectures: Introduction to the AdS/CFT
Correspondence'', hep-th/0009139.

\bibitem{M2} J.M. Maldacena, ``TASI 2003 Lectures on AdS/CFT'',
hep-th/0309246.

\bibitem{BMN} D. Berenstein, J. Maldacena, H. Nastase, ``Strings in flat
space and pp waves from $\mathcal{N}=4$ Super Yang Mills'', JHEP \textbf{0204%
} (2002) 013, hep-th/0202021.

\bibitem{BFHP1} M. Blau, J. Figueroa-O'Farill, C. Hull, G. Papadopoulos,
``Penrose limits and maximal supersymmetry'', Class. Quant. Grav. \textbf{19}
(2002) L87, hep-th/0201081.

\bibitem{BFHP2} M. Blau, J. Figueroa-O'Farill, C. Hull, G. Papadopoulos, ``A
new maximally supersymmetric background of type IIB superstring theory'',
JHEP \textbf{0201} (2001) 047, hep-th/0110242.

\bibitem{IKM} N. Itzhaki, I.R. Klebanov, S. Mukhi, ``PP Wave Limit and
Enhanced Supersymmetry in Gauge Theories'', JHEP \textbf{0203} (2002) 048,
hep-th/0202153.

\bibitem{GO} J. Gomis, H. Ooguri, ``Penrose Limit of N=1 Gauge Theories'',
Nucl. Phys. \textbf{B 635} (2002) 106, hep-th/0202157.

\bibitem{RT} J.G. Russo, A.A. Tseytlin, ``On solvable models of type IIB
superstring in NS-NS and R-R plane wave backgrounds'', JHEP \textbf{0204}
(2002) 021, hep-th/0202179.

\bibitem{CLP} M. Cvetic, H. Lu, C.N. Pope, ``Penrose Limits, PP-Waves and
Deformed M2-branes'', Phys. Rev. \textbf{D 69} (2004) 046003,
hep-th/0203082; ``M-theory PP-waves, Penrose Limits and Supernumerary
Supersymmetries'', Nucl. Phys. \textbf{B 644} (2002) 65, hep-th/0203229.

\bibitem{GNS} U. Gursoy, C. Nunez, M. Schvellinger, ``RG flows from Spin(7),
CY 4-fold and HK manifolds to AdS, Penrose limits and pp waves'', JHEP
\textbf{0206} (2002) 015, hep-th/0203124.

\bibitem{DGR} S.R. Das, C. Gomez, S.-J. Rey, ``Penrose limit, Spontaneous
Symmetry Breaking and Holography in PP-Wave Background'', Phys. Rev. \textbf{%
D 66} (2002) 046002, hep-th/0203164.

\bibitem{DP} A. Dabholkar, S. Parvizi, ``Dp Branes in PP-wave Background'',
Nucl. Phys. \textbf{B 641} (2002) 223, hep-th/0203231.

\bibitem{S} E.M. Sahraoui, ``Superstring Theory on AdS Spaces in The Penrose
Limit'', African Jour. Math. Phys. \textbf{1} (2004) 53.

\bibitem{HS} M. Hssaini, M.B. Sedra, ``Type IIB String Backgrounds on
Parallelizable PP-Waves and Conformal Liouville Theory'', African Jour.
Math. Phys. \textbf{1} (2004) 91.

\bibitem{KPRT} N. Kim, A. Pankiewicz, S.-J. Rey, S. Theisen, ``Superstring
on pp-wave orbifold from large-N quiver gauge theory'', Eur. Phys. J.
\textbf{C 25} (2002) 327, hep-th/0203080.

\bibitem{AFM} O. Aharony, A. Fayyazuddin, J. Maldacena, ``The large L limit
of N=1,2 from three-banes in F-theory'', JHEP \textbf{9807} (1998) 013,
hep-th/9807159.

\bibitem{BGMNN} D. Berenstein, E. Gava, J. Maldacena, K.S. Narain, H.
Nastase, ``Open strings on plane waves and their Yang-Mills duals'',
hep-th/0203249.

\bibitem{AS} M. Alishahiha, M.M. Sheikh-Jabbari, ``The pp wave limits of
orbifolded $AdS_{5}\times S^{5}$'', Phys. Lett. \textbf{B 535} (2002) 328,
hep-th/0203018.

\bibitem{SS} E.M. Sahraoui, E.H. Saidi, ``Metric Building of pp Wave
Orbifold Geometries'', Phys. Lett. \textbf{B 558} (2003) 221, hep-th/0210168.

\bibitem{work} Work in progress.

\bibitem{LNV} A. Lawrence, N. Nekrasov, C. Vafa, ``On conformal theories in
4 dimension'', Nucl. Phys. \textbf{B 533} (1998) 199, hep-th/9803015.

\bibitem{KS} S. Kachru, E. Silverstein, ``4d conformal field theories and
strings on orbifolds'', Phys. Rev. Lett. \textbf{80} (1998) 4855,
hep-th/9802183.

\bibitem{FHHP1} B. Feng, A. Hanany, Y.-H. He, N. Prezas, ``Discrete Torsion,
Covering Groups and Quiver Diagrams'', JHEP \textbf{0104} (2001) 037,
hep-th/0011192.

\bibitem{FHHP2} B. Feng, A. Hanany, Y.-H. He, N. Prezas, ``Discrete Torsion,
non-Abelian Orbifolds and the Schur Multiplier'', JHEP \textbf{0101} (2001)
033, hep-th/0010023.

\bibitem{KMV} S. Katz, P. Mayr, C. Vafa, ``Mirror symmetry and exact
solution of 4d N=2 gauge theories I'', Adv. Theor. Math. Phys. \textbf{1}
(1998) 53, hep-th/9706110.

\bibitem{BFS} A. Belhaj, A E. Fallah, E.H. Saidi, ``On the non-simply laced
mirror geometries in type II strings'', Class. Quant. Grav. \textbf{17}
(2000) 515.

\bibitem{BS1} A. Belhaj, E.H. Saidi, ``Toric geometry, enhanced non simply
laced gauge symmetries in superstrings and F-theory compactifications'',
hep-th/0012131.

\bibitem{BS2} A. Belhaj, E.H. Saidi, ``Non Simply Laced Quiver gauge
Theories in Superstrings Compactifications'', African Jour. Math. Phys
\textbf{1} (2004) 29.

\bibitem{ABS1} M. Ait Ben  Haddou, A. Belhaj, E.H. Saidi, ``Classification of N=2
supersymmetric CFT$_4$s: Indefinite Series'', J. Phys. {\bf A 38} (2005) 1793,  
 hep-th/0308005.

\bibitem{ABS2} M. Ait Ben Haddou, A. Belhaj, E.H. Saidi, ``Geometric Engineering
of N=2 CFT$_4$s based on Indefinite Singularities: Hyperbolic Case'', Nucl.
Phys. \textbf{B 674} (2003) 593, hep-th/0307244.

\bibitem{KKV} S. Katz, A. Klemm, C. Vafa, ``Geometric Engineering of Quantum
Field Theories'', Nucl. Phys. \textbf{B 497} (1997) 173, hep-th/9609239.

\bibitem{HOV} K. Hori, H. Ooguri, C. Vafa, ``Non-Abelian Conifold
Transitions and N=4 Dualities in Three Dimensions'', Nucl. Phys. \textbf{B
504} (1997) 147, hep-th/9705220.

\bibitem{Kac} V.G. Kac, \textquotedblleft Infinite dimensional Lie
algebras\textquotedblright, 3rd edition (Cambridge University Press, 1990).

\bibitem{malika}  M. Ait Ben Haddou, E.H. Saidi, ``Hyperbolic
Invariance'', hep-th/0405251.

\bibitem{iran} E.H. Saidi, ``Hyperbolic Invariance in type II Superstrings'',
hep-th/0502176.

 \bibitem{AABDS}  
  R. Ahl Laamara, M. Ait Ben Haddou, A. Belhaj, L.B. Drissi,
 E.H. Saidi, ``RG Cascades in Hyperbolic Quiver Gauge Theories'',
    Nucl. Phys. \textbf{B 702} (2004) 163, hep-th/0405222.
    
\end{thebibliography}
\end{document}